\documentclass{article}
\usepackage[T1]{fontenc}
\usepackage[latin9]{inputenc}
\usepackage[a4paper]{geometry}
\usepackage{fancyhdr}
\usepackage{color}
\usepackage{array}
\usepackage{float}
\usepackage{multirow}
\usepackage{amsmath}
\usepackage{amsthm}
\usepackage{amssymb}
\usepackage{stmaryrd}
\usepackage{graphicx}
\usepackage[numbers]{natbib}
\usepackage[unicode=true,
 bookmarks=true,bookmarksnumbered=true,bookmarksopen=true,bookmarksopenlevel=2,
 breaklinks=false,pdfborder={0 0 0},pdfborderstyle={},backref=section,colorlinks=true]
 {hyperref}
\hypersetup{pdftitle={Introducing Geometric Algebra to Geometric Computing Software Developers: A Computational Thinking Approach},
 pdfauthor={Ahmad Hosny Eid},
 pdfsubject={Geometric Algebra},
 pdfkeywords={Computational Thinking, Geometric Algebra, Geometric Computing, Software Engineering}}

\makeatletter

\providecommand{\tabularnewline}{\\}
\floatstyle{ruled}
\newfloat{algorithm}{tbp}{loa}
\providecommand{\algorithmname}{Algorithm}
\floatname{algorithm}{\protect\algorithmname}

\numberwithin{equation}{section}
\numberwithin{figure}{section}
\newcommand{\lyxaddress}[1]{
\par {\raggedright #1
\vspace{1.4em}
\noindent\par}
}
\theoremstyle{plain}
\newtheorem{thm}{\protect\theoremname}
  \theoremstyle{definition}
  \newtheorem{defn}[thm]{\protect\definitionname}

\usepackage{textcomp}

\makeatother

  \providecommand{\definitionname}{Definition}
\providecommand{\theoremname}{Theorem}

\begin{document}

\title{Introducing Geometric Algebra to Geometric Computing Software Developers:
A Computational Thinking Approach}

\author{Ahmad Hosny Eid}
\maketitle

\lyxaddress{Department of Electrical Engineering,\\
Faculty of Engineering,\\
Port-Said University,\\
Egypt}
\begin{abstract}
Designing software systems for Geometric Computing applications can
be a challenging task. Software engineers typically use software abstractions
to hide and manage the high complexity of such systems. Without the
presence of a unifying algebraic system to describe geometric models,
the use of software abstractions alone can result in many design and
maintenance problems. Geometric Algebra (GA) can be a universal abstract
algebraic language for software engineering geometric computing applications.
Few sources, however, provide enough information about GA-based software
implementations targeting the software engineering community. In particular,
successfully introducing GA to software engineers requires quite different
approaches from introducing GA to mathematicians or physicists. This
article provides a high-level introduction to the abstract concepts
and algebraic representations behind the elegant GA mathematical structure.
The article focuses on the conceptual and representational abstraction
levels behind GA mathematics with sufficient references for more details.
In addition, the article strongly recommends applying the methods
of Computational Thinking in both introducing GA to software engineers,
and in using GA as a mathematical language for developing Geometric
Computing software systems. 
\end{abstract}
Keywords: Computational Thinking, Geometric Algebra, Geometric Computing,
Software Engineering

\section{Introduction}

Geometric Algebra (GA) is an expressive algebraic framework capable
of unifying many mathematical tools that engineers and scientists
use to model their ideas \citep{Hestenes.1984,Hestenes.2002,Dorst.2009}.
GA can be used for unified algebraic representation and manipulation
of multidimensional Euclidean and non-Euclidean geometries in a consistent
manner \citep{Sommer.2010,Sobczyk.2004,9780521715959,Gunn.2016}.
Many good sources exist that explain the mathematics behind GA and
explore some of its possible applications \citep{Nilsson.2002,Pozo.2002,Dorst.2009,Perwass.2009,Lundholm.20090730,Dorst.2010,Dorst.2011,1453854932,Chisolm.20120527,Hitzer_2013,Chappell.2014,Kanatani.2015,Macdonald.2017}.
These sources vary in their scope, intended audience, goals, level
of details, and mathematical rigor. Few sources investigate the concepts,
options, and issues software engineers need to understand and study
when designing practical GA-based software systems for Geometric Computing
applications \citep{Nilsson.2002,Zaharia.2002b,Fontijne.2007,Dorst.2009,Hildenbrand.2013,Ab_amowicz_2014,Luo.2016,Hildenbrand_2016,Breuils.2017,Benger_2017}.
This led to less attention given to GA-based models simply because
software engineers don't have enough GA material targeting their domain
of knowledge. The software engineering domain has quite different
thought process characteristics from that of non-software oriented
engineers, mathematicians, and physicists typically producing the
GA models. Without sufficient attention from the developers of Geometric
Computing software implementations, many of the good GA models would
be trapped inside the limited academic circle of the GA community. 

\subsection{Geometric Algebra and Geometric Computing}

In many areas of computer science, engineering, mathematics, physics,
biology, and chemistry we find common geometric ideas defining, relating,
and manipulating objects in space and time. In addition, there is
a prevalent use of modern computing environments to implement geometric
algorithms and to process geometric information \citep{9781848001152}.
Many researchers informally use the term ``Geometric Computing''
(GC) to express this intersection between classical geometry and modern
computation. To the best of my knowledge there is no solid definition
of this term in modern literature. Some researchers even use the term
Geometric Computing to actually refer to Computational Geometry \citep{9780521563291,9789810218768},
which is just one application area that requires GC. As an attempt
to make the meaning of this term clear as I understand and use it
in this work, I will adopt the following definition, which is a modification
of the term ``Computing'' in the 1989 ACM report on \textquotedbl{}Computing
as a Discipline\textquotedbl{} \citep{Denning.1989}:
\begin{defn}
The discipline of \textbf{Geometric Computing} is the systematic study
of algorithmic processes that describe and transform geometric information:
their theory, analysis, design, efficiency, implementation, and application.
The fundamental question underlying all geometric computing is \textquotedblleft What
(and how) geometric processes can be efficiently automated?\textquotedblright{}
\end{defn}
An essential ingredient in creating GC applications is the use of
symbolic algebraic tools, in the mathematical sense, to express and
manipulate abstract geometric objects, spaces, and processes. Many
such tools exist from diverse areas of mathematics; for example matrix
algebra, 3D vector algebra, quaternions, complex numbers, several
kinds of hyper-complex numbers, and many more. The use of so many
conceptually and computationally incompatible algebraic tools to express
geometric ideas results in various problems. Such problems manifest
in multiple levels and forms including: 
\begin{itemize}
\item The difficulty of expressing geometrically intuitive ideas in an algebraically
consistent manner. 
\item The need to learn many distinct algebraic representations in order
to model the geometry of relatively complex problems.
\item The need for many conversions between algebraic frameworks within
the context of the same problem domain. 
\item The awkward isolation of people working in areas of research that
essentially depend on the same set of geometric ideas primarily because
such groups tend to use isolated algebraic frameworks. 
\end{itemize}
The prevalent state in developing GC applications is to rely on software
abstractions \citep{Jackson.2012} to unify the interface between
the users and the GC software infrastructure. For example, in a typical
GC software implementation the software engineer creates a set of
classes, implementing a unified software interface, to represent primitive
geometric objects like points, lines, spheres, circles, planes, etc.
The software engineer would then implement transformations on all
these geometric objects using specialized hand-written subroutines
for each class; an exhausting and difficult task for large systems.
The situation gets even worse when implementing geometric operations
involving multiple objects like an intersections, collision detection,
or distance computations \citep{9788184895001,9788184894936}. Such
approach eventually creates many problems in GC software design, complexity,
maintenance, and cost. A much better approach is to rely instead on
higher-level algebraic abstractions to unify the mathematical base
of many geometric objects. This is partially done in computer graphics
and robotics, for example, when implementing 3D affine transformations
using $4\times4$ homogeneous matrices \citep{9781439803349}. 

There has been a search going on for decades to find a unifying algebraic
framework capable of expressing geometric ideas in a universal, consistent,
dimension-independent, and coordinates-independent manner. Recent
research and numerous applications have proven Geometric Algebra to
be a powerful algebraic framework that is capable of providing such
features. GA-based algebraic abstractions enable domain specific optimizations,
provide unification of geometric representations, and clarify expression
of geometric ideas \citep{Dorst.2009,Gunn.2011,Yuan_2012}. In addition,
GA can replace and extend most of the distinct algebraic frameworks
we use in practice. Thus we can learn a single algebraic framework
and uniformly apply it to more domains with minimum need for representational
conversions. This would also remove many of the communication boundaries
between scientific and engineering fields that have a common base
of geometric ideas. For more information about the historical developments
that led to modern GA the reader can refer to \citep{Chappell_2016,Hestenes.2017}.

\begin{figure}
\noindent \begin{centering}
\includegraphics[width=5in]{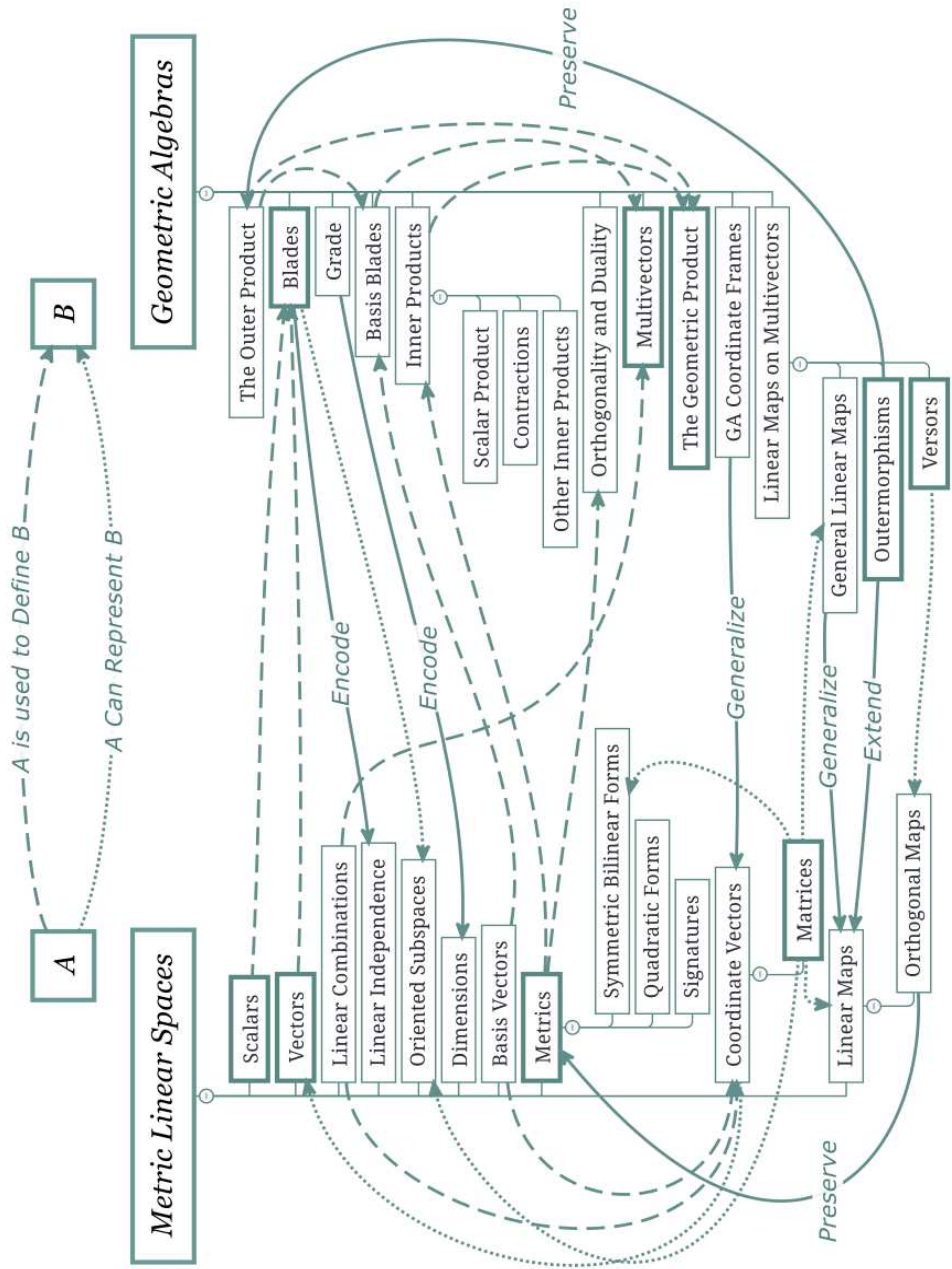}
\par\end{centering}
\caption{Main GA abstractions and their relations}
\label{fig:ga-concepts}
\end{figure}

\begin{figure}
\noindent \begin{centering}
\includegraphics[width=5in]{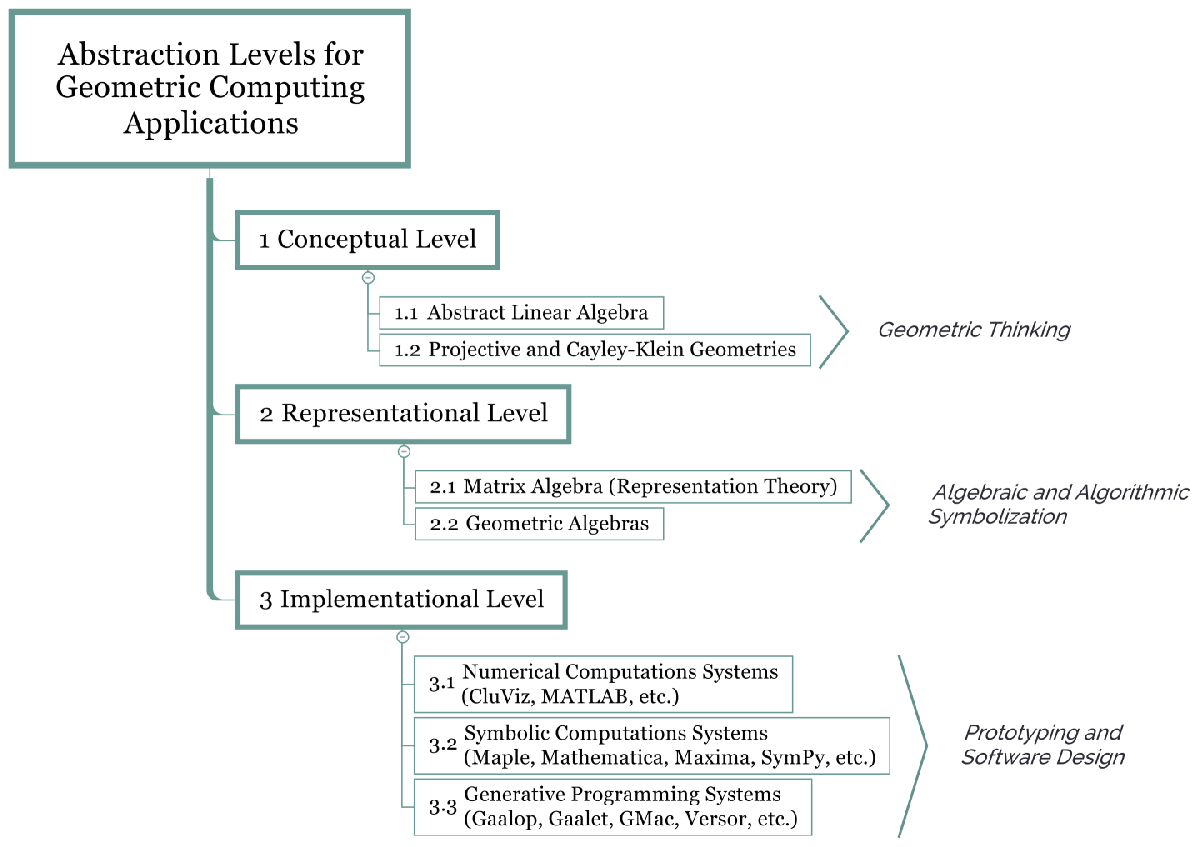}
\par\end{centering}
\caption{Abstraction levels and their tools for Geometric Computing applications}
\label{fig:gc-abstractions}
\end{figure}

\subsection{GA as a Language for Computational Thinking}

Computational Thinking (CT) complements critical thinking as a way
of reasoning to understand and solve problems, take proper actions,
and interact with our surroundings. The concepts and techniques of
CT are drown from computer and information science while having broad
application in the arts, sciences, engineering, humanities and social
sciences \citep{Kules_2016}. One definition of CT is as follows \citep{selby2014refining}:
\begin{defn}
\textbf{Computational Thinking} is a brain-based activity that enables
problems to be resolved, situations better understood, and values
better expressed through systematic application of abstraction, decomposition,
algorithmic design, generalization, and evaluation in the production
of an automation implementable by a digital or human computing device.
\end{defn}
CT relies on using abstraction and decomposition when attacking a
large complex task or designing a large complex system; it requires
thinking at multiple levels of abstraction \citep{Wing.2006}. Geometric
Algebra can be a valuable mathematical language to acquire and develop
such CT skills for handling Geometric Computing problems. As illustrated
in Figure \ref{fig:ga-concepts}, a Geometric Algebra is an abstract,
elegant, and sophisticated mathematical structure with many integrating
components. In order to fully appreciate all aspects of GA-based software
implementations, the team containing GA-model developers and software
engineers should collectively think on 3 integrating levels of abstraction,
as required by sound CT and shown in Figure \ref{fig:gc-abstractions}:
\begin{enumerate}
\item \textbf{The Conceptual Level.} On the most abstract level, we find
mathematical algebraic\textbackslash{}geometric concepts like scalar
fields, linear spaces, subspaces, linear maps, orthogonal operators,
metrics, duality, and space embeddings. Each of these concepts have
particular roles to play, and must have a specific set of features
to be able to play its roles. Fully understanding these concepts,
in the most abstract sense possible, is essential to avoid many bad
design decisions, and to take full advantage of GA's powerful unifying
language. The team should also appreciate the strong relations between
abstract linear algebra and projective and Cayley-Klein geometries
\citep{onishchik.2006,9783642172854}.
\item \textbf{The Representational Level.} On an intermediate level of abstraction,
we find that each concept of the above can have many representations
in less abstract mathematical domains. For example, a linear space
can be an abstraction of signals (as in electrical engineering), polynomials,
geometric directions in Euclidean space, or tuples of coordinate vectors.
An orthogonal map can be represented using an orthogonal matrix, a
GA versor, or a Discrete Fourier Transform. Understanding the commonalities
and relations between such representations and the limitations of
each is essential for the selection of the best representations for
a particular application domain. GA can provide many advantages over
matrix algebra in this level, while still integrating will with established
matrix representations.
\item \textbf{The Implementational Level.} On the lowest level of abstraction
we find software abstractions for representing all the above inside
a computational environment. Here we find elements such as floating
point numbers, combinatorial representations (for example classes
and structures), programming paradigms, software interfaces, subroutines,
memory hierarchies, and computer networks. Computers impose many physical
constraints on the above two levels of abstraction that must be taken
into consideration when addressing practical GA-based software implementations.
Many GA-based software tools are currently present to be used at this
level including numerical, symbolic, and Generative Programming-based
systems.
\end{enumerate}
These three levels are familiar to software engineers in other domains
of application. For example, in database systems design we find three
analogous levels of Conceptual Design, Logical Design, and Physical
Design \citep{9781305627482}. The role GA plays in Geometric Computing
applications can be thought to be analogous to the role of Relational
Algebra in relational database systems design. The study of the mathematics
behind Relational Algebra alone is not sufficient to produce successful
database applications, however. Software engineers must address other
complementary aspect of the design related to user interaction with
data (using SQL as a Domain Specific Language for example), physical
storage and transfer of data, optimization of data query executions,
data visualization and presentation, scalability, and many more. Without
addressing such aspects, Relational Algebra wouldn't have become a
basic part of computer science curricula worldwide. We must address
similar complementary aspects for Geometric Algebra in order to achieve
its rightful place in the scientific, educational, and industrial
fields. 

Whenever possible, expressing our ideas at the top level of abstraction
is very powerful conceptually. At this level we can understand and
relate many application areas at a fundamental level. We can communicate
ideas and transfer knowledge between them more easily. Sadly, many
people don't have access to this level of abstraction in practice.
We are taught to think about our mathematical tools starting from
the second intermediate level of abstraction, not the first top level.
The benefits of eliminating this serious problem appears in all areas
in which GA can be applied; for example:
\begin{itemize}
\item Many transformations we apply in signal and image processing are just
instances of abstract orthogonal linear maps, with more unifying common
properties than initially perceived. Such transforms include continuous
and discrete Fourier transforms, Laplace, z-, Walsh-Hadamard, slant,
Haar, Karhunen-Loeve, and wavelet transforms \citep{Wang.2011}. Using
GA to represent and apply these transforms can lead to new applications
and insights \citep{BayroCorrochano.2010b,Hitzer.2013b}, and eventually
to new unified architectures for multi-dimensional signal processing
software systems with modeling and processing capabilities well beyond
the current systems.
\item In geometric modeling and geometric reasoning, Euclidean, Hyperbolic,
and Elliptic geometries have a common algebraic foundation within
GA. This enables us to create GA-based universal geometric constructions
and apply them to specific problems with any desired geometry of these
three \citep{Sommer.2010,Sobczyk.2004,9780521715959,Gunn.2016}. Some
dynamic geometry software systems already apply this approach, like
Cinderella \citep{Hitzer_2003,cinderella} that internally models
the general Cayley-Klein geometry using complex numbers \citep{9783642172854}.
\item Many algebras that are very useful in practice are actually sub-algebras
of some GA. The list include the algebra of real numbers, complex
numbers, n-D Euclidean vectors, quaternions, dual quaternions, spinors,
Clifford's dual numbers, and Grassmann numbers. GA can unify and convert
these numbers within the context of a single problem, engineering
discipline, or scientific field.
\end{itemize}
Another anti-CT pattern facing most software engineers in designing
GC applications results from not having a clear separation between
those three levels of abstraction. In many cases, intermediate representational
abstractions are incorrectly perceived to be identical to the conceptual
abstractions. As an example, consider the default use of matrices
to represent linear maps in GC applications. There are other intermediate
representations that are better than matrices in modeling certain
geometric aspects with better computational properties. For example,
it's much easier to extract the axis and angle of rotation of a 3D
general rotation linear map if we use a quaternion to represent the
linear map. Quaternions require less memory, less processing, and
are numerically more stable compared to rotation matrices \citep{Dorst.2009}.
As another example for incorrectly mixing levels of abstraction, many
programmers blindly use floating point numbers as a perfect representation
of real numbers, not taking into consideration some of their problematic
features \citep{Goldberg_1991}. 

Clear separation of the first two abstraction levels can result from
studying a course in projective geometry \citep{9783642172854}, abstract
algebra \citep{9780471433347}, and abstract linear spaces \citep{golan.2007,Axler.2015}
in addition to the classical coordinate based linear algebra courses
\citep{Poole.2015}. GA can be very helpful in this regard as it contains
enough mathematical abstraction and generality to provide clear understanding
and separation of abstract levels of thinking. This skill is typically
available to mathematicians and physicists, but less so for computer
scientists and engineers. Separation of the third level requires careful
study of the physical limitations of representing and communicating
information inside computational environments. In addition, a through
understanding of capabilities of modern programming languages and
programming paradigms is necessary to design better implementations
\citep{9780123745149,9781848829138}. This skill is typically available
to computer scientists and software engineers, but less so for mathematicians
and physicists.

From another angle, learning GA can take much time and effort. Applying
Computational Thinking to the GA learning problem can reduce time
and effort considerably. Because GA is relevant to so many areas in
science and engineering, its presentation should be formulated to
each specific discipline. A very good example for presenting GA to
electrical and electronic engineers, for example, is \citep{Chappell.2014}.
Similar efforts are needed to properly introduce GA to software engineers
and software developers. Presenting GA to a software engineer is different
from presenting to a mathematician or physicist. The mindset of a
software engineer prefers dealing with diagrams, specifications, relations,
and algorithms rather than axioms, theorems, proofs, and equations
\citep{Nilsson.2002}. Such efforts also include designing easy to
use domain-specific GA-based software systems for educational and
prototyping purposes in addition to production purposes for Geometric
Computing applications.

\subsection{Aims and Contents}

This article is intended as a Computational Thinking driven exposition
of GA for software engineers interested in creating GA-based GC software
systems. I attempt to emphasize the conceptual and representational
abstraction levels related to each mathematical element of Geometric
Algebra, leaving the implementational level to future articles. The
conceptual level is purely mathematical and is independent of any
particular software implementation. The representational level is
also mathematical but typically defines the high-level design of the
GA-based software system. My main intention here is to provide a unified
entry point for facilitating further study of the mathematics behind
the concepts summarized here that is suitable for software engineers.

The main body of this article consists of 3 parts. In the first part
of this article in section 2, I summarize the main abstract and algebraic
concepts of Metric Linear Spaces, the base on which GAs are constructed.
In the second part in section 3, I build on the concepts of section
2 to explain the elegant mathematical structure of Geometric Algebra
with references to additional information sources for the interested
reader. Since I'm mainly interested here in the most computationally-significant
algebraic constructions of GA, I will not discuss GA's numerous geometric
interpretations found in the literature. In the third part in section
4, I focus on defining GA Coordinate Frames and how to use them for
computing linear operations, products, and maps on GA multivectors.
This is the mathematical base for the symbolic computations infrastructure
layer in GMac, a universal GA-based implementation generator system
I designed \citep{Eid.20160716,gmac}. Finally, in section 5 I provide
some concluding remarks and suggestions.

\section{Metric Linear Spaces}

\subsection{Scalar Fields}

Many number systems exist in mathematics with varying properties and
applications. In practice, however, we tend to concentrate on a few
of them: rational numbers $\mathbb{Q}$ , real numbers $\mathbb{R}$
, and complex numbers $\mathbb{C}$ . Such numbers are also called
scalars to distinguish them from vectors in linear spaces. There are
common properties of these number systems that, when abstracted into
algebraic relations, give us the concept of a scalar field \citep{9780471433347}.
On the top conceptual level of abstraction, a field $\mathbb{F}$
is a set of ``scalars'' closed under two operations called addition
and multiplications satisfying some familiar properties like associativity
and commutativity of addition and multiplication, presence of unique
additive and multiplicative identities and inverses, and distributivity
of multiplication over addition. From these simple properties many
features, theorems, and operations can be defined and deduced based
on these abstract concepts without having to give concrete examples
like the real or complex number systems. Some roles of scalars in
Geometric Computing applications include:
\begin{itemize}
\item Used as abstractions of physical measurements like mass, velocity,
length, area, etc..
\item Used to encode, quantify, sort, and compare geometric objects and
their properties. 
\item Used as construction elements in Linear Combinations and other related
combinations over Vectors.
\end{itemize}
Mathematically, we can construct a linear space, hence a GA, over
any scalar field; including finite fields \citep{9780521460941}.
Linear spaces over finite fields have interesting properties that
could be investigated using Geometric Algebra especially for digital
and discrete geometry applications \citep{9781439873786,Aveneau_2014}.
In the GA literature there exist strong assertions that only real
numbers should be used as a base for constructing GAs \citep{Hestenes.2017}.
This point of view is mainly based on the existence of isomorphisms
between complex numbers-based GAs and real numbers-based GAs; so the
use of complex-based GAs is mathematically redundant and geometrically
more complex for modeling the physical space and time we live in.
This is certainly a respectable point of view, especially in physics.
From a software engineering and educational point of view, however,
I recommend to leave the door open for using the most suitable number
system for a particular problem at hand. I believe many problems can
be more easily transformed from the classical representations into
GA-based representations if we are flexible about the choice of the
number system we use \citep{Lundholm.20090730}.

\subsection{Linear Combinations and Abstract Vectors}

At the base of the elegant GA mathematical structure we find the abstract
concept of Linear Spaces; also commonly called Vector Spaces \citep{golan.2007,Axler.2015,Poole.2015}.
Many study linear spaces because of their basic role in encoding the
Superposition Principle; a cornerstone in modern science and engineering.
Typical mathematical introductions to linear spaces concentrate on
the abstract algebraic properties of vectors and their two main operations
of vector addition and scalar multiplication. From a computational
point of view, however, the central concept in linear spaces is the
Linear Combination. A linear combination is an expression of the form
$a_{1}v_{1}+a_{2}v_{2}+\cdots+a_{k}v_{k}\equiv\sum_{i=1}^{k}a_{i}v_{i}$
where $v_{i}$ are ``vectors'' and $a_{i}$ are scalars not all
zero. A linear space is simply any set of ``vectors'' that is closed
under linear combinations over a given scalar field; i.e. any linear
combination of any collection of vectors is also a vector in the same
set. The familiar algebraic properties of vector addition and scalar
multiplication are necessary to perform linear combinations consistently.
This very abstract concept has so many manifestations in science and
engineering that it is a central concept in many applications. All
other main concepts of linear spaces are derived from linear combinations;
for example:
\begin{itemize}
\item \textbf{Span:} The span of a given set of vectors $span\left(v_{1},v_{2},\ldots,v_{k}\right)$
is the set of vectors resulting from all possible linear combinations
of these vectors. Here the vectors $v_{i}$ are fixed while the scalars
$a_{i}$ can take any possible values from their field. 
\item \textbf{Subspace:} A linear subspace $W$ of a larger linear space
$V$, denoted here as $W\leq V$ , is a subset of the linear space
$V$ that is closed under linear combinations. The span of any set
of vectors from $V$ is always a subspace of $V$ .
\item \textbf{Linear Independence:} A collection of vectors are called Linearly
Dependent when we can express any of them as a linear combination
of the others; else they are Linearly Independent (LID) vectors. These
two are basic conceptual relations among any given collection of vectors.
\item \textbf{Basis:} A basis is a LID set of vectors $\left\{ e_{1},e_{2},\ldots,e_{n}\right\} $
that spans the whole linear space. Any vector in the linear space
can be expressed as a unique linear combination of the basis vectors.
A linear space can have an infinite number of basis sets, but they
all contain the same number of vectors $n$. This number $n$ is the
dimension of the linear space denoted by $dim\left(V\right)$. In
all the following discussions, the basis is assumed to be an ordered
set, not a general set; denoted here as $\left\langle e_{1},e_{2},\ldots,e_{n}\right\rangle $.
\item \textbf{Coordinate Vector:} Given a fixed ordered basis $\boldsymbol{E}=\left\langle e_{1},e_{2},\ldots,e_{n}\right\rangle $
we can express any abstract vector $v$ as a linear combination of
the basis vectors $v=\sum_{i=1}^{k}a_{i}e_{i}$. The scalar coefficients
$a_{i}\in\mathbb{F}$ can be written as a tuple $v_{E}=\left(a_{1},a_{2},\ldots,a_{n}\right)\in\mathbb{F}^{n}$
that is called the coordinate vector representation of $v$;. The
abstract vector $v$ and its coordinate vector $v_{E}$ are two conceptually
distinct entities, but have a linear isomorphism between them; so
we can compute with coordinate vectors and interpret the results in
the context of the abstract linear space. Sometimes we prefer to express
the coordinate vector in matrix form as a column vector holding the
same scalars. I will denote the column vector representation of an
abstract vector $v$ on the basis $\boldsymbol{E}$ as: $\left[v\right]_{\boldsymbol{E}}=\left[\begin{array}{cccc}
a_{1} & a_{2} & \cdots & a_{n}\end{array}\right]^{T}$.
\item \textbf{Linear Map:} A linear map is a map between two linear spaces
$\mathbf{f}:V\rightarrow W$ that preserves linear combinations $\mathbf{f}[\sum_{i=1}^{k}a_{i}v_{i}]=\sum_{i=1}^{k}a_{i}\mathbf{f}[v_{i}]$.
When the two linear spaces are the same, its is called a linear operator. 
\item \textbf{Other Combinations:} Imposing constraints on the scalar coefficients
of linear combinations leads to theoretically and practically significant
concepts with many important geometric interpretations like Affine
Combinations $\sum_{i=1}^{k}a_{i}=1$ \citep{9781439803349}, Conical
Combinations $a_{i}\geq0$, and Convex Combinations $a_{i}\geq0,\:\sum_{i=1}^{k}a_{i}=1$
\citep{Tuy_2016}. 
\end{itemize}
It is important to note that we are not yet talking about distances
and angles between vectors or orthogonality of vectors because such
concepts require the more fundamental concept of metric defined later.
The main relation between vectors in non-metric abstract linear spaces
is the Linear Dependence\textbackslash{}Independence relation. The
main construction operation is the Linear Combination. We can ``divide''
two vectors (i.e. compare their relative scale) but only if one of
them is a linear combination (i.e. a scaled version) of the other.
Generally, this is not how engineers are usually taught linear spaces
in undergraduate courses, but a clear understanding and separation
of these fundamental concepts is necessary to correctly understand
and use the mathematical structure of Geometric Algebra that is based
on abstract linear spaces.

\subsection{Abstract Vectors and Coordinate Vectors}

In order to use computers for dealing with abstract concepts of linear
spaces, we need an equivalent intermediate representation that only
uses numbers and their basic operations of addition and multiplication.
Mathematics provide a base for such representation through coordinate
vectors. Without loss of generality I will concentrate on the field
of real numbers $\mathbb{R}$ as the scalar field for all the following
discussions. Having an n-dimensional abstract linear space $V$ on
$\mathbb{R}$ with basis $\boldsymbol{E}=\left\langle e_{1},e_{2},\ldots,e_{n}\right\rangle $
we can set up a linear isomorphism (i.e. one-to-one linear map) $\phi$
defined with its inverse map $\phi^{-1}$ as follows:

\begin{align}
\phi:\mathbb{R}^{n} & \rightarrow V,\;\phi:\left(a_{1},a_{2},\ldots,a_{n}\right)\mapsto\sum_{i=1}^{n}a_{i}e_{i};\\
\phi^{-1}:V & \rightarrow\mathbb{R}^{n},\;\phi^{-1}:v\mapsto v_{\boldsymbol{E}}
\end{align}

This way, linear combinations on the coordinate vectors of the real
linear space $\mathbb{R}^{n}$ are equivalent representations of the
same linear combinations on the abstract linear space $V$. Now we
can add two vectors in $V$ by simply adding the real components of
their coordinate vectors in $\mathbb{R}^{n}$ and apply the linear
isomorphism to get the final result in $V$. We can do the same for
scalar multiplication by multiplying the scalar with the components
of the coordinate vector. All derived linear operations on $V$ can
be formulated ``numerically'' on the equivalent real linear space
$\mathbb{R}^{n}$. This is the playground of matrix algebra \citep{Poole.2015},
the typical starting point where most engineers learn about linear
spaces. The n-dimensional real coordinate vectors space $\mathbb{R}^{n}$
is a linear space that is equivalent to all n-dimensional abstract
linear spaces; $\mathbb{R}^{n}$ is a universal intermediate representation
for all abstract linear spaces. 

One important point to realize is that by changing the basis of $V$
we are also changing the linear isomorphism $\phi$ because the same
abstract vector has a different linear combination on a different
basis. To make our computations consistent we must use the same basis
for all related computations. In addition, some facts should remain
the same regardless of the used basis and isomorphism. For example,
linear independence of a set of vectors should remain the same regardless
of the selected basis. Such properties are called coordinate-independent
or basis-independent. GA can provide many coordinate-independent formulations
for properties of linear spaces and at the same time act as an excellent
intermediate representation through its multivectors and products.
Because a GA is itself a linear space, as will be explained later,
we can always represent all GA multivectors and operations using matrix
algebra. This is the approach used in some GA software systems like
the Clifford Multivector Toolbox for MATLAB \citep{Sangwine_2016,matlab.mv.2016}
for example.

\subsection{Metrics and Their Representations}

A metric linear space is just a linear space with an additional bilinear
map, called the metric, that associates a scalar with each pair of
vectors \citep{roman.2007}. The objective of defining a metric is
to enable comparing vectors and subspaces of different attitude in
space using scalars. Many familiar concepts we use are actually based
on the more fundamental metric concept. Such concepts include distance,
length, area, angle, orthogonality, orthogonal maps, projections,
rotations, and many others. In GA the definition of a metric is based
on the concept of a symmetric bilinear form and the associated concept
of a quadratic form. A symmetric bilinear form $\mathbf{B}$ on the
real linear space $V$ is a mapping $\mathbf{B}:V\times V\rightarrow\mathbb{R}$
that is linear in both arguments (i.e. bilinear) and symmetric $\mathbf{B}(u,v)=\mathbf{B}(v,u)\,\,\,\forall u,v\in V$.
A related concept is the quadratic form that is related to a symmetric
bilinear form by: $\mathbf{Q}(u)=\frac{1}{2}\mathbf{B}(u,u),\,\mathbf{B}(u,v)=\mathbf{Q}(u+v)-\mathbf{Q}(u)-\mathbf{Q}(v)\,\,\forall u,v\in V$.
The quadratic form satisfies the relation $Q(av)=a^{2}Q(v)\,\,\forall v\in V,a\in\mathbb{R}$.

The metric also associates each vector in the linear space with some
scalar by putting the vector in both inputs of the metric. This scalar
is called the norm $\left\Vert v\right\Vert \equiv v^{2}\equiv\mathbf{B}(v,v)$
of the vector $v\in V$ and is equal to double the quadratic form
of the vector $\left\Vert v\right\Vert =2\mathbf{Q}\left(v\right)$\footnote{This is different from classical literature where the norm of a Euclidean
vectors $x$ is the square root of the inner product $\sqrt{x\cdot x}$
equivalent to its length. I will use here the notation $\left|x\right|=\sqrt{x\cdot x}$
assuming $x\cdot x\geq0$.}. If two vectors are associated with the same scalar they are of equal
norm, and null vectors are vectors having zero norm. In this context
the norm is any general real number; even zero and negative numbers
are allowed for non-zero vectors in GA. This is one important generalization
different from metrics in classical linear algebra that are usually
restricted to being positive definite. One of the common interpretations
of vector norm in the special case of Euclidean linear spaces is the
the squared length of a direction vector.

If the linear space has the basis $\left\langle e_{1},e_{2},\cdots,e_{n}\right\rangle $
then we can construct a bilinear form matrix $\mathbf{A_{B}}=\left[a_{ij}\right],a_{ij}=\mathbf{B}(e_{i},e_{j})$
, also called the metric matrix on this basis. This matrix is a real
symmetric matrix that we can use to compute the bilinear form of any
two vectors $u,v\in V$ given their representation on the basis as
follows:

\begin{eqnarray}
u & = & u_{1}e_{1}+\cdots+u_{n}e_{n},\nonumber \\
v & = & v_{1}e_{1}+\cdots+v_{n}e_{n}\nonumber \\
\mathbf{\Rightarrow B}(u,v) & = & \left(\begin{array}{ccc}
u_{1} & \cdots & u_{n}\end{array}\right)\mathbf{A_{B}}\left(\begin{array}{ccc}
v_{1} & \cdots & v_{n}\end{array}\right)^{T}
\end{eqnarray}

Using bilinear forms the concept of orthogonality of vectors can be
defined as follows: two vectors $u,v$ are called orthogonal iff $\mathbf{B}(u,v)=0$.
The inner product of two vectors is simply the bilinear form of the
vectors $u\cdot v\equiv\mathbf{B}(u,v)$, and the norm is the inner
product of a vector with itself $v^{2}=v\cdot v$; thus justifying
the use of the name Inner Product Matrix (IPM) for the symmetric bilinear
form matrix. The IPM $\mathbf{A_{B}}$, being a real symmetric matrix,
can be diagonalized using a Change of Basis Matrix (CBM) $\mathbf{P}$
to obtain a diagonal matrix $\mathbf{D}=\mathbf{P}^{T}\mathbf{\mathbf{A_{B}}P}$
where $\mathbf{P}$ is an orthogonal matrix $\mathbf{P}^{-1}=\mathbf{P}^{T}$.
The columns of $\mathbf{P}$ are orthogonal eigen vectors for $\mathbf{A_{B}}$.
The diagonalization can always be performed such that the numbers
on the diagonal (called the eigen values) are either -1, 0, or +1.
The number of eigen values that are 1,-1, and 0 are characteristics
for a given metric and define what is called the signature of the
bilinear\textbackslash{}quadratic form. A bilinear\textbackslash{}quadratic
form is said to have the signature ($p,q,r$) if there exists a diagonalization
of the IPM having $p$ eigen values with value 1, $q$ eigen values
with value -1 and $r$ eigen values with value zero. If the IPM is
singular (i.e. has no inverse which is equivalent to $r>0$) the bilinear
form is called degenerate. If all the eigen values are positive the
IPM is positive definite and the space is a Euclidean space; there
exists a basis with all basis vectors norms equal to +1. A mixed-signature
metric space has some non-zero vectors with norm equal to zero. Such
vectors are called null vectors and only exist in mixed-signature
spaces (spaces having a bilinear form with $p>0$ and $q>0$) in addition
to degenerate spaces. The signature of the IPM extends to the signature
of the whole GA that we construct using the IPM. By combining the
concept of metric and the concept of space embedding, discussed later,
we can consistently model Euclidean and non-Euclidean geometries using
GAs of various signatures. 

\begin{table}

\caption{The concept of a unit circle in different metrics on 2D linear space}
\label{tbl:metric-circle}
\noindent \centering{}%
\begin{tabular}{|c|c|c|}
\hline 
IPM & Equation & Geometric Shape\tabularnewline
\hline 
\hline 
$\left(\begin{array}{cc}
a & 0\\
0 & b
\end{array}\right)$ & $ax^{2}+by^{2}=1$ & $\begin{cases}
\textnormal{Circle} & a=b>0\\
\textnormal{Imaginary Circle} & a=b<0\\
\textnormal{Ellipse} & a>0,b>0\\
\textnormal{Imaginary Ellipse} & a<0,b<0\\
\textnormal{Hyperbola} & ab<0
\end{cases}$\tabularnewline
\hline 
$\left(\begin{array}{cc}
1 & 0\\
0 & 0
\end{array}\right)$ & $x^{2}=1$ & Straight Line\tabularnewline
\hline 
$\left(\begin{array}{cc}
-1 & 0\\
0 & 0
\end{array}\right)$ & $x^{2}=-1$ & Imaginary Straight Line\tabularnewline
\hline 
$\left(\begin{array}{cc}
0 & 0\\
0 & 0
\end{array}\right)$ & $0=1$ & No Geometry Defined\tabularnewline
\hline 
$\left(\begin{array}{cc}
0 & a\\
a & 0
\end{array}\right)$ & $xy=\dfrac{1}{2a}$ & Rectangular Hyperbola\tabularnewline
\hline 
\end{tabular}
\end{table}

To illustrate how a metric effects the geometry of the space, Table
\ref{tbl:metric-circle} shows some possible metrics of a 2D linear
space with basis $\left\langle e_{1},e_{2}\right\rangle $. Using
this general definition of the unit circle ``The set of position
vectors having unit norm $\left\{ v:v=xe_{1}+ye_{2},\left\Vert v\right\Vert =1,x,y\in\mathbb{R}\right\} $''
we get the general equation $x^{2}e_{1}^{2}+2xy\left(e_{1}\cdot e_{2}\right)+y^{2}e_{2}^{2}=1$.
We see that only in Euclidean space $e_{i}\cdot e_{j}=\delta_{i}^{j}$
where we get the familiar circle equation, where in other metrics
we get totally different geometries. The same goes for the geometric
meanings of other metric-dependent concepts like orthogonality, angle,
rotation, distance, area, projection, etc.

\subsection{Linear Maps and Their Representations}

Linear maps are a central concept for creating Geometric Computing
applications. One of the main reasons is that linear maps have a direct
relation to multi-dimensional Projective Geometry \citep{9783642172854,Gunn.2011},
which is the base for all Euclidean and non-Euclidean geometries,
and has many applications in computer graphics, computer vision, robotics,
and image processing, for example. I will denote the effect of a linear
map $\mathbf{f}:V\rightarrow W$ on a vector $x\in V$ and on a subspace
$X=span\left(x_{1},x_{2},\ldots x_{k}\right)\leq V$ as $\mathbf{f}\left[x\right]\in W$
and $\mathbf{f}\left[X\right]\leq W$ respectively. Classically the
concept of a linear map is associated with matrix algebra through
the following construction: assuming the real linear spaces $V,W$
with bases $\boldsymbol{A}=\left\langle a_{1},a_{2},\cdots,a_{n}\right\rangle ,\boldsymbol{B}=\left\langle b_{1},b_{2},\cdots,b_{m}\right\rangle $
respectively; we can define a linear map $\mathbf{f}:V\rightarrow W$
such that the effect of $\mathbf{f}$ on each basis vector in $\boldsymbol{A}$
is known and expressed as a linear combination of the basis vectors
in $\boldsymbol{B}$ (i.e. the column vectors $m_{i}=[\mathbf{f}[a_{i}]]_{\boldsymbol{B}}$
are known for all $a_{i}$), the matrix of $\mathbf{f}$ acting on
$\boldsymbol{A}$ with respect to $\boldsymbol{B}$ is defined as
the $m\times n$ matrix:

\begin{equation}
\mathbf{M_{f}}=\left[\mathbf{f}\right]_{\boldsymbol{A,B}}\equiv\left[\begin{array}{cccc}
m_{1} & m_{2} & \cdots & m_{n}\end{array}\right],\,\,m_{i}=[\mathbf{f}[a_{i}]]_{\boldsymbol{B}}
\end{equation}

We can then compute a coordinate representation of the transformation
$\mathbf{f}[x]\in W$ of any vector $x=\sum_{i=1}^{n}x_{i}a_{i}\in V$
through a simple matrix multiplication operation:

\begin{eqnarray}
[\mathbf{f}[x]]_{\boldsymbol{B}} & = & [x_{1}\mathbf{f}[a_{1}]+x_{2}\mathbf{f}[a_{2}]+\cdots+x_{n}\mathbf{f}[a_{n}]]_{\boldsymbol{B}}\nonumber \\
 & = & x_{1}[\mathbf{f}[a_{1}]]_{\boldsymbol{B}}+x_{2}[\mathbf{f}[a_{2}]]_{\boldsymbol{B}}+\cdots+x_{n}[\mathbf{f}[a_{n}]]_{\boldsymbol{B}}\nonumber \\
 & = & \left[\mathbf{f}\right]_{\boldsymbol{A},\boldsymbol{B}}\left[x\right]_{\boldsymbol{A}}
\end{eqnarray}

When the two linear spaces are the same $V=W$ then $\mathbf{f}$
is a linear operator on $V$. When the two basis are also the same
$\boldsymbol{A}=\boldsymbol{B}$ these relations become: 
\begin{equation}
\mathbf{M_{f}}=\left[\mathbf{f}\right]_{\boldsymbol{A}}\equiv\left[\begin{array}{cccc}
m_{1} & m_{2} & \cdots & m_{n}\end{array}\right],\,\,m_{i}=[\mathbf{f}[a_{i}]]_{\boldsymbol{A}},\,\,[\mathbf{f}[v]]_{\boldsymbol{A}}=\left[\mathbf{f}\right]_{\boldsymbol{A}}\left[v\right]_{\boldsymbol{A}}\forall v\in V
\end{equation}
The unique matrix $\left[\mathbf{f}\right]_{\boldsymbol{A,B}}$ is
called the matrix representation of $\mathbf{f}$ with respect to
basis $\boldsymbol{A}$ and $\boldsymbol{B}$. We can then find many
properties of the linear map by applying matrix algebra operations
on its matrix. For example: 
\begin{itemize}
\item The dimensions of the domain $V$ and co-domain $W$ of a linear map
$\mathbf{f}:V\rightarrow W$ are respectively equal to the number
of columns and rows of its matrix $\left[\mathbf{f}\right]_{\boldsymbol{A,B}}$
on any two basis. In addition, we can apply a composition of linear
maps between linear spaces using matrix multiplication of their matrices.
This is extensively used in computer graphics and robotics for composing
sequences of motions expressed as linear maps.
\item The adjoint linear operator $\mathbf{f}^{T}$ associated with a linear
operator $\mathbf{f}$ defined on a real metric linear space $V$
with bilinear form $\mathbf{B}$ is the operator that satisfies $\mathbf{B}(\mathbf{f}[x],y)=\mathbf{B}(x,\mathbf{f}^{T}[y])\,\,\forall x\in V$.
The matrix representation of $\mathbf{f}^{T}$ is the transpose of
the matrix representation of $\mathbf{f}$, but only on the same basis:
$\left[\mathbf{f}^{T}\right]_{\boldsymbol{A}}=\left[\mathbf{f}\right]_{\boldsymbol{A}}^{T}$. 
\item An isomorphism is a linear map defined on two linear spaces of the
same dimension that is also invertible. The matrix of an isomorphism
on any two bases is non-singular. In addition, the matrix of an invertible
linear operator $\mathbf{f}$ and the matrix of its inverse $\mathbf{f}^{-1}$
are inverses to each other, but only on the same basis $\left[\mathbf{f}\right]_{\boldsymbol{A}}\left[\mathbf{f}^{-1}\right]_{\boldsymbol{A}}=\mathbf{I}$. 
\item Any isomorphism has a unique basis-independent scalar associated with
it called its determinant $\left|\mathbf{f}\right|$ \citep{Shafarevich_2013}.
The determinant of the isomorphism is equal to the determinant of
its matrix on any bases $\left|\mathbf{f}\right|=\left|\left[\mathbf{f}\right]_{\boldsymbol{A},\boldsymbol{B}}\right|$.
The geometric significance of the determinant is more apparent within
the context of GA's outermorphisms discussed later.
\item We can define a unique Change of Basis isomorphism between two linear
spaces of the same dimension $\mathbf{g}:V\rightarrow W$ that takes
$\boldsymbol{A}$ to $\boldsymbol{B}$ such that $\mathbf{g}\left[a_{i}\right]=b_{i}\:\forall i=1,2,\ldots,n$..
This isomorphism also has a unique invertible matrix $\left[\mathbf{g}\right]_{\boldsymbol{A,B}}$
called a Change of Basis Matrix (CBM). This means that the same invertible
matrix may represent an invertible linear operator on the same basis,
or a change of basis linear map on two different bases. This is one
instance of mixing of conceptual abstractions that is common in matrix
algebra formulations. Such issue might lead to confusions in algebraic
formulations when using matrix algebra to represent abstract linear
maps. 
\item Two matrices $\mathbf{\mathrm{\boldsymbol{M}}},\mathrm{\boldsymbol{N}}$
are called similar $\mathrm{\boldsymbol{M}}\sim\mathrm{\boldsymbol{N}}$
if they represent the same linear operator on different bases, or
equivalently if there is a CBM $\mathrm{\boldsymbol{C}}$ such that
$\mathrm{\boldsymbol{M}}=\boldsymbol{\mathrm{C}}^{-1}\mathrm{\boldsymbol{N}}\boldsymbol{\mathrm{C}}$.
Similarity between square matrices is an equivalence relation. Many
invariant properties of similar matrices are actually properties of
their common abstract linear map. Most notably, spectral analysis
and invariant subspace techniques in linear algebra \citep{089871608X}
depend on this relation between an abstract linear map and its infinite
number of representation matrices. These techniques are very important
in many scientific and engineering applications.
\item An invertible linear operator $\mathbf{f}$ that satisfies $\mathbf{B}(\mathbf{f}[x],\mathbf{f}[y])=\mathbf{B}(x,y)\,\,\forall x,y\in V$,
where $\mathbf{B}$ is the bilinear form on $V$, is called an orthogonal
linear operator; it preserves the metric between vectors. This means
that $\mathbf{f}$ preserves many metric-dependent properties and
operations like the inner product, norm, orthogonality, and angle
between vectors. In addition, its adjoint is equal to its inverse:
$\mathbf{f}^{-1}=\overline{\mathbf{f}}$. For non-degenerate metrics,
the matrix of an orthogonal operator is invertible, has $\pm1$ determinant,
and has columns that represent orthonormal vectors; i.e. each two
column vectors are orthogonal and have unit (i.e. $\pm1$) norm. These
matrices are called orthogonal matrices and are very important in
many practical applications. We can analyze\textbackslash{}construct
any such map as a composition of a series of geometric reflections
in homogeneous hyper-planes (i.e. $(n-1)$-dimensional subspaces)
of the linear space. The Householder operator \citep{Davis_2006,0521880688,9781107004122},
one of the most important computational tools in numerical matrix
algebra, is based on this conceptual construction. GA provide a better
algebraic alternative using its Versors and Versor Product.
\item The Kernel $ker_{\mathbf{f}}$ or Null Space of a linear map $\mathbf{f}:V\rightarrow W$
is the set of vectors that transform to the zero vector of $W$ under
$\mathbf{f}$: $ker_{\mathbf{f}}=\left\{ v:v\in V,\mathbf{f}\left[v\right]=0_{W}\right\} $.
The range $range_{\mathbf{f}}$ or Image of $\mathbf{f}$ is the set
of vectors in $W$ that are transformations under $\mathbf{f}$ of
vectors in $V$: $range_{\mathbf{f}}=\left\{ w:\exists v\in V\,\mathbf{f}\left[v\right]=w\right\} $.
These two sets are subspaces satisfying the relations $ker_{\mathbf{f}}\leq V$,
$range_{\mathbf{f}}\leq W$, and $dim\left(V\right)=dim\left(ker_{\mathbf{f}}\right)+dim\left(range_{\mathbf{f}}\right)$
where $null_{\mathbf{f}}\equiv dim\left(ker_{\mathbf{f}}\right)$
is also called the nullity of$\mathbf{f}$ and $rank_{\mathbf{f}}\equiv dim\left(range_{\mathbf{f}}\right)$
is called the rank of$\mathbf{f}$. We can use matrix algebra to find
the kernel of $\mathbf{f}$ using its representation matrix $\left[\mathbf{f}\right]_{\boldsymbol{A,B}}$
on any two bases by solving the matrix equation $\left[\mathbf{f}\right]_{\boldsymbol{A,B}}v=0$
for coordinate vectors $v$ to find a set of LID spanning coordinate
vectors for the kernel. We can also represent the range of a linear
map using $\left[\mathbf{f}\right]_{\boldsymbol{A,B}}$ by viewing
its column vectors as representing abstract vectors in $W$ that span
$range_{\mathbf{f}}$. This leads to the familiar matrix rank of $\left[\mathbf{f}\right]_{\boldsymbol{A,B}}$
being equal to the rank of the linear map it represents $rank\left(\left[\mathbf{f}\right]_{\boldsymbol{A,B}}\right)=rank_{\mathbf{f}}$.
These linear spaces and their relations are an important part of the
Fundamental Theorem of Linear Algebra \citep{Strang_1993} usually
expressed using matrices not abstract linear maps.
\end{itemize}
It is very important when designing Geometric Computing applications
in a Computational Thinking sound manner to have clear conceptual
distinction between an abstract linear map and its infinite number
of possible matrix representations. In GC applications it is typical
that the choice of basis is not arbitrary or even unique. The same
problem may need many bases to be used, as in the case of robotics
and computer graphics for example. Because matrices can also represent
subspaces (as lists of column vectors) and metrics (as IPMs), matrix
algebra formulations can hide the abstract geometric meaning behind
the clutter of its less abstract and basis-implicit representations.
The use of GA formulations instead of matrix algebra can, in many
cases, enforce a clear separation of basic abstract concepts from
their representations.

\subsection{Oriented Subspaces}

When we use matrix algebra to represent linear spaces, we have a well-developed
set of tools to algebraically represent and manipulate abstract vectors.
In many applications in science and engineering, however, we often
need to algebraically represent and manipulate whole subspaces in
addition to vectors. Some common subspace manipulations we use include:
\begin{itemize}
\item To construct a subspace given a set of vectors that spans the subspace;
the vectors may or may not be linearly independent. 
\item We may need to extract information about a subspace such as its dimensionality,
its relation to fixed subspaces in the problem, and the ``best''
basis of vectors we can use to span the subspace. Here the word ``best''
is context-dependent. We may prefer a basis for getting more numerically-stable
computations, or perhaps for having a better correspondence with actual
physical elements of our model.
\item To apply a linear map to a whole subspace and get another.
\item To operate on two or more subspaces in order to get another subspace
as output. For example, to find the common subspace of two subspaces,
to find the smallest subspace containing two subspaces, to project
one subspace on another, to find a subspace that complements another
into a bigger subspace, and to reflect one subspace on another. 
\item To compare two subspaces having different attitudes in space. This
includes, for example, finding the angle of a single rotation operation
that takes one subspace into another, or finding if two subspaces
are orthogonal to each other in the sense that each vector in the
first is orthogonal to all vectors in the second subspace.
\end{itemize}
Any single vector $v$ actually represents a 1-dimentional subspace
$\overleftrightarrow{v}$ through its span: $\overleftrightarrow{v}=span\left(v\right)$.
Extending this to more dimensions we can use the span of k LID vectors
to represent their k-dimensional subspace $W=span\left(v_{1},v_{2},\ldots,v_{k}\right)$.
A matrix $\boldsymbol{\mathbf{A}}_{W}$ can represent an ordered set
of vectors by putting their equivalent coordinate representations
on some basis $\boldsymbol{E}$ as rows or columns in the matrix:
$\boldsymbol{\mathbf{A}}_{W}=\left[\begin{array}{cccc}
\left[v_{1}\right]_{\boldsymbol{E}} & \left[v_{2}\right]_{\boldsymbol{E}} & \cdots & \left[v_{k}\right]_{\boldsymbol{E}}\end{array}\right]$. This way we can use matrix algebra and matrix operations to manipulate
this ``list of coordinate vectors'' as an indirect (and mostly awkward)
computational representation of abstract linear subspaces. This kind
of representation has disadvantages for practical Geometric Computing
applications. Matrix algebra is a suitable mathematical abstraction
for low-level computations inside machines, but is not an intuitive
modeling abstractions when designing GC models and algorithms. Much
geometric information get scattered among the numbers of the matrix,
and we need significant effort to extract such information. In addition,
matrix algebra-based formulations are often basis-dependent and metric-dependent.
As I will explain in the next section, Geometric Algebra can provide
more powerful and geometrically significant representations for subspaces
using GA's Blades. GA-based formulations are found to be significantly
more compact and basis-independent for many applications.

While the set intersection $U\cap W$ of two subspaces $U,W\leq V$
is also a subspace in $V$, their set union $U\cup W$ is not guaranteed
to be a linear space. An analogous operation to set union that guarantees
a subspace result is called the sum of subspaces defined as $W+U=\{x::x=w+u;\,\,w\in W,u\in V\},\,\,W,V\leq V$.
Having a set of mutually disjoint subspaces $W_{1},W_{2},\cdots,W_{k}\leq V$
(i.e. $W_{i}\cap W_{j}=\{0\}\,\,\forall i,j=1,2,\cdots k,\,\,i\neq j$)
the subspace sum of $W_{i}$ is called the direct sum of the disjoint
subspaces and is denoted by $\oplus_{i=1}^{k}W_{i}\equiv W_{1}\oplus W_{2}\oplus\cdots\oplus W_{k}$.
The dimension of the direct sum of disjoint subspaces is equal to
the numerical sum of their respective dimensions $dim\left(\oplus_{i=1}^{k}W_{i}\right)=\sum_{i=1}^{k}dim\left(W_{i}\right)$.
We often use this notation to construct a larger linear space, like
the linear Grassmann space of multivectors, out of a number of mutually
disjoint linear spaces. This conceptual construction is metric-independent
and basis-independent.

Another important concept is the orthogonal complement of a metric
subspace $W\leq V$ defined by $W^{\perp}=\{x:\in V:\,y\perp x\,\forall y\in W\}$.
The orthogonal complement of a subspace $W\leq V$ has the following
properties:

\begin{eqnarray}
V & = & W\oplus W^{\perp}\\
x & \perp & y\,\,\forall x\in W,y\in W^{\perp}\\
(W^{\perp})^{\perp} & = & W
\end{eqnarray}

The classical treatment of subspaces in linear algebra mostly deals
with un-oriented subspaces, were a subspace is just a set of vectors
closed under linear combinations. In many practical scientific and
engineering applications, however, we need to distinguish between
two opposite orientations for any subspace. This orientation concept
is particularly useful in applications involving Projective and Cayley-Klein
Geometries \citep{onishchik.2006}. We can mathematically define the
concept of orientation for linear spaces as follows \citep{Shafarevich_2013}:
Let $V$ be a finite-dimensional real linear space and let $\boldsymbol{E}=\left\langle e_{1},e_{2},\ldots,e_{n}\right\rangle $
and $\boldsymbol{F}=\left\langle f_{1},f_{2},\ldots,f_{n}\right\rangle $
be two ordered bases for $V$ with a Change of Basis isomorphism $\mathbf{g}:V\rightarrow V$.
The bases $\boldsymbol{E}$ and $\boldsymbol{F}$ are said to have
the same orientation iff $\mathbf{g}$ has a positive determinant;
otherwise they have opposite orientations, meaning that $\mathbf{g}$
involves a geometric reflection. The property of having the same orientation
defines an equivalence relation on the set of all ordered bases for
$V$. There are only two equivalence classes determined by this relation.
An orientation on $V$ is an assignment of $+1$ to one equivalence
class and $-1$ to the other. Blades in Geometric Algebra can naturally
represent oriented subspaces as I will explain later in the next section.

From the previous discussions we can see that matrix algebra is a
good intermediate representation capable of representing metrics,
linear maps, and subspaces; but we need to be extra careful about
the selection of basis and abstract meanings behind matrix operations.
However, GA provides better basis-independent, metric-independent,
and dimension-independent alternatives for studying and extending
oriented linear subspaces and linear maps without the explicit need
to use matrices. Most notably here, GA Blades can naturally represent
not only oriented subspaces, but weighted oriented subspaces as I
will explain in the next section.

\subsection{Space Embeddings}

The abstract concepts I described in earlier subsections are necessary
tools that enable the use of the powerful conceptual idea of Space
Embedding \citep{Timashev_2011}. In the study of 3D Euclidean space,
for example, simple geometric concepts like points, general lines,
and planes can\textquoteright t be mathematically represented as elements
of a 3D linear space; they simply don\textquoteright t satisfy the
abstract axioms of 3D linear spaces. In 1827, August Ferdinand Möbius
introduced homogeneous coordinates, or projective coordinates, to
solve this problem by embedding 3D Euclidean space into a 4D projective
space. Using this embedding we could easily model additional geometric
concepts as 4D vectors and subspaces. This algebraic construction
has greatly impacted many applications in engineering and computer
science including robotics, computer graphics, computer vision, computer-aided
design, and more. By extending this idea to larger dimensions and
using various metrics, we can embed a smaller space of interest, linear
or not, into a larger metric linear space. Then we can use the algebraic
tools of the larger linear space to represent and manipulate the objects
of the smaller space. This is one kind of linearization that scientists
and engineers should exploit more in their work. Expressing this is
possible, in principle, using matrix algebra; but it's much better
to use Geometric Algebra to express Space Embeddings. Many GAs are
already applied for representing mathematical and geometric spaces
in this way including:
\begin{itemize}
\item Among the first, and most important GAs comes the Space-Time Algebra
(STA) \citep{Hestenes_2015,Dressel_2015}, a GA of signature (1,3,0)
that provides a unified, coordinate-free mathematical framework for
both classical and quantum physics. STA is particularly important
for electrical engineers as it combines the electric and magnetic
fields into a single complex and frame-independent bivector field,
and reduces electrodynamics to a single Maxwell equation on multivectors
with explicit kinship to Dirac's equation.
\item The 3D Euclidean GA with metric of signature (3,0,0) is a simple space
to express rotations on homogeneous lines and planes \citep{Dorst.2009,Kanatani.2015}.
The algebra of quaternions is a sub-algebra of this GA.
\item The 4D Homogeneous GA with metric of signature (4,0,0) is a GA extension
of Möbius's homogeneous coordinates mentioned above \citep{Dorst.2009,Kanatani.2015}.
Some of the Euclidean transforms are linear orthogonal maps in this
space, while others are non-orthogonal linear maps. 
\item Most notably, the 5D Conformal GA (CGA) \citep{Dorst.2009,Hildenbrand.2013,Perwass.September2003}
is the most applied GA with too many practical GC applications to
reference here. This space has a metric with the signature (4,1,0).
Some of the objects we can linearize with CGA vectors and subspaces
include spheres, circles, point-pairs, general lines and planes, tangent
lines and planes, and many more. All conformal and similarity transforms
(translations, reflections, rotations, uniform scalings, inversions,
etc.) are linear orthogonal maps in this space. In addition, perspective
projection could be represented using rotations of this space \citep{Goldman_2014}.
\item Projective GA (PGA) is a class of degenerate GAs of signatures (n,0,1)
that provides a powerful efficient model for n-dimensional Euclidean
geometry. \citep{Gunn.2011,Gunn.2011b,Gunn.2016,Gunn.2016b}, especially
for applications in kinematics and rigid body mechanics. For classical
flat euclidean geometry, PGA exhibits distinct advantages over the
alternative approaches. PGA serves also as an ideal stepping-stone
both scientifically and pedagogically to more complex GAs such as
CGA.
\item Recently, the 10D Double Conformal GA (DCGA) with metric of signature
(8,2,0) \citep{Easter_2017} is used to represent points and general
(quartic) Darboux cyclide surfaces in Euclidean 3D space, including
circular tori and all quadrics, and all surfaces formed by their inversions
in spheres. In addition to representing Dupin cyclides, which are
quartic surfaces formed by inversions in spheres of torus, cylinder,
and cone surfaces; and parabolic cyclides which are cubic surfaces
formed by inversions in spheres that are centered on points of other
surfaces. All DCGA entities can be transformed by orthogonal maps
of this space, and reflected in spheres and planes.
\end{itemize}
More GAs are also under study for other purposes \citep{Perwass.2009,Klawitter_2014,Goldman.2015,Dorst.2015,Klawitter.2016}.
The list will probably grow over time requiring efficient software
implementations to computationally realize the potentials of such
GA-based space embeddings.

\section{Geometric Algebras}

The previous discussion about scalars and metric linear spaces introduced
many familiar concepts of linear algebra in a way to be suitable for
constructing Geometric Algebras. The cornerstone in the GA structure
is the concept of Blade and the operation of Outer Product. All concepts
in metric linear spaces can be generalized, in geometrically significant
ways, to handle blades rather than just vectors. Blades are excellent
representations for oriented linear subspaces, and adding them to
metrics and space embeddings gives GA its representational and computational
power. To really understand and appreciate the power of GA as a mathematical
language, a software engineer, as a good Computational Thinker, has
to investigate GA on 3 levels:
\begin{itemize}
\item The abstract level including the defining mathematical axioms and
main algebraic properties. Understanding this level is more important
to GA model developers, but it's also important for GC software engineers
for communicating with the developers of GA models, and for having
a solid mathematical base for GA-based computations. I recommend starting
with simple GA introductions, for example \citep{Dorst.2009,1453854932,Hildenbrand.2013,Kanatani.2015}.
\item The representational level where GC software engineers study examples
for geometric entities and processes they can represent and manipulate
with elements of GA. The GA literature is the best place to develop
a good understanding of GA at this level for any particular fields
of study.
\item The computational level including how to use elements of GA Coordinate
Frames to perform and interpret useful computations. I will provide
more details on this level in section 4. The best way to appreciate
GA on this level is to learn by doing: by selecting some GA software
system, like CLUViz \citep{Perwass.2009,Hildenbrand.2013,cluviz},
and actually computing with and visualizing GA elements. 
\end{itemize}
In this section, I attempt to briefly discuss the mathematical GA
structure through a gradual construction Computational Thinking-based
process. My intention is not to provide much mathematical details,
but to prepare for the discussion about the last computational level
in the following section about GA Coordinate Frames. The mathematics
in this section mostly follows the first 7 chapters of \citep{Dorst.2009}
which contains more mathematical details, discussions, and very good
practical programming examples.

\subsection{Blades and The Outer Product}

In 3D Euclidean space we are taught a number of products involving
Euclidean direction vectors expressed on an orthogonal basis:
\begin{itemize}
\item The scalar multiplication between a scalar and a vector $av$ that
changes the length of the vector $v$ by the scalar factor $a$.
\item The dot product of two vectors $u\cdot v$ that produces a scalar
proportional to cosine the angle between two vectors and their lengths
$u\cdot v=\left\Vert u\right\Vert \left\Vert v\right\Vert \cos\left(\theta\right)$.
\item The cross product of two vectors $u\times v$ that produces a third
vector orthogonal to both vectors with a length proportional to the
sine of the angle between them and their lengths $\left\Vert u\times v\right\Vert =\left\Vert u\right\Vert \left\Vert v\right\Vert \sin\left(\theta\right)$.
\end{itemize}
These operations along with vector addition construct the core of
classical vector algebra \citep{9781439803349}, a basic mathematical
tool in science and engineering historically emerging from a war among
mathematicians and engineers \citep{Chappell_2016}. In mathematics,
however, there are many other products between vectors with significant
geometric interpretations and much better universal representative
capabilities. One such products is called the Exterior or Outer Product
of vectors $x\wedge y$, a cornerstone in the structure of Geometric
Algebra \citep{Dorst.2009,Kanatani.2015}. 

We can use an abstract vector in a n-dimensional linear space with
Euclidean metric (n,0,0) as a representation of an nD Euclidean direction
vector $v$. We can also think about $v$ as a generator of all the
points on a homogeneous line; a line passing through the origin of
Euclidean nD space, parallel to $v$. We can generate all the points
on that line by a scalar multiplication $av$ with any real number
$a$. Then $v$ represents a 1-dimensional subspace having some attitude
in space, a length or weight, and an orientation. The outer product
can algebraically extend this simple geometric construction to more
dimensions. If $x$ and $y$ are two LID vectors, then $x\wedge y$
is a distinct algebraic entity that can represent a Directed Area
in nD Euclidean space. This directed area, called a bivector, has
an attitude determined by the combined attitudes of its 2 vectors,
a weight proportional to their lengths, and an orientation resulting
from their order in the outer product. In addition, this bivector
represents a homogeneous plane in nD Euclidean space. 

We can extend this even more by taking the outer product of $k$ LID
vectors, where $k\leq n$, to obtain a new class of algebraic entities
called Blades. As we can represent the same homogeneous line using
many vectors differing only by their lengths or orientations, we can
represent any k-dimensional subspace using an infinite number of blades
differing only in their weights or orientations. This construction
also has similar representational roles in other metric spaces, but
the metric defines the ``geometric shape'' that the blade represents.
This is where the concept of subspace with non-Euclidean metric differs
from our intuitive flat hyperplane geometry of multi-dimensional Euclidean
spaces. One important characteristic of the outer product is that
it's a metric-independent concept. The algebraic axioms of the outer
product do not depend on the selected metric of the linear space,
only the interpretation of the resulting blades do. 

In other space embeddings, Blades have a surprising capability to
linearly represent many geometric objects we need in practical applications.
For example in the 5D Conformal GA, 4-blades can represent points,
spheres, and general planes. This unifies the geometry of points,
spheres, and planes by algebraically treating a plane as a sphere
with infinite radius, and a point as a sphere with zero radius, enabling
interesting interpolations between them. In addition, we can represent
spheres with positive or negative squared radii using 4-blades in
CGA, i.e. we can represent a sphere with imaginary radius. This adds
more geometric freedom and algebraic consistency to many CGA-based
models by removing many special cases that we need to explicitly address
while developing GA-based geometric models.

Because the outer product is metric-independent, without loss of generality
I will concentrate in this section on the simple real Euclidean linear
spaces $\mathbb{R}^{\mathrm{n}}$ with a basis $\left\langle e_{1},e_{2},\cdots,e_{n}\right\rangle $
as they are isomorphic to all other real Euclidean linear spaces of
the same dimension. The focus is on all subspaces of $\mathbb{R}^{\mathrm{n}}$
of dimensions $k$ where $0\leq k\leq n$. The geometric meaning of
any such subspace is a k-dimensional homogeneous flat (the origin,
a line through the origin, a plane through the origin, etc.) in $\mathbb{R}^{\mathrm{n}}$.
The \textbf{Outer Product} of an ordered set of $k$ LID vectors $\left\langle a_{1},a_{2},\cdots,a_{k}\right\rangle $
is used to define algebraic objects, called k-blades in GA, that can
be used to represent subspaces algebraically with four main characteristics
for each subspace:
\begin{enumerate}
\item The dimensionality of a subspace $k$: This is represented by the
Grade k of the k-blade, the number of LID vectors in the outer product
producing the blade.
\item The attitude of the subspace: this is equivalent to the traditional
un-oriented span in classical linear algebra of the set of vectors
$\{a_{1},a_{2},\cdots,a_{k}\}$.
\item The orientation of the subspace: which is a sign (+1 or -1) associated
with the subspace to define the relative orientation or handedness
of its basis.
\item The weight of the subspace: which is a real number associated with
the attitude (and it also includes the sign i.e. the orientation\textbackslash{}handedness
of the subspace).
\end{enumerate}
The simplest subspace is the 0-dimensional subspace spanned by no
vectors (i.e. it only contains the zero vector) with a corresponding
0-blade that is simply a scalar $\lambda\in\mathbb{R}$; this subspace
will be denoted by $B_{0}^{n}=\mathbb{R}$. Any vector $x\in\mathbb{R}^{\mathrm{n}}$
is a 1-blade by definition and it corresponds to a 1-dimensional subspace
spanned by that vector alone; the space of 1-blades will be denoted
by by $B_{1}^{n}=\mathbb{R}^{\mathrm{n}}$. The set of k-blades for
any value of $k\in\{0,1,\cdots,n\}$ is denoted by $B_{k}^{n}$ and
the set of all blades is denoted here by $B^{n}=\bigcup_{i=0}^{n}B_{i}^{n}$,
the set union of blades of all grades. I will use the notation $grade(A)$
to refer to the grade of a blade $A$. I will denote that a blade
$A$ represents an oriented subspace $W$ as: $A\propto W$. I will
also use the notation $\overleftrightarrow{A}$ to indicate the oriented
subspace spanned by a blade $A$.

The Outer Product is an associative grade-rising bilinear map used
to construct higher-grade blades from lower-grade ones $\wedge:B_{r}^{n}\times B_{s}^{n}\rightarrow B_{r+s}^{n}\,\,,r,s,r+s\in\{0,1,\cdots,n\}$.
The basic properties of the outer product of scalars (0-blades), vectors
(1-blades), bivectors (2-blades), and general k-blades are as follows:

\begin{eqnarray}
\alpha\wedge\beta & = & \alpha\beta\\
\alpha\wedge x=x\wedge\alpha & = & \alpha x\\
x\wedge y & = & -y\wedge x\label{eq:op_asym_vec}\\
X\wedge(Y+Z) & = & X\wedge Y+X\wedge Z\\
A\wedge(B\wedge C) & = & (A\wedge B)\wedge C\\
A\wedge(\alpha B) & = & \alpha(A\wedge B)
\end{eqnarray}

\begin{eqnarray*}
\alpha,\beta & \in & B_{0}^{n};\\
x,y,z & \in & B_{1}^{n};\\
X,Y,Z,(Y+Z) & \in & B_{k}^{n};\\
A,B,C & \in & B^{n}
\end{eqnarray*}

In addition, the anti-symmetry property (\ref{eq:op_asym_vec}) is
a special case of a more general relation $X\wedge Y=(-1)^{rs}Y\wedge X,\,\,X\in B_{r}^{n},Y\in B_{s}^{n}$.
The anti-symmetry property (\ref{eq:op_asym_vec}) leads to the important
relation $x\wedge x=-x\wedge x=0$. This means that the Outer Product
of a collection of linearly dependent vectors is always zero. A a
non-zero blade algebraically encodes the relation of linear independence
among vectors. This is one major difference between the use of Blades
vs matrices for representing linear subspaces that has many conceptual,
representational, and computational consequences. Having the r-blade
$X=x_{1}\wedge x_{2}\wedge\cdots\wedge x_{r}\in B_{r}^{n}$ and the
s-blade $Y\in B_{s}^{n},s\geq r$ we say that $X$ is a sub-blade
of $Y$ denoted by $X\leq Y$ iff $x_{i}\wedge Y=0,\forall i=0,1,\ldots,r$.
If $X\propto\overleftrightarrow{X}$ and $Y\propto\overleftrightarrow{Y}$
then $X\leq Y\Leftrightarrow\overleftrightarrow{X}\leq\overleftrightarrow{Y}$
justifying the use of the same notation. When we write $x\leq X$
we imply that the vectors $x,x_{1},x_{2},\ldots,x_{r}$ are linearly
dependent, meaning that $x\in\overleftrightarrow{X}$; or equivalently
that $\overleftrightarrow{x}\leq\overleftrightarrow{X}$. The pseudo-scalar
of the linear space is defined as $I=e_{1}\wedge e_{2}\wedge\cdots\wedge e_{n}$.
The pseudo-scalar blade represents the full linear space and is of
special importance in may applications because it contains all other
blades $A\leq I\,\forall A\in B^{n}$.

Two additional computationally useful operations can be defined on
blades: the Reverse of a blade $A=a_{1}\wedge a_{2}\wedge\cdots\wedge a_{k}$
defined as $\widetilde{A}\equiv A^{\sim}\equiv a_{k}\wedge a_{k-1}\wedge\cdots\wedge a_{1}=(-1)^{k(k-1)/2}A$
and its Grade Involution $\widehat{A}\equiv A^{\wedge}\equiv\left(-1\right)^{k}A$.
These two operations are used to make many algebraic formulations
involving blades more compact. For linear spaces with other metrics,
the above relations are exactly the same because the Outer Product
is a metric-independent concept, their interpretations are different,
however, from the Euclidean case depending on the used metric.

We must take care that algebraically adding two k-blades can result
in a non-blade; the result can't be expressed as the outer product
of LID vectors, and thus doesn't represent a subspace. This means
that the sets $B_{k}^{n}$ are not linear spaces, neither is their
union $B^{n}$. I will come back to this new algebraic entity when
discussing multivectors later. Because blades algebraically represent
subspaces, we can generalize operations such as the inner product
and linear maps to take blades rather than only vectors. This is the
next step in constructing the full GA mathematical structure.

\subsection{Generalizing the Inner Product\label{subsec:Metric-Products-on}}

In nD Euclidean spaces, we can define useful geometric operations
on vectors using the inner product. For example the squared length
of a vector $\left\Vert x\right\Vert =x\cdot x$ and the angle between
two vectors $\cos(\theta)=u\cdot v\diagup\left(\left|u\right|\left|v\right|\right)$.
We can extend the bilinear form of any metric linear space to operate
on k-blades of any grade, not just vectors. We can use this extended
bilinear form as a product to define similar geometrically significant
operations for higher-grade blades. This product is called the \textbf{Scalar
Product} of blades \citep{Dorst.2002,Dorst.2009}. The scalar product
can be defined as follows:

\begin{eqnarray}
*:B_{k}^{n}\times B_{k}^{n} & \rightarrow & B_{0}^{n}\nonumber \\
\alpha*\beta & = & \alpha\beta,\nonumber \\
where\,\,\alpha,\beta & \in & B_{0}^{n}\nonumber \\
X*Y & = & (-1)^{k(k-1)/2}\left|\begin{array}{cccc}
\mathbf{B}(x_{1},y_{1}) & \mathbf{B}(x_{1},y_{2}) & \cdots & \mathbf{B}(x_{1},y_{k})\\
\mathbf{B}(x_{2},y_{1}) & \mathbf{B}(x_{2},y_{2}) & \cdots & \mathbf{B}(x_{2},y_{k})\\
\vdots & \vdots & \ddots & \vdots\\
\mathbf{B}(x_{k},y_{1}) & \mathbf{B}(x_{k},y_{2}) & \cdots & \mathbf{B}(x_{k},y_{k})
\end{array}\right|\nonumber \\
 & = & \left|\begin{array}{cccc}
x_{1}\cdot y_{k} & x_{1}\cdot y_{k-1} & \cdots & x_{1}\cdot y_{1}\\
x_{2}\cdot y_{k} & x_{2}\cdot y_{k-1} & \cdots & x_{2}\cdot y_{1}\\
\vdots & \vdots & \ddots & \vdots\\
x_{k}\cdot y_{k} & x_{k}\cdot y_{k-1} & \cdots & x_{k}\cdot y_{1}
\end{array}\right|\\
where\,\,X & = & x_{1}\wedge x_{2}\wedge\cdots\wedge x_{k},Y=y_{1}\wedge y_{2}\wedge\cdots\wedge y_{k}\nonumber \\
X*Y & = & 0\,\,{\textstyle otherwise}\nonumber 
\end{eqnarray}

From the symmetry of the definition we can deduce the following property:
$A*B=B*A=\widetilde{A}*\widetilde{B}$. Using the scalar product we
can extend the norm of vectors to a k-blade $A$ as: $\left\Vert A\right\Vert =A*\widetilde{A}$
and define $\left|A\right|=\sqrt{A*\widetilde{A}}$ but only if $A*\widetilde{A}\geq0$.
A blade with zero norm is called a null blade. In nD Euclidean spaces
this norm is equal to the squared area of of 2-blades, the squared
volume of 3-blades, etc. In addition, the angle $\theta$ between
two non-zero Euclidean k-blades $A,B$ of the same grade $k$ can
be defined as $\cos(\theta)=\dfrac{A*\widetilde{B}}{\left|A\right|\left|B\right|}$.
Reinterpreting a zero cosine within this larger context, it either
means that two blades are geometrically perpendicular in the usual
sense (i.e. it takes a right turn to align them); or that they are
algebraically orthogonal in the sense of being independent; i.e.,
not having enough in common in terms of dimension or attitude such
that there is no single rotation with any angle that can make them
identical. For two blades of different grades, the scalar product
has a zero value by definition; it can only relate subspaces of the
same dimension. 

To compare subspaces of different dimensions another bilinear product
is required that should be universally applicable to all blades. The
\textbf{Left Contraction} of blades \citep{Dorst.2002,Dorst.2009}
is one such product having geometrically significant interpretations.
The Left Contraction Product is denoted by $A\rfloor B$ and pronounced
``$A$ contracted on $B$'' where $\rfloor:B_{r}^{n}\times B_{s}^{n}\rightarrow B_{s-r}^{n}\,\,,r,s,s-r\in\{0,1,\cdots,n\}$
is a grade-lowering bilinear map on blades. This product was introduced
by Lounesto as the adjoint of the Outer Product under the extended
bilinear form expressed here as the Scalar Product \citep{Lounesto_1993}.
The Left Contraction Product is bilinear and distributive over addition,
but not associative; this is apparent from comparing the grade of
$(A\rfloor B)\rfloor C$ and $A\rfloor(B\rfloor C)$ that are generally
not equal. The Left Contraction is identical to the Scalar Product
of two same-grade blades $A\rfloor B=A*B,\,\,\forall A,B\in B_{k}^{n}$.
Having $A\propto\overleftrightarrow{A},\,B\propto\overleftrightarrow{B},\,A\in B_{r}^{n},B\in B_{s}^{n}$,
the geometric meaning of $A\rfloor B$ is the (s-r)-blade $C\propto\overleftrightarrow{B}\cap(\overleftrightarrow{A})^{\perp}$.
If the subspace $\overleftrightarrow{C}=\overleftrightarrow{B}\cap(\overleftrightarrow{A})^{\perp}$has
a dimension other than $s-r$ the result of $A\rfloor B$ is considered
zero by definition to preserve its linearity. A constructive explicit
definition of the left contraction is as follows \citep{Dorst.2009}:

\begin{eqnarray}
\alpha\rfloor\beta & = & \alpha\beta\\
\alpha\rfloor A & = & \alpha A\\
A\rfloor B & = & 0,\,\,grade(A)>grade(B)\\
a\rfloor b & = & \mathbf{B}(a,b)=a\cdot b\\
a\rfloor(B\wedge C) & = & (a\rfloor B)\wedge C+(-1)^{grade(B)}B\wedge(a\rfloor C)\\
(A\wedge B)\rfloor C & = & A\rfloor(B\rfloor C)\label{eq:dual_eqn1}\\
\alpha,\beta & \in & B_{0}^{n},\nonumber \\
A,B,C & \in & B^{n}\nonumber 
\end{eqnarray}

The relation (\ref{eq:dual_eqn1}) is valid for any three blades $A,B,C$
whereas the following relation of the three blades is only valid under
a certain condition:

\begin{equation}
(A\rfloor B)\rfloor C=A\wedge(B\rfloor C),\,\,A\leq C\label{eq:dual_eqn_2}
\end{equation}

Equations (\ref{eq:dual_eqn1}) and (\ref{eq:dual_eqn_2}) are called
the duality formulas that link the Outer and Contraction products
on blades. One more useful property of the contraction is given by:

\begin{eqnarray}
x\rfloor(a_{1}\wedge a_{2}\wedge\cdots\wedge a_{k}) & = & \sum_{i=1}^{k}a_{1}\wedge a_{2}\wedge\cdots\wedge(x\rfloor a_{i})\wedge\cdots\wedge a_{k}\\
\Rightarrow x\rfloor(a\wedge b) & = & (x\cdot a)b-(x\cdot b)a
\end{eqnarray}

Geometrically when $A,B$ are blades, $A\rfloor B$ is another blade
contained in $B$ and perpendicular to $A$ with a norm proportional
to the norms of $A,B,$ and the projection of $A$ on $B$. In addition,
the following relation between a vector and a blade is important:
$x\rfloor A=0\Leftrightarrow x\perp y,\,\,\forall y\leq A$; meaning
that $x\rfloor A=0$ iff $x$ is orthogonal to all vectors contained
in the subspace $\overleftrightarrow{A}$. Another computationally
useful version of the Left Contraction can be defined that is called
the \textbf{Right Contraction} product, denoted by $B\lfloor A$ and
pronounced as ``$B$ contracted by $A$'' where $\lfloor:B_{r}^{n}\times B_{s}^{n}\rightarrow B_{r-s}^{n}\,\,,r,s,r-s\in\{0,1,\cdots,n\}$.
The right contraction is related to the left contraction by:

\begin{eqnarray}
B\lfloor A & = & \left(\widetilde{A}\rfloor\widetilde{B}\right)^{\sim}=(-1)^{a(b+1)}A\rfloor B,\\
 &  & a=grade\left(A\right),b=grade\left(B\right)\nonumber 
\end{eqnarray}

The duality formulas (\ref{eq:dual_eqn1}) and (\ref{eq:dual_eqn_2})
can be written for the right contraction as:

\begin{eqnarray}
C\lfloor\left(B\wedge A\right) & = & \left(C\lfloor B\right)\lfloor A,\,\,\forall A,B,C\in B^{n}\\
C\lfloor\left(B\lfloor A\right) & = & \left(C\lfloor B\right)\wedge A\,\,\forall A,B,C\in B^{n},A\leq C
\end{eqnarray}

\subsection{Orthogonality and Duality of Blades}

Any non-null blade $A\in B_{k}^{n},\left\Vert A\right\Vert \neq0$
can have an inverse blade $A^{-1}$ with respect to the left contraction
product (i.e. $A\rfloor A^{-1}=1$) defined as: 
\begin{equation}
A^{-1}=\dfrac{\widetilde{A}}{\left\Vert A\right\Vert }=\dfrac{(-1)^{k(k-1)/2}}{A*\widetilde{A}}A,\,k=grade(A)
\end{equation}
This inverse is not unique with respect to the left contraction, but
is always present for non-null blades. A special case is the inverse
of a non-null vector given by $a^{-1}=\dfrac{a}{\left\Vert a\right\Vert }$.
When combined with the geometric product in the next subsection, this
inverse defines a geometrically meaningful ``division'' by non-null
blades and vectors for the first time. For any blade with unit norm
like the pseudo-scalar of a Euclidean space the inverse of the blade
is its reverse $I^{-1}=I^{\sim},\left\Vert I\right\Vert =1$. For
a mixed-signature metric space with signature $(p,q,0)$ the inverse
of the pseudo-scalar is given by $I^{-1}=(-1)^{q}I^{\sim}$. For degenerate
metric spaces the inverse of the pseudo-scalar is not defined. 

Using the inverse of a blade a very important operation on blades
can be defined that is called the dual of a blade $A\in B_{r}^{n}$
with respect to a larger containing blade $X\in B_{s}^{n},A\leq X$
that is a linear mapping $*:B_{r}^{n}\times B_{s}^{n}\rightarrow B_{s-r}^{n}$
that acts as follows:

\begin{equation}
A^{*X}=A\rfloor X^{-1},\,\,\forall A\leq X
\end{equation}

When the larger blade is the space pseudo-scalar $I$ the dual is
simply written as $A^{*}=A\rfloor I^{-1}$. The geometric meaning
of the dual $A^{*}$ is simply a blade orthogonal to the original
blade $A$ such that they together complete the space; i.e. $A\propto\overleftrightarrow{A}\Leftrightarrow A^{*}\propto(\overleftrightarrow{A})^{\perp}$.
This means that any blade $A\in B_{r}^{n}$ can computationally represent
two subspaces \citep{Dorst.2009,Perwass.2009}: 
\begin{itemize}
\item The $r$-Blade $A$ directly represents the $r$-dimensional subspace
$X=\left\{ x:x\wedge A=0\right\} $; this is denoted here as $A\propto X$.
In this case, the subspace $X$ is called the Outer Product Null Space
(OPNS) of the blade $A$.
\item The $r$-Blade $A$ dually represents the $\left(n-r\right)$-dimensional
subspace $Y=\left\{ y:y\rfloor A=0\right\} $; this is denoted here
as $A\overset{\bot}{\propto}Y$. In this case, the subspace $Y$ is
called the Inner Product Null Space (IPNS) of the blade $A$. 
\end{itemize}
These two representation methods will need special attention when
consistently applying linear maps on subspaces using outermorphisms
of blades in subsection \ref{subsec:outermorphisms}. In 3D Euclidean
spaces we use the IPNS in the form of normal vectors computed from
the cross product. We can then replace and generalize the cross product
using the relation $u\times v=\left(u\wedge v\right)^{\ast}\in B_{n-2}^{n}$.

By applying relation (\ref{eq:dual_eqn_2}), we find that taking the
dual of a blade two times results in the same blade with a weight
change: 

\begin{eqnarray}
(A^{*X})^{*X} & = & (-1)^{s(s-1)/2}\dfrac{1}{\left\Vert X\right\Vert }A\\
 &  & \forall A\in B_{r}^{n},X\in B_{s}^{n},A\leq X\nonumber 
\end{eqnarray}

Another related operation on a blade $A\leq X$ called the un-dualization
of the blade $A$ with respect to the blade $X$ can be defined as
follows:

\begin{equation}
A^{\odot X}=A\rfloor X,\,\,\forall A\leq X
\end{equation}

Applying the un-dualization after the dualization (and similarly applying
the dualization after the un-dualization) results in the original
blade with no weight change: $(A^{*X})^{\odot X}=(A^{\odot X})^{*X}=A$.
Using the duality formulas a duality relation can be found between
the contraction products and the outer product for any two blades:
\begin{equation}
(A\wedge B)^{*X}=A\rfloor B^{*X},\:(A\rfloor B)^{*X}=A\wedge B^{*X}\:\forall A,B\leq X
\end{equation}

A useful application on the concepts in this subsection is the typical
need is to express a vector $x\in\mathbb{R}^{n}$ as a linear combination
of general (i.e. not necessarily orthogonal) basis vectors $\left\langle b_{1},b_{2},\cdots,b_{n}\right\rangle $
\citep{Dorst.2009}. First an association of each basis vector $b_{i}$
with a reciprocal vector is done, defined as $c_{i}=(-1)^{i-1}(b_{1}\wedge b_{2}\wedge\cdots\wedge b_{i-1}\wedge b_{i+1}\wedge\cdots\wedge b_{n})\rfloor I^{-1},\,\,i=1,2,\cdots,n,\,\,I=b_{1}\wedge b_{2}\wedge\cdots\wedge b_{n}$.
The basis $\left\langle b_{1},b_{2},\cdots,b_{n}\right\rangle $ and
$\left\langle c_{1},c_{2},\cdots,c_{n}\right\rangle $ are easy to
be shown mutually orthogonal $b_{i}\cdot c_{j}=\delta_{i}^{j},\,\forall i,j=1,2,\cdots n$.
The geometric meaning of a reciprocal basis vector $c_{i}$ is the
orthogonal complement of the span of all basis vectors except the
basis vector $b_{i}$. To determine the coefficients $x_{i}$ such
that $x=x_{1}b_{1}+x_{2}b_{2}+\cdots x_{n}b_{n}$ the relation $x_{i}=x\cdot c_{i}$
(i.e. $x=\sum_{i=1}^{n}(x\cdot c_{i})b_{i}$) is used. If the linear
space is Euclidean with orthonormal basis then all basis vectors have
a norm of $\left\Vert b_{i}\right\Vert =1$ hence the reciprocal basis
vector $c_{i}$ is the same as the basis vector $b_{i}$. Generally,
two reciprocal basis vectors are not co-linear $b_{i}\wedge c_{i}\neq0$
however the following relation holds: $\sum_{i=1}^{n}b_{i}\wedge c_{i}=0$.

\subsection{Multivectors and The Geometric Product}

Having a mathematical structure consisting of an n-dimensional real
linear space $V$ with basis $\boldsymbol{E}=\left\langle e_{1},e_{2},\cdots,e_{n}\right\rangle $,
and associated bilinear form $\mathbf{B}$ with signature $(p,q,r)$,
up until this point we can perform the following algebraic operations
using the scalars and vectors of this structure:
\begin{enumerate}
\item Create vectors using linear combinations of other vectors. This involves
the operations of scalar multiplication and vector addition. We can
also represent any vector as a linear combination of the basis vectors
$e_{i}$.
\item Apply the bilinear form to vectors as an inner product $x\cdot y$
to get a geometrically significant scalar value.
\item Construct k-blades from LID vectors using the outer product where
$k=0,1,\ldots,n$. 
\item Extend the bilinear form to blades as a Scalar Product $s=A\ast B$
having a geometrically significant scalar value $s$.
\item Apply the Left Contraction as a dual operation to the Outer Product
on blades to obtain a geometrically significant blade $C$ from two
blades $C=A\rfloor B$.
\end{enumerate}
What remains to reach the full Geometric Algebra structure is the
following steps. These steps are easy to formulate mathematically,
but they create the surprisingly elegant and universal GA structure:
\begin{enumerate}
\item Create a total of $2^{n}$different Basis Blades by taking all possible
non-zero outer products of the basis vectors in $\boldsymbol{E}$.
\item Create linear combinations of the basis blades to get new algebraic
entities called k-vectors and multivectors. This leads to the construction
of a non-metric graded linear Grassmann Space $\bigwedge^{n}$ from
the base linear space $\mathbb{R}^{n}$.
\item Extend the metric of the base linear space $V$ to act on multivectors.
This leads to a metric graded linear Geometric Algebra $\mathcal{G}^{p,q,r}$. 
\item Define a universal bilinear Geometric Product (GP) between multivectors
based on the Outer Product and the Bilinear Form between vectors.
This product actually contains all other bilinear products as special
cases. Physicists and pure mathematicians usually start with this
step backwards and deduce the other products from the GP. However,
for software developers this construction sequence could be more suitable
for their create\textbackslash{}refactor Computational Thinking mental
process.
\end{enumerate}
\begin{table}
\caption{Example for constructing $n+1$ basis sets for k-vectors $\boldsymbol{E}_{k}^{n}$
from the set of basis vectors $\boldsymbol{E}=\left\langle e_{1},e_{2},\ldots,e_{n}\right\rangle $
using the outer product}
\label{tbl:basis-blades}
\noindent \centering{}%
\begin{tabular}{|c|c|c|c|}
\hline 
Grade & Dimension & Name & Basis Blades\tabularnewline
\hline 
\hline 
0 & 1 & $\boldsymbol{E}_{0}^{4}$ & $\left\langle 1\right\rangle $\tabularnewline
\hline 
1 & 4 & $\boldsymbol{E}_{1}^{4}$ & $\boldsymbol{E}=\left\langle e_{1},e_{2},e_{3},e_{4}\right\rangle $\tabularnewline
\hline 
2 & 6 & $\boldsymbol{E}_{2}^{4}$ & $\left\langle e_{1}\wedge e_{2},e_{1}\wedge e_{3},e_{2}\wedge e_{3},e_{1}\wedge e_{4},e_{2}\wedge e_{4},e_{3}\wedge e_{4}\right\rangle $\tabularnewline
\hline 
3 & 4 & $\boldsymbol{E}_{3}^{4}$ & $\left\langle e_{1}\wedge e_{2}\wedge e_{3},e_{1}\wedge e_{2}\wedge e_{4},e_{1}\wedge e_{3}\wedge e_{4},e_{2}\wedge e_{3}\wedge e_{4}\right\rangle $\tabularnewline
\hline 
4 & 1 & $\boldsymbol{E}_{4}^{4}$ & $\left\langle e_{1}\wedge e_{2}\wedge e_{3}\wedge e_{4}\right\rangle $\tabularnewline
\hline 
\end{tabular}
\end{table}

\textbf{In the first step} of this construction, the 0-grade basis
blade is the scalar $1$ by definition. There are $n$ 1-blades that
are the basis vectors themselves $e_{i}$. We can create $\left(\begin{array}{c}
n\\
2
\end{array}\right)=n\left(n-1\right)$ basis 2-blades (bivectors) using the basis vectors $e_{i}$. Note
that $e_{i}\wedge e_{j}=-e_{j}\wedge e_{i}\:\forall i\neq j$, so
we can only consider one of them to be a basis 2-blade and the other
just one of its scalar multiples. I will select the basis 2-blade
such that $i<j$ to get a canonical ordering of the basis 2-blades
based on the ordering of the basis vectors. Generally, for all k-blades
we can extend this construction to obtain canonically ordered $\left(\begin{array}{c}
n\\
k
\end{array}\right)=\dfrac{n!}{k!\left(n-k\right)!}$ basis k-blades of the form $e_{j_{1}}\wedge e_{j_{2}}\wedge\cdots\wedge e_{j_{k}},\,j_{1}<j_{2}<\cdots<j_{k}$
for each $k=0,1,2,\ldots n$. This leads to a total of $\sum_{k=0}^{n}\left(\begin{array}{c}
n\\
k
\end{array}\right)=2^{n}$ basis blades. I will denote the $n+1$ sets of basis k-blades as$\boldsymbol{E}_{k}^{n}$
and the set of all basis blades as $\boldsymbol{E}^{n}=\bigcup_{k=0}^{n}\boldsymbol{E}_{k}^{n}$.
Table \ref{tbl:basis-blades} shows an example for constructing the
basis blades of the 4D Euclidean Geometric Algebra.

\textbf{Going to the second step}, it is now natural to try to apply
linear combinations to basis blades to get other elements. Taking
a linear combination of basis k-blades in$\boldsymbol{E}_{k}^{n}$
dosn't generally produce a k-blade. For example, the algebraic element
$3e_{1}\wedge e_{2}-2e_{1}\wedge e_{3}=e_{1}\wedge(3e_{2}-2e_{3})$
is a 2-blade since it is the outer product of two vectors $e_{1}$and
$3e_{2}-2e_{3}$, while $e_{1}\wedge e_{2}+e_{3}\wedge e_{4}$ can
never be expressed as an outer product of vectors so it's not a blade.
This new kind of element is actually called a k-vector, or a Homogeneous
Multivector. Such element is a member of the Clifford Algebra \citep{Hestenes.1984}
on which the GA is based. This simply means that basis k-blades span
the $\left(\begin{array}{c}
n\\
k
\end{array}\right)$-dimensional linear space of k-vectors denoted here by $\bigwedge_{k}^{n}$,
so that all k-blades are k-vectors $B_{k}^{n}\subseteq\bigwedge_{k}^{n}$,
but not the other way around. The only 4 values for $k$ where both
k-blades and k-vectors are identical are $k=0,1,n-1,n$. The elements
of these four spaces are called scalars, vectors, pseudo-vectors,
and pseudo-scalars respectively.

We can now complete this step by taking linear combinations of basis
blades of different grades and identifying the zero scalar with all
zero k-vectors as a single algebraic entity for convenience. This
is the most general case by which we get a full $2^{n}$-dimensional
linear space called the Grassmann Space of Multivectors and denoted
by $\bigwedge^{n}=\bigoplus_{k=0}^{n}\bigwedge_{k}^{n}$. In this
way scalars, vectors, k-blades, and k-vectors are all special cases
of these multivectors.

A useful metric-independent operator to define on multivectors is
the Grade Extraction operator $\left\langle \right\rangle _{k}:\bigwedge^{n}\rightarrow\bigwedge_{k}^{n}$
that extracts the k-vector component from any multivector. For example,
if $A=e_{3}+3e_{1}\wedge e_{2}-2e_{3}\wedge e_{4}-e_{1}\wedge e_{2}\wedge e_{3}\wedge e_{4}\in\bigwedge^{4}$
is a multivector, then $\left\langle A\right\rangle _{0}=0$, $\left\langle A\right\rangle _{1}=e_{3}$,
$\left\langle A\right\rangle _{2}=3e_{1}\wedge e_{2}-2e_{3}\wedge e_{4}$,
$\left\langle A\right\rangle _{3}=0$, and $\left\langle A\right\rangle _{4}=-e_{1}\wedge e_{2}\wedge e_{3}\wedge e_{4}$.
If $A\in\bigwedge_{k}^{n}$ is a k-vector then $\left\langle A\right\rangle _{k}=A,\:\left\langle A\right\rangle _{r}=0\,\forall r\neq k$.
This way, we can symbolically express any multivector $A\in\bigwedge^{n}$
as the sum of its k-vectors: $A=\sum_{k=0}^{n}\left\langle A\right\rangle _{k}$.
I will also denote the sum of even grade k-vectors in a multivector
$A$ as $\left\langle A\right\rangle _{even}\equiv\sum_{r}\left\langle A\right\rangle {}_{2r}$,
and the sum of its odd grade k-vectors as $\left\langle A\right\rangle _{odd}\equiv\sum_{r}\left\langle A\right\rangle {}_{2r+1}$
so that any multivector can also be expressed as $A=\left\langle A\right\rangle _{even}+\left\langle A\right\rangle _{odd}$.
If the multivector only contains k-vectors of even grade $A=\left\langle A\right\rangle _{even}$
it is called an even multivector. If it only contains k-vectors of
odd grades $A=\left\langle A\right\rangle _{odd}$ it is called an
odd multivector. We can then define a useful Grade Parity operator
on multivectors as:

\begin{eqnarray*}
grade(A) & = & \begin{cases}
1 & A=\left\langle A\right\rangle _{odd}\\
0 & A=\left\langle A\right\rangle _{even}\\
undefined & otherwise
\end{cases}\\
 &  & \forall A\in\bigwedge^{n}
\end{eqnarray*}

\textbf{Now for the third step} to construct a Geometric Algebra $\mathcal{G}^{p,q,r}$
from a base n-dimensional metric linear space $\mathbb{R}^{n}$ with
signature $(p,q,r)$, we just need to generalize the linear products
and operations of $\mathbb{R}^{n}$ to multivectors to obtain a full
Geometric Algebra $\mathcal{G}^{p,q,r}$ out of the non-metric Grassmann
Space of multivectors $\bigwedge^{n}$ where $n=p+q+r$. Because all
basic algebraic products and operations are linear the generalizations
are straight forward as follows: 

\begin{eqnarray}
\widetilde{A} & = & \sum_{r=0}^{n}(-1)^{r(r-1)/2}\left\langle A\right\rangle {}_{r}\\
\widehat{A} & = & \sum_{r=0}^{n}(-1)^{r}\left\langle A\right\rangle {}_{r}\\
A\wedge B & = & \sum_{r=0}^{n}\sum_{s=0}^{n-r}\left\langle A\right\rangle {}_{r}\wedge\left\langle B\right\rangle {}_{s}\\
A\ast B & = & \sum_{r=0}^{n}\left\langle A\right\rangle {}_{r}\ast\left\langle B\right\rangle {}_{r}\\
A\rfloor B & = & \sum_{s=0}^{n}\sum_{r=0}^{s}\left\langle A\right\rangle {}_{r}\rfloor\left\langle B\right\rangle {}_{s}\\
A\lfloor B & = & \sum_{r=0}^{n}\sum_{s=0}^{r}\left\langle A\right\rangle {}_{r}\lfloor\left\langle B\right\rangle {}_{s}\\
 &  & \forall A,B\in\mathcal{G}^{p,q,r}\nonumber 
\end{eqnarray}

The above relations are mathematically useful, but computationally
inefficient for computing with multivectors. I will give much better
formulations in the section 4 when talking about computing with GA
Coordinate Frames. In addition, not all multivectors have a geometrically
significant meaning in a given problem domain. We must be careful
to clearly distinguish between algebraic computations on multivectors
from the actual geometric meaning they represent. This issue is generally
less sever in GA than in matrix algebra due to the richer and more
geometric significant structure of GA.

\textbf{The fourth and final step} is to define the closed bilinear
universal Geometric Product of multivectors, the following is not
an axiomatic definition, but more like a listing of the main properties
of the GP. First of all, the GP is associative (\ref{eq:gp_associative1}),
bilinear (\ref{eq:gp_bilinear1}, \ref{eq:gp_bilinear2}), and distributive
over addition (\ref{eq:gp_dist1}, \ref{eq:gp_dist2}):

\begin{eqnarray}
X(YZ) & = & (XY)Z\label{eq:gp_associative1}\\
(aX+bY)Z & = & a(XZ)+b(YZ)\label{eq:gp_bilinear1}\\
Z(aX+bY) & = & a(ZX)+b(ZY)\label{eq:gp_bilinear2}\\
(X+Y)Z & = & XZ+YZ\label{eq:gp_dist1}\\
Z(X+Y) & = & ZX+ZY\label{eq:gp_dist2}\\
 &  & \forall X,Y,Z\in\mathcal{G}^{p,q,r},\:a,b\in\mathbb{R}\nonumber 
\end{eqnarray}

On scalars and vectors the GP is defined using the multiplication
of real numbers and the scalar multiplication of vectors and scalars:

\begin{eqnarray}
ab=ba & \equiv & \textnormal{The same as real numbers multiplication}\\
ax=xa & \equiv & \textnormal{The same as scalar multiplication}\\
 &  & \forall x\in\mathbb{R}^{n},\:a,b\in\mathbb{R}\nonumber 
\end{eqnarray}

By assuming an orthonormal basis $\boldsymbol{E}=\left\langle e_{1},e_{2},\cdots,e_{n}\right\rangle $
for $\mathbb{R}^{n}$, i.e. $e_{i}\cdot e_{j}=0\:\forall i\neq j,\:e_{i}^{2}\in\left\{ 1,-1,0\right\} $then
on vectors the GP is defined using the outer and inner products as
follows:

\begin{eqnarray}
xx=x^{2} & \equiv & \mathbf{B}(x,x)=x\cdot x=\left\Vert x\right\Vert \\
xy & = & x\cdot y+x\wedge y\\
e_{i}e_{j} & = & -e_{j}e_{i}\label{eq:gp_basis_vector}\\
 &  & \forall x,y\in\mathbb{R}^{n},e_{i}e_{j}\in\boldsymbol{E},i\neq j\nonumber 
\end{eqnarray}

Using relation (\ref{eq:gp_basis_vector}) we can now compute the
GP of any two basis blades easily. Then we can use the other relations
to compute the GP on general multivectors of any kind as long as they
are expressed on the orthonormal basis $\boldsymbol{E}$. Note that
the GP is niter commutative nor anti-commutative for general multivectors.
With the GP any non-null vector $a\in\mathbb{R}^{n}$ has the unique
inverse: $a^{-1}=\dfrac{1}{\left\Vert a\right\Vert }a$. The inverse
$a^{-1}$ is a vector in the same direction of $a$ but properly scaled
to make $aa^{-1}=1$. We can prove that the main products, with one
vector argument, are related to the GP using the following relations
on Blades, then extend them by linearity to multivectors:

\begin{eqnarray}
v\wedge X & = & \frac{1}{2}(vX+\widehat{X}v)\\
X\wedge v & = & \frac{1}{2}(Xv+v\widehat{X})\\
v\rfloor X & = & \frac{1}{2}(vX-Xv)\\
X\lfloor v & = & \frac{1}{2}(Xv-v\widehat{X})\\
 &  & \forall v\in V,X\in\mathcal{G}^{p,q,r}\nonumber 
\end{eqnarray}

We can also compute the main products on blades using the GP:

\begin{eqnarray}
A\wedge B & = & \left\langle AB\right\rangle _{r+s}\,\,r+s\leq n\\
A\rfloor B & = & \left\langle AB\right\rangle _{s-r}\,\,0\leq s-r\leq n\\
A\lfloor B & = & \left\langle AB\right\rangle _{r-s}\,\,0\leq r-s\leq n\\
 &  & \forall A\in B_{r}^{n},B\in B_{s}^{n}\nonumber \\
A*B & = & \left\langle AB\right\rangle _{0}\\
 &  & \forall A\in B_{k}^{n},B\in B_{k}^{n}\nonumber 
\end{eqnarray}

Then we can use linearity to generalize these relations to multivectors:

\begin{eqnarray}
A\wedge B & = & \sum_{r=0}^{n}\sum_{s=0}^{n-r}\left\langle \left\langle A\right\rangle _{r}\left\langle B\right\rangle _{s}\right\rangle _{r+s}\\
A\rfloor B & = & \sum_{s=0}^{n}\sum_{r=0}^{s}\left\langle \left\langle A\right\rangle _{r}\left\langle B\right\rangle _{s}\right\rangle _{s-r}\\
A\lfloor B & = & \sum_{r=0}^{n}\sum_{s=0}^{r}\left\langle \left\langle A\right\rangle _{r}\left\langle B\right\rangle _{s}\right\rangle _{r-s}\\
A*B & = & \sum_{r=0}^{n}\left\langle \left\langle A\right\rangle _{r}\left\langle B\right\rangle _{r}\right\rangle _{0}\\
 &  & \forall A,B\in\mathcal{G}^{p,q,r}\nonumber 
\end{eqnarray}

These last relations are useful mathematically for expressing the
bilinear products using the GP, but they are also computationally
inefficient. I will explain the more efficient method for computing
the GP and all the bilinear products in section 4.

\subsection{Linear Maps on Multivectors}

The construction of a GA is based on a linear Grassmann Space. When
we use GA to model some practical GC problem, we might need several
GA spaces each representing one aspect of the problem. We could also
need to define several linear maps to transform multivectors between
the GAs. We can always define a general linear map between two Grassmann
Spaces $\mathbf{T}:\bigwedge^{n}\rightarrow\bigwedge^{m},\:\mathbf{T}\left[aX+bY\right]=a\mathbf{T}\left[X\right]+b\mathbf{T}\left[Y\right]\:\forall X,Y\in\bigwedge^{n}$.
Such linear map can have a $2^{m}\times2^{n}$ representation matrix
$\mathbf{M_{T}}$ on two sets of bases blades. This map could transform,
for example, a vector into a 3-vector, or a bivector into a mixed-grade
multivector. Because most GA operations on multivectors are linear
or bilinear, we can exploit this representation for applying many
numerical linear algebra techniques to multivectors using matrix algebra
in the background \citep{Perwass.2009,Sangwine_2016}. The actual
interpretation and possible applications associated with such linear
maps are not discussed here.

A general linear map on multivectors is only required to preserve
linear combinations of multivectors. In many practical GC applications,
however, we need to impose more restrictions on general linear maps.
Some of the most applied restrictions are: 
\begin{itemize}
\item The preservation of the Outer Products $\mathbf{T}\left[A\wedge B\right]=\mathbf{T}\left[A\right]\wedge\mathbf{T}\left[B\right]\:\forall A,B\in\bigwedge^{n}$.
Such linear maps are called Outermorphisms. An important class of
outermorphisms are invertible outermorphisms, which can be used as
Change of Basis Outermorphisms (CBO) between GA Coordinate Frames
as discussed later in subsection \ref{subsec:gacf-components}. 
\item The preservation of the Geometric Products $\mathbf{T}\left[AB\right]=\mathbf{T}\left[A\right]\mathbf{T}\left[B\right]\:\forall A,B\in\mathcal{G}^{p,q,r}$.
These linear maps are called Automorphisms. Every automorphism $\mathbf{T}$
also preserves all bilinear products on multivectors including the
outer and inner products $\mathbf{T}\left[A\star B\right]=\mathbf{T}\left[A\right]\star\mathbf{T}\left[B\right]\:\forall A,B\in\mathcal{G}^{p,q,r}$.
This means that an automorphism is also an outermorphism and an orthogonal
linear map on vectors and multivectors in general.
\end{itemize}

\subsection{Outermorphisms\label{subsec:outermorphisms}}

The concept of a linear map on vectors $\mathbf{f}:\mathbb{R}^{n}\rightarrow\mathbb{R}^{m}$
can be extended to act on a whole subspace $S=span\left(x_{1},x_{2},\ldots,x_{k}\right)\leq\mathbb{R}^{n}$
by applying $\mathbf{f}$ to the spanning vectors of the subspace
and reconstructing the transformed subspace afterwards $\mathbf{f}\left[S\right]=span\left(\mathbf{f}\left[x_{1}\right],\mathbf{f}\left[x_{2}\right],\ldots,\mathbf{f}\left[x_{k}\right]\right)\leq\mathbb{R}^{m}$.
An alternative approach is possible using the algebraic constructions
of GA through extending the linear map to act on arbitrary blades,
by constructing what is called an Outermorphism $\overline{\mathbf{f}}$
based on $\mathbf{f}$ as follows:

\begin{eqnarray}
\overline{\mathbf{f}}:\bigwedge^{n} & \rightarrow & \bigwedge^{m}\\
\overline{\mathbf{f}}\left[a\right] & = & a\\
\overline{\mathbf{f}}\left[x\right] & = & \mathbf{f}\left[x\right]\\
\overline{\mathbf{f}}\left[aX+bY\right] & = & a\overline{\mathbf{f}}\left[X\right]+b\overline{\mathbf{f}}\left[Y\right]\\
\overline{\mathbf{f}}\left[X\wedge Y\right] & = & \overline{\mathbf{f}}\left[X\right]\wedge\overline{\mathbf{f}}\left[Y\right]\\
 &  & \forall a,b\in B_{0}^{n},x\in B_{1}^{n},X,Y\in\bigwedge^{n}\nonumber 
\end{eqnarray}

An extension of a map of ``vectors to vectors'' in this manner to
the whole of the Grassmann Algebra is called extension as a linear
outermorphism, since its last property shows that a morphism (i.e.,
a mapping) is obtained that commutes with the outer product. Outermorphisms
have nice algebraic properties that are essential to their geometrical
usage \citep{Dorst.2009}:
\begin{itemize}
\item Blades Remain Blades: Geometrically, oriented subspaces are transformed
to oriented subspaces of the same grade: $grade(A)=grade(\overline{\mathbf{f}}\left[A\right])\,\forall A\in B^{n}$.
This means that the dimensionality of subspaces do not change under
a linear transformation.
\item Preservation of Factorization. If two blades $A,B$ have a blade $C$
in common then the blades $\overline{\mathbf{f}}\left[A\right],\overline{\mathbf{f}}\left[B\right]$
have $\overline{\mathbf{f}}\left[C\right]$ in common. 
\end{itemize}
The determinant of a linear operator $\mathbf{f}$ is a fundamental
scalar property of $\mathbf{f}$ defined using its outermorphism as:
$\overline{\mathbf{f}}\left[I\right]=det\left(\mathbf{f}\right)I$\footnote{Here I use $det\left(\mathbf{f}\right)$ instead of $det\left(\overline{\mathbf{f}}\right)$
because the determinant is a property of $\mathbf{f}$ that can be
defined through its extension $\overline{\mathbf{f}}$.}. It signifies the change in weight between the pseudo-scalar of the
space $I$ and its transformed version under $\overline{\mathbf{f}}$
which is the original definition of determinants in abstract linear
algebra. Using this definition it is easy to show properties of determinants
of linear transforms such as $det\left(\mathbf{g}\circ\mathbf{f}\right)=det\left(\mathbf{g}\right)det\left(\mathbf{f}\right)$
without using matrices and coordinates as usually done in linear algebra
texts. Another important concept in linear algebra is the adjoint
of a linear operator $\mathbf{f}$ denoted here by $\mathbf{f}^{T}$
. For any linear operator $\mathbf{f}:\mathbb{R}^{n}\rightarrow\mathbb{R}^{n}$
defined on a real linear space $\mathbb{R}^{n}$ having arbitrary
(not necessarily orthogonal) basis $\left\langle b_{1},b_{2},\cdots,b_{n}\right\rangle $
the adjoint operator is defined using the reciprocal basis $\left\langle c_{1},c_{2},\cdots,c_{n}\right\rangle $as: 

\begin{eqnarray}
\mathbf{f}^{T}:\mathbb{R}^{n} & \rightarrow & \mathbb{R}^{n}\nonumber \\
\mathbf{f}^{T}\left[x\right] & = & \sum_{i=1}^{n}(x\cdot\mathbf{f}\left[b_{i}\right])c_{i}\,\forall x\in\mathbb{R}^{n}
\end{eqnarray}

The outermorphism of the adjoint can be constructed as above. The
adjoint outermorphism satisfies the following relations for all blades:

\begin{eqnarray}
\overline{\mathbf{f}}\left[A\right]*B & = & A*\overline{\mathbf{f}}^{T}\left[B\right]\,\,\forall A,B\in B^{n}\\
\left(\overline{\mathbf{f}}^{T}\right)^{T} & = & \overline{\mathbf{f}}\\
\left(\overline{\mathbf{f}}^{T}\right)^{-1} & = & \left(\overline{\mathbf{f}}^{-1}\right)^{T}\equiv\overline{\mathbf{f}}^{-T}
\end{eqnarray}

Applying an outermorphism to the scalar product is simple since it
always produces a scalar: $\overline{\mathbf{f}}[A*B]=A*B$. For the
left contraction product the relation is: $\overline{\mathbf{f}}[A\rfloor B]=\overline{\mathbf{f}}^{-T}[A]\rfloor\overline{\mathbf{f}}[B]$,
and in the case that $\mathbf{f}$ is an orthogonal operator the relation
becomes simpler: $\overline{\mathbf{f}}\left[A\rfloor B\right]=\overline{\mathbf{f}}\left[A\right]\rfloor\overline{\mathbf{f}}\left[B\right]$
because $\overline{\mathbf{f}}^{-T}=\overline{\mathbf{f}}$ in this
case. Actually, for orthogonal isomorphisms $\overline{\mathbf{f}}^{-T}=\overline{\mathbf{f}}$
any bilinear product on multivectors $\star$ satisfies the relation
$\overline{\mathbf{f}}\left[A\star B\right]=\overline{\mathbf{f}}\left[A\right]\star\overline{\mathbf{f}}\left[B\right]$
including the outer and geometric products.

We know that in 3D Euclidean space transforming a normal vector $w=u\times v$
using some linear map $\mathbf{f}$ will generally not preserve its
orthogonality property $\mathbf{f}\left[w\right]\cdot\mathbf{f}\left[u\right]\neq0$,
$\mathbf{f}\left[w\right]\cdot\mathbf{f}\left[v\right]\neq0$ $\forall u,v\in\mathbb{R}^{n}$,
$w=u\times v$. To correctly transform $w$ as a normal vector we
need to use $\mathbf{f}^{-T}\left[w\right]$ not $\mathbf{f}\left[w\right]$.
This is because $w$ is a dual representation for the subspace $span\left(u,v\right)$not
a direct representation like $u\land v$. This idea can be generalized
in GA for any blade $A\overset{\bot}{\propto}X$. When applying an
outermorphism to $A$ we need to use $Y=det(\mathbf{f})\overline{\mathbf{f}}^{-T}\left[A\right]\overset{\bot}{\propto}\mathbf{f}\left[X\right]$,
not the usual $\overline{\mathbf{f}}\left[A\right]$, in order to
ensure the consistency of transforming the represented subspace under
the linear map $\mathbf{f}$. Only for orthogonal outermorphisms $\overline{\mathbf{f}}^{-T}=\overline{\mathbf{f}}$
with $det\left(\mathbf{f}\right)=1$ that we can use $\overline{\mathbf{f}}\left[A\right]\overset{\bot}{\propto}\mathbf{f}\left[X\right]$
for consistently transforming a blade $A\overset{\bot}{\propto}X$.
This also means that only invertible linear maps can be used to consistently
transform blades like $A$. If the linear map is not invertible we
can only consistently transform blades $A\propto X$ but not $B\overset{\bot}{\propto}Y$.
This includes normal vectors in 3D Euclidean geometry as a special
case.

We can write an expression for the inverse of an outermorphism, if
it exists, as follows:

\begin{eqnarray}
\mathbf{\overline{\mathbf{f}}^{-1}}\left[A\right] & = & \dfrac{1}{det\left(\mathbf{f}\right)}\left(\overline{\mathbf{f}}^{T}\left[A^{*}\right]\right)^{\odot}\nonumber \\
 & = & \dfrac{1}{det\left(\mathbf{f}\right)}\overline{\mathbf{f}}^{T}\left[A\rfloor I^{-1}\right]\rfloor I
\end{eqnarray}

Although this expression uses metric-dependent dualities it is actually
a metric-independent expression because the two dualities cancel each
other. Hence any metric can be assumed for computing the inverse outermorphism,
preferably a simple Euclidean metric. In section 4 I will explain
how to represent and compute with outermorphisms on GA Coordinate
Frames using matrices.

\subsection{Representing Orthogonal Operators with Versors}

Using the geometric product of non-null vectors, a definition for
a powerful GA-based representation for linear orthogonal maps can
be made. This representation, alternative to real orthogonal matrices,
is called a Versor. According to the Cartan-Dieudonné Theorem \citep{Gallier_2001},
any orthogonal transformation in $\mathbb{R}^{n}$ is equivalent to
a composition of simple reflections on $(n-1)$-dimensional subspaces.
Algebraically, a reflection of a single vector $a\in\mathbb{R}^{n}$
on a $(n-1)$-dimensional subspace dually represented by a non-null
vector $v\in\mathbb{R}^{n}$ can be defined using the geometric product
as the simple linear expression $-vav^{-1}$. In this expression the
actual norm of $v$ is irrelevant since it is canceled by the inverse
in $v^{-1}$. We can simply extend this to an outermorphism on $r$-blades
as:

\begin{eqnarray}
\mathbf{L}_{v}\left[A\right] & = & \left(-1\right)^{r}vAv^{-1}\\
 & = & v\widehat{A}v^{-1}\\
 &  & \forall A\in B_{r}^{n}\nonumber 
\end{eqnarray}

We can further extend this as a composition of simple reflections
for the blade $A\propto X$ on $k$ $(n-1)$-dimensional subspaces
dually represented by non-null vectors $v_{1},v_{2},\cdots,v_{k}$
can be written as:

\begin{eqnarray}
\mathbf{L}_{V}\left[A\right] & = & \left(-1\right)^{kr}v_{k}\cdots v_{2}v_{1}Av_{1}^{-1}v_{2}^{-1}\cdots v_{k}^{-1}\nonumber \\
 & = & \left(-1\right)^{kr}VAV^{-1}\label{eq:dual-hyperplane-blade-reflection}\\
 &  & V=v_{k}\cdots v_{2}v_{1}\in\mathcal{V},\,A\in B_{r}^{n}\nonumber \\
\Rightarrow\mathbf{L}_{V}\left[a\right] & = & \left(-1\right)^{k}VaV^{-1}\\
 & = & \widehat{V}aV^{-1}\label{eq:dual-hyperplane-vector-reflection}\\
 &  & \forall a\in\mathbb{R}^{n}\nonumber 
\end{eqnarray}

The multivector $V=v_{k}\cdots v_{2}v_{1}$ is called a Versor and
is essentially an even or odd multivector created by the geometric
product of the non-null vectors $v_{i}$. I will denote the set of
even versors as $\mathcal{V}_{+}\subset\mathcal{G}^{p,q,r}$, the
set of odd versors as $\mathcal{V}_{-}\subset\mathcal{G}^{p,q,r}$,
and the set of all versors as $\mathcal{V}=\mathcal{V}_{+}\cup\mathcal{V}_{-}\subset\mathcal{G}^{p,q,r}$.
In addition, an important class of versors is the set of non-null
blades $\mathcal{B}=\left\{ A:A\in B^{n},\left\Vert A\right\Vert \neq0\right\} \subseteq\mathcal{V}$,
as any non-null blade can be expressed as the geometric product of
non-null orthogonal vectors. 

Using this construction we can define a new bilinear product $V\varovee A\equiv\mathbf{L}_{V}\left[A\right]=\left(-1\right)^{kr}VAV^{-1}\,\forall V\in\mathcal{V}^{p,q,r},A\in B_{r}^{n}$
called the \textbf{Versor Product}. For some fixed $V$, the versor
product is an orthogonal outermorphism extending the orthogonal linear
map on vectors in equation (\ref{eq:dual-hyperplane-vector-reflection}).
We can extend the versor product to handle any general multivector
$X=\left\langle X\right\rangle _{even}+\left\langle X\right\rangle _{odd}\in\mathcal{G}^{p,q,r}$
as follows:

\begin{eqnarray}
V\varovee X & = & V\left\langle X\right\rangle _{even}V^{-1}+\left(-1\right)^{k}V\left\langle X\right\rangle _{odd}V^{-1}\\
 & = & \begin{cases}
VXV^{-1} & V\in\mathcal{V}_{+}\\
V\widehat{X}V^{-1} & V\in\mathcal{V}_{-}
\end{cases}
\end{eqnarray}

Versors and the versor product construct a very powerful representational
component of Geometric Algebra. For example, we can use the versor
product to orthogonally transform other orthogonal maps $V\varovee X\in\mathcal{V}\,\forall V,X\in\mathcal{V}$.
Orthogonal maps are themselves objects to be transformed by other
orthogonal maps using versors. We can then create an arbitrary hierarchy
of orthogonal maps acting on subspaces to express a sophisticated
geometric process on subspaces. In addition, this naturally leads
to a powerful algebraic representation for Orthogonal Groups \citep{9780821820193}.

Any even versor $V\in\mathcal{V}_{+}$ represents a rotation, which
is an orthogonal map that has a determinant of $1$ and preserves
orientation (handedness) of a subspace it transforms. Any odd versor
$V\in\mathcal{V}_{-}$ represents an anti-rotation (i.e. a composition
of a rotation and a single reflection), which is an orthogonal map
that has a determinant of $-1$ and changes orientation of a subspace
it transforms. This result is independent of the used metric, basis,
or space dimension. If an orthogonal outermorphism $\mathbf{L}$ is
represented by a versor $V$, the inverse outermorphism $\mathbf{L}^{-1}$is
represented by $V^{-1}$. In addition, the composition of two orthogonal
outermorphisms $\mathbf{L}_{V_{2}},\mathbf{L}_{V_{1}}$ respectively
represented by versors $V_{2},V_{1}$ is represented by the geometric
product of the two versors $\left(\mathbf{L}_{V_{2}}\circ\mathbf{L}_{V_{1}}\right)\left[X\right]\equiv\mathbf{L}_{V_{2}}\left[\mathbf{L}_{V_{1}}\left[X\right]\right]=\mathbf{L}_{V_{2}V_{1}}\left[X\right]$. 

The versor product $V\varovee X$, being both an outermorphism and
an innermorphism, preserves all GA bilinear products $\star$ including
the outer and geometric products:

\begin{eqnarray}
V\varovee\left(aX+bY\right) & = & a\left(V\varovee X\right)+b\left(V\varovee Y\right)\\
V\varovee\left(X\star Y\right) & = & \left(V\varovee X\right)\star\left(V\varovee Y\right)\\
 &  & \forall V\in\mathcal{V},X,Y\in\mathcal{G}^{p,q,r},a,b\in\mathbb{R}\nonumber 
\end{eqnarray}

This is a very important property of the versor product. Any algebraic
construction based on the above operations can be transformed directly
under an orthogonal map in a structure-preserving manner. Meaning
that transforming the components and then creating the structure is
equivalent to creating the structure and then applying the orthogonal
map to the whole geometric structure; may it be an oriented subspace
or an orthogonal map by itself. 

\subsection{Computing with Oriented Subspaces}

The above discussion on versors is based on a single type of reflections:
to reflect an oriented subspace directly represented by some blade
in a $(n-1)$-dimensional subspaces dually represented by a non-null
vector. We can also study reflections of arbitrary oriented subspaces
in other oriented subspaces. We can assume any of the two kinds of
representations for the reflected subspace $W$ and the reflection
subspace $V$ resulting in 4 computational possibilities. The mathematical
details are presented in \citep{Dorst.2009} and I will only show
the final results here. The reflection formulas in the 4 cases take
the general form $\mathbf{F}_{A}\left[X\right]=\left(-1\right)^{s}AXA^{-1}$
where $s$ is an integer dependent on the case and the grades of the
blades $A\in B_{a}^{n}$ and $X\in B_{x}^{n}$ representing $V$ and
$W$ respectively as shown in Table \ref{tbl:subspace-reflections}.
In all 4 cases for a fixed $A$ this expression defines an invertible
outermorphism $\mathbf{F}_{A}\left[X\right]$ on blades that can be
extended to act on general multivectors $X\in\mathcal{G}^{p,q,r}$.
The 3rd case is where we can extend $A$ to be a versor, not just
a non-null blade, and obtain a geometrically significant interpretation
using the versor product and the the Cartan-Dieudonné Theorem. In
addition, the sign factor can be ignored if the orientation of the
resulting subspace is not relevant for a particular problem so we
can just use $AXA^{-1}$ in all 4 cases.

\begin{table}

\caption{Values of the sign factor $s$ in the expression$\left(-1\right)^{s}AXA^{-1}$
used for computing the reflection of an oriented subspace $W$ in
an oriented subspace $V$}
\label{tbl:subspace-reflections}
\noindent \begin{centering}
\begin{tabular}{|c|c|c|c|}
\hline 
Case & Blade Representing $V$ & Blade Representing $W$ & Sign Factor $s$\tabularnewline
\hline 
\hline 
1 & $A\propto V$ & $X\propto W$ & $x\left(a+1\right)$\tabularnewline
\hline 
2 & $A\propto V$ & $X\overset{\bot}{\propto}W$ & $\left(x+1\right)\left(a+1\right)+n-1$\tabularnewline
\hline 
3 & $A\overset{\bot}{\propto}V$ & $X\propto W$ & $xa$\tabularnewline
\hline 
4 & $A\overset{\bot}{\propto}V$ & $X\overset{\bot}{\propto}W$ & $\left(x+1\right)a$\tabularnewline
\hline 
\end{tabular}
\par\end{centering}
\end{table}

We can also use a blade $A\in B_{a}^{n}$ to construct a projection
outermorphisms $\mathbf{P}_{A}\left[X\right]$ using the following
equivalent relations:

\begin{eqnarray}
\mathbf{P}_{A}\left[X\right] & = & \left(-1\right)^{x\left(a+1\right)}A\lfloor\left(X\rfloor A^{-1}\right)\\
 & = & (X\rfloor A)\rfloor A^{-1}\\
 & = & (X\rfloor A)A^{-1}
\end{eqnarray}

In this case the blade $A$ directly represents an oriented subspace
$A\propto V$ on which we can project another subspace $W$ directly
represented by $X\in B_{x}^{n}$. One important difference between
a reflection outermorphism $\mathbf{F}_{A}\left[X\right]$ and a projection
outermorphism $\mathbf{P}_{A}\left[X\right]$ is that $\mathbf{F}_{A}\left[\mathbf{F}_{A}\left[X\right]\right]=X$
(i.e. a double reflection is an identity map) while $\mathbf{P}_{A}\left[\mathbf{P}_{A}\left[X\right]\right]=\mathbf{P}_{A}\left[X\right]$
meaning that applying the same projection is equivalent to a single
projection of the projected subspace. These constructions add more
representational power to blades. A blade can directly or dually represent
a weighted oriented subspace. In addition, a non-null blade $A$ can
represent reflection outermorphisms $\mathbf{F}_{A}\left[X\right]=\left(-1\right)^{s}AXA^{-1}$,
a projection outermorphism $\mathbf{P}_{A}\left[X\right]=(X\rfloor A)A^{-1}$,
a dualization outermorphism $X^{\ast A}=X\rfloor A^{-1}\,\forall X\leq A$,
or an orthogonal outermorphism $\mathbf{L}_{A}\left[X\right]=A\varovee X$.

We can define additional computations on oriented subspaces using
blades. Having two disjoint subspaces $V\cap W=\left\{ \phi\right\} $
directly represented by two blades $A\propto V,B\propto W$ we can
construct the smallest subspace containing both of them, called their
Join, as $A\sqcup B\equiv A\land B\propto V\oplus W$. This is mainly
because $A$ and $B$ have no vectors in common so their outer product
is not zero. If the two subspaces are not disjoint this expression
will give a zero blade and can't be used to compute the geometric
Join of the subspaces.

There exists a related difference between the geometric meaning of
a projection outermorphism and the classical geometric meaning of
projection of subspaces. For example in 3D Euclidean space, if we
geometrically project a homogeneous line on a homogeneous plane the
result is not always a line in the projection plane but sometimes
a point. This degenerate case means that geometric projections do
not preserve the dimensionality of the projected subspace like projection
outermorphisms do. For an important class of geometric operations
on subspaces, outermorphisms are not suitable representations, and
we generally need an algorithmic approach for computing them. Such
geometric operations include:
\begin{itemize}
\item Factoring a given blade $A$ into a set of vectors $v_{i}$ such that
$A=v_{1}\wedge v_{1}\wedge\cdots\wedge v_{r}$. This may be a metric-dependent
or independent operation according to the conditions we assume on
$v_{i}$.
\item Factoring a given versor $V$ into a set of non-null vectors $v_{i}$
such that $V=v_{r}\cdots v_{2}v_{1}$. his is a metric-dependent operation
by nature.
\item Finding the blade $J$ that directly represents the smallest subspace
containing two blades $J=A\sqcup B\propto\{x::x=a+b;\,a\in\overleftrightarrow{A},b\in\overleftrightarrow{B}\}$.
This operation is called the Join of two blades and can be a metric-dependent
or independent operation on subspaces according to the properties
we need $J$ to satisfy.
\item Finding the blade $M$ that directly represents the largest subspace
common to two blades $M=A\sqcap B\propto\overleftrightarrow{A}\cap\overleftrightarrow{B}$.
This operation is called the Meet of two blades and can be a metric-dependent
or independent operation on subspaces according to the properties
we need $M$ to satisfy.
\item Geometrically projecting a subspace on another using the blades they
are represented by.
\end{itemize}
The interested reader can find detailed information on how to algorithmically
perform these subspace computations using GA operations in many sources
including \citep{Bouma_2002,Zaharia.2002b,Dorst.2009,Perwass.2009,Fontijne_2010}.

\section{Computing with GA Coordinate Frames}

When introducing Geometric Algebra to software developers it is much
better to follow a method that builds gradual construction of concepts
as done in the previous two sections. From a computational point of
view, however, the opposite approach is much more suitable. In this
section I explain the mathematics behind practical computing with
a GA Coordinate Frame (GACF). This explanation is an extension and
reformulation of the additive representation of multivectors described
in \citep{Zaharia.2002b,Fontijne.2007,Dorst.2009}. The symbolic computations
layer in GMac \citep{Eid.20160716,gmac} is mainly based on this formulation.

\subsection{Components of a GACF\label{subsec:gacf-components}}

\begin{figure}
\noindent \begin{centering}
\includegraphics[width=5in]{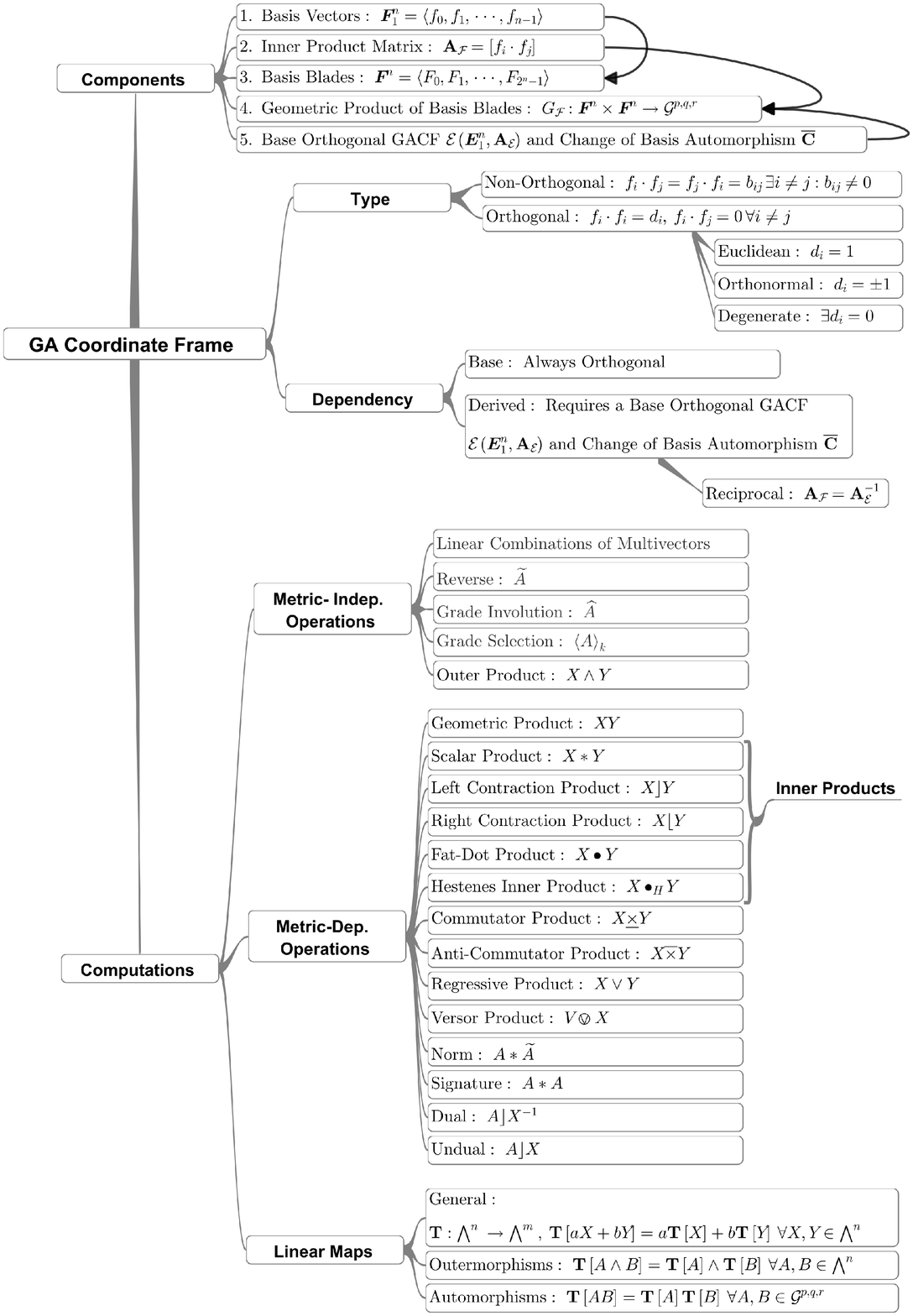}
\par\end{centering}
\caption{Elements of a GA Coordinate Frame}
\label{fig:gacf-components}

\end{figure}

A GACF $\mathcal{F}\left(\boldsymbol{F}_{1}^{n},\mathbf{A}_{\mathcal{F}}\right)$
is the mathematical structure used to define all basic computations
of a Geometric Algebra $\mathcal{G}^{p,q,r}$ in terms of the more
basic scalar coordinates often used to write a program on a computer.
A GACF has contains several components, can be of several types, and
can be used to perform GA computations as illustrated in Figure \ref{fig:gacf-components}.
A GACF can be completely defined using two components:
\begin{enumerate}
\item An ordered set of $n$ basis vectors that determine the dimensionality
of the GACF's base vector space: $\boldsymbol{F}_{1}^{n}=\left\langle f_{0},f_{1},\cdots,f_{n-1}\right\rangle $.
\item A symmetric real bilinear form $\mathbf{B}:\boldsymbol{F}_{1}^{n}\times\boldsymbol{F}_{1}^{n}\rightarrow\mathbb{R},\,\mathbf{B}\left(f_{i},f_{j}\right)=\mathbf{B}\left(f_{j},f_{i}\right)=f_{i}\cdot f_{j}$
to determine the inner product of basis vectors usually given by the
symmetric $n\times n$ bilinear form matrix $\mathbf{A}_{\mathcal{F}}=\left[f_{i}\cdot f_{j}\right]$;
also called the Inner Product Matrix (IPM) of the GACF. According
to the general structure of the IPM $\mathbf{A}_{\mathcal{F}}$ a
GACF $\mathcal{F}$ can be of any of the types listed in Table \ref{tbl:gacf-classes1}.
\end{enumerate}
From these two components, we can automatically construct three additional
ones to serve important purposes for GA computations within the GACF:
\begin{enumerate}
\item The ordered set of $2^{n}$ basis blades of all grades $\boldsymbol{F}^{n}=\left\langle F_{0},F_{1},\cdots,F_{2^{n}-1}\right\rangle $.
This set is automatically determined by the set of basis vectors $\boldsymbol{F}_{1}^{n}$.
This component is independent of the metric represented by $\mathbf{A}_{\mathcal{F}}$.
\item The bilinear multivector coordinates map $G_{\mathcal{F}}:\boldsymbol{F}^{n}\times\boldsymbol{F}^{n}\rightarrow\mathcal{G}^{p,q,r}$
that defines the geometric product of basis blades as a multivector
expressed on the same basis blades $G_{\mathcal{F}}(F_{i},F_{j})=F_{i}F_{j}=\sum_{k=0}^{2^{n}-1}m_{k}F_{k},\,m_{k}\in\mathbb{R}$.
This bilinear map is automatically determined by the set of basis
vectors $\boldsymbol{F}_{1}^{n}$ and the bilinear form $\mathbf{B}$.
\item If the bilinear form is not orthogonal (i.e. $\mathbf{A}_{\mathcal{F}}$
is not diagonal), a base orthogonal GACF $\mathcal{E}\left(\boldsymbol{E}_{1}^{n},\mathbf{A}_{\mathcal{E}}\right)$
of the same dimension is needed, in addition to an orthogonal Change-of-Basis
Matrix (CBM) $\mathbf{C}$. The orthogonal CBM is used to express
the basis vectors of $\mathcal{F}$ as linear combinations of the
basis vectors of $\mathcal{E}$, and defines a Change of Basis Automorphism
(CBA) $\overline{\mathbf{C}}$ that can safely transform linear operations
on multivectors between $\mathcal{E}$ and $\mathcal{F}$. This component
is required for the computation of the geometric product of basis
blades $G_{\mathcal{F}}$ for non-orthogonal bilinear forms. We can
either define $\mathbf{C}$ implicitly from the orthonormal eigen
vectors of $\mathbf{A}_{\mathcal{F}}$, or the user can directly supply
$\mathcal{E}\left(\boldsymbol{E}_{1}^{n},\mathbf{A}_{\mathcal{E}}\right)$
and $\mathbf{C}$ to define the IPM of $\mathcal{F}$. The details
of this component are described in subsections \ref{subsec:derived-gacf}
and \ref{subsec:non-orthogonal-gacf}.
\end{enumerate}

\begin{table}
\caption{Classification of GA Coordinate Frames According to their IPM}
\label{tbl:gacf-classes1}
\noindent \centering{}%
\begin{tabular}{|c|c|}
\hline 
GACF Type & IPM Form\tabularnewline
\hline 
\hline 
\multirow{2}{*}{Euclidean} & Identity matrix\tabularnewline
 & $f_{i}\cdot f_{i}=1$, $f_{i}\cdot f_{j}=0\,\forall i\neq j$\tabularnewline
\hline 
\multirow{2}{*}{Orthonormal} & Invertible, diagonal, with $\pm1$ entries\tabularnewline
 & $f_{i}\cdot f_{i}=\pm1$, $f_{i}\cdot f_{j}=0\,\forall i\neq j$\tabularnewline
\hline 
\multirow{2}{*}{Orthogonal} & Diagonal\tabularnewline
 & $f_{i}\cdot f_{i}=d_{i},\:f_{i}\cdot f_{j}=0\,\forall i\neq j$\tabularnewline
\hline 
\multirow{2}{*}{Degenerate} & Non-invertible, diagonal, with some zeros on diagonal\tabularnewline
 & $f_{i}\cdot f_{i}=d_{i}$$\,\exists d_{i}=0$, $f_{i}\cdot f_{j}=0\,\forall i\neq j$\tabularnewline
\hline 
\multirow{2}{*}{Non-Orthogonal} & Invertible, symmetric, non-diagonal\tabularnewline
 & $f_{i}\cdot f_{j}=f_{j}\cdot f_{i}=b_{ij}\,\exists i\neq j:b_{ij}\neq0$\tabularnewline
\hline 
\end{tabular}
\end{table}

Using these five components any multivector $X$ can be represented
by a column vector of real coefficients $\left[x_{i}\right]_{\mathcal{F}}$
where $X=\sum_{k=0}^{2^{n}-1}x_{k}F_{k},\,x_{k}\in\mathbb{R}$ and
the geometric product of two multivectors $X,Y$ can be easily computed
as: 
\begin{equation}
XY=\sum_{r=0}^{2^{n}-1}\sum_{s=0}^{2^{n}-1}x_{r}y_{s}G_{\mathcal{F}}(F_{r},F_{s})\label{eq:gp-gacf-expr}
\end{equation}

We can then formulate the remaining GA bilinear products using a basis-selection
mechanism from the general geometric product expression (\ref{eq:gp-gacf-expr}).

\subsection{Representing GACF Basis Blades}

Basis vectors and blades are abstract mathematical entities defined
only by their relations to each other. To represent such abstract
entities inside computers we usually use symbolic representations
like assigning a unique ID for each basis blade. We then implement
computational processes that are closely analogous to the abstract
relations between these entities. In order to define the basis blades
$\boldsymbol{F}^{n}=\left\langle F_{0},F_{1},\cdots,F_{2^{n}-1}\right\rangle $
for a GACF of any type, a canonical ID representation is defined based
on the basis vectors $\boldsymbol{F}_{1}^{n}=\left\langle f_{0},f_{1},\cdots,f_{n-1}\right\rangle $.
First we introduce the general Ordered Subset Selection (OSS) operator
$\prod_{\oplus}\left(S,i\right)$ that applies any associative binary
operator $\oplus$ with the identity element $I_{\oplus}$ to a subset
of an ordered set of elements $S=\left\langle s_{0},s_{1},\cdots,s_{k-1}\right\rangle $
selected according to the integer index $i$ as follows: 
\begin{equation}
\prod_{\oplus}\left(S,i\right)=\begin{cases}
I_{\oplus} & ,i=0\\
s_{m} & ,i=2^{m},m\in\{0,1,\cdots,k-1\}\\
s_{i_{1}}\oplus s_{i_{2}}\oplus\cdots\oplus s_{i_{r}} & ,\begin{array}{c}
i=2^{i_{1}}+2^{i_{2}}+\cdots+2^{i_{r}},\\
i_{i}<i_{2}<\cdots<i_{r}
\end{array}
\end{cases}
\end{equation}
The OSS operator basically expresses the integer $i$ as a binary
number $\left(i\right)_{2}$ and selects elements from $S$ based
on the 1s positions in $\left(i\right)_{2}$. The OSS then applies
the associative binary operator $\oplus$ to the selected elements.
Using the OSS operator, we can define the basis blades from basis
vectors as follows:

\begin{eqnarray}
F_{i} & = & \prod_{\wedge}\left(\boldsymbol{F}_{1}^{n},i\right)\\
 & = & \begin{cases}
1 & ,i=0\\
f_{m} & ,i=2^{m},m\in\{0,1,\cdots,n-1\}\\
f_{i_{1}}\wedge f_{i_{2}}\wedge\cdots\wedge f_{i_{r}} & ,\begin{array}{c}
i=2^{i_{1}}+2^{i_{2}}+\cdots+2^{i_{r}},\\
i_{i}<i_{2}<\cdots<i_{r}
\end{array}
\end{cases}\nonumber 
\end{eqnarray}

We can now represent any multivector $M=\sum_{i=0}^{2^{n}-1}m_{i}F_{i}$
using a column vector of real coefficients $[M]_{\mathcal{F}}=\left[\begin{array}{cccc}
m_{0} & m_{1} & \cdots & m_{2^{n}-1}\end{array}\right]^{T}$. This column vector is called the additive representation of (or
the coordinates multivector of) a multivector $M$ on basis $\mathcal{F}$.
The subset selection operator $\prod_{\wedge}\left(\boldsymbol{F}_{1}^{n},k\right)$
defines a bijective map between basis blades and n-bit binary patterns.
The integer $i\in\{0,1,\cdots,2^{n}-1\}$ of the basis blade $F_{i}$
expressed as an n-bit binary number $\left(i\right)_{2}^{n}$ uniquely
defines the structure of the basis blade $F_{k}$ . This n-bit binary
pattern is called the ID of the basis blade $id\left(F_{i}\right)$
and its n-bits binary form is denoted as $id\left(F_{i}\right)_{2}^{n}$
. The grade of the basis blade $g=grade\left(F_{i}\right)$ is then
equal to the number of 1s in $id\left(F_{i}\right)_{2}^{n}$. 

Table \ref{tbl:basis-id-example} illustrates this correspondence
on a 5D linear space with basis vectors $\boldsymbol{F}_{1}^{5}=\left\langle f_{0},f_{1},f_{2},f_{3},f_{4}\right\rangle $.
Any multivector can be stored in computer memory as an array (or perhaps
for efficiency reasons as a dictionary or hash table) of $2^{n}$
scalars representing the coefficients of the basis blades with respect
to the given GACF. A pair (ID, scalar) is called a Term, and represents
a weighted basis blade. A multivector is represented as a sum of terms
with different IDs ranging from $0$ to $2^{n}-1$.

\begin{table}
\caption{Example for representing a basis blade $f_{0}\wedge f_{2}\wedge f_{3}$
using its integer ID of value 13}
\label{tbl:basis-id-example}
\noindent \centering{}$(13)_{10}=(01101)_{2}$$\Longleftrightarrow$%
\begin{tabular}{|c|c|c|c|c|}
\hline 
$2^{4}$ & $2^{3}$ & $2^{2}$ & $2^{1}$ & $2^{0}$\tabularnewline
\hline 
$f_{4}$ & $f_{3}$ & $f_{2}$ & $f_{1}$ & $f_{0}$\tabularnewline
\hline 
\hline 
$0$ & $1$ & $1$ & $0$ & $1$\tabularnewline
\hline 
 & $f_{3}$ & $f_{2}$ &  & $f_{0}$\tabularnewline
\hline 
\end{tabular}$\Longleftrightarrow$$f_{0}\wedge f_{2}\wedge f_{3}=F_{13}$
\end{table}

Another important property is the order of the basis blade among its
$g$-vector basis blades of the same grade $g$ . This property is
called here the Index of the basis blade $index(F_{i})$. In addition,
we can define a useful integer operator called the ``ID from grade-index''
operator $id\left(g,k\right)$ that retrieves the ID of a basis blade
given its grade $g=grade(F_{i})$ and index $k=index\left(F_{i}\right)$:

\begin{eqnarray}
id\left(g,k\right) & \equiv & id\left(F_{i}\right)=i:\\
 &  & g=grade(F_{i}),\,k=index(F_{i})\nonumber 
\end{eqnarray}

The $id\left(g,k\right)$ operator is useful when defining outermorphisms
as I will described later. In addition, we can use the $id\left(g,k\right)$
operator to describe the ordered set of basis k-vectors of the same
grade $g\in\{0,1,\cdots,n\}$ which is a subset of the basis blades
set $\boldsymbol{F}^{n}$ as follows:

\begin{eqnarray}
\boldsymbol{F}_{g}^{n} & = & \left\langle F_{i_{0}},F_{i_{1}},\cdots,F_{i_{r-1}}\right\rangle \subset\boldsymbol{F}^{n}\\
 & = & \left\langle f_{0}^{g},f_{1}^{g},\cdots,f_{r-1}^{g}\right\rangle \\
 &  & i_{k}=id\left(g,k\right)\,\forall k\in\left\{ 0,1,\cdots,r-1\right\} ,r=\left(\begin{array}{c}
n\\
g
\end{array}\right)\nonumber 
\end{eqnarray}

These integer operators create a symbolic metric-independent representation
for basis blades having an important property of being representationally
consistent across multiple metrics and dimensions. Having two basis
sets $\mathcal{F}\left(\boldsymbol{F}_{1}^{m},\mathbf{A}_{\mathcal{F}}\right)$and
$\mathcal{E}\left(\boldsymbol{E}_{1}^{n},\mathbf{A}_{\mathcal{E}}\right)$
for two different GAs of dimensions $m$ and $n$ prescriptively with
$m<n$, we find that $id\left(F_{i}\right)=i=id\left(E_{i}\right)\,\forall i\in\left\{ 0,1,\ldots2^{m-1}\right\} $.
We can directly compute metric-independent and dimension-independent
properties of a basis blade $F_{i}$ only using information about
its ID $i$, grade $g$, and index $k$. For example, we can compute
the $\pm$ signs associated with its reversal $\widetilde{F_{i}}$
and grade involution $\widehat{F_{i}}$ respectively as $sign\left(\widetilde{F_{i}}\right)\equiv\dfrac{\widetilde{F_{i}}}{F_{i}}$=$\left(-1\right)^{g\left(g-1\right)/2}$
and $sign\left(\widehat{F_{i}}\right)\equiv\dfrac{\widehat{F_{i}}}{F_{i}}=\left(-1\right)^{g}$.
We can automatically create a universal lookup table like the following
one to store all these metric independent information for any GACF
of dimension less than or equal to a maximum dimension $n_{max}$.
We can then use this global table for computing such metric-independent
properties of basis blades on any GACF of any metric of dimension
$n\leq n_{max}$.

\begin{table}
\caption{Example for a global lookup table for metric-independent operations
on basis blades of dimensions $\leq4$}

\noindent \centering{}%
\begin{tabular}{|c|c|c|c|c|c|c|}
\hline 
{\footnotesize{}$i=id\left(F_{i}\right)$} & {\footnotesize{}$F_{i}$} & {\footnotesize{}$id\left(F_{i}\right)_{2}^{4}$} & {\footnotesize{}$grade(F_{i})$} & {\footnotesize{}$index(F_{i})$} & {\footnotesize{}$sign\left(\widetilde{F_{i}}\right)$} & {\footnotesize{}$sign\left(\widehat{F_{i}}\right)$}\tabularnewline
\hline 
\hline 
{\footnotesize{}0} & {\footnotesize{}$1$} & {\footnotesize{}$0000$} & {\footnotesize{}0} & {\footnotesize{}0} & {\footnotesize{}$+1$} & {\footnotesize{}$+1$}\tabularnewline
\hline 
{\footnotesize{}1} & {\footnotesize{}$f_{0}$} & {\footnotesize{}$0001$} & {\footnotesize{}1} & {\footnotesize{}0} & {\footnotesize{}$+1$} & {\footnotesize{}$-1$}\tabularnewline
\hline 
{\footnotesize{}2} & {\footnotesize{}$f_{1}$} & {\footnotesize{}$0010$} & {\footnotesize{}1} & {\footnotesize{}1} & {\footnotesize{}$+1$} & {\footnotesize{}$-1$}\tabularnewline
\hline 
{\footnotesize{}3} & {\footnotesize{}$f_{0}\wedge f_{1}$} & {\footnotesize{}$0011$} & {\footnotesize{}2} & {\footnotesize{}0} & {\footnotesize{}$-1$} & {\footnotesize{}$+1$}\tabularnewline
\hline 
{\footnotesize{}4} & {\footnotesize{}$f_{2}$} & {\footnotesize{}$0100$} & {\footnotesize{}1} & {\footnotesize{}2} & {\footnotesize{}$+1$} & {\footnotesize{}$-1$}\tabularnewline
\hline 
{\footnotesize{}5} & {\footnotesize{}$f_{0}\wedge f_{2}$} & {\footnotesize{}$0101$} & {\footnotesize{}2} & {\footnotesize{}1} & {\footnotesize{}$-1$} & {\footnotesize{}$+1$}\tabularnewline
\hline 
{\footnotesize{}6} & {\footnotesize{}$f_{1}\wedge f_{2}$} & {\footnotesize{}$0110$} & {\footnotesize{}2} & {\footnotesize{}2} & {\footnotesize{}$-1$} & {\footnotesize{}$+1$}\tabularnewline
\hline 
{\footnotesize{}7} & {\footnotesize{}$f_{0}\wedge f_{1}\wedge f_{2}$} & {\footnotesize{}$0111$} & {\footnotesize{}3} & {\footnotesize{}0} & {\footnotesize{}$-1$} & {\footnotesize{}$-1$}\tabularnewline
\hline 
{\footnotesize{}8} & {\footnotesize{}$f_{3}$} & {\footnotesize{}$1000$} & {\footnotesize{}1} & {\footnotesize{}3} & {\footnotesize{}$+1$} & {\footnotesize{}$-1$}\tabularnewline
\hline 
{\footnotesize{}9} & {\footnotesize{}$f_{0}\wedge f_{3}$} & {\footnotesize{}$1001$} & {\footnotesize{}2} & {\footnotesize{}3} & {\footnotesize{}$-1$} & {\footnotesize{}$+1$}\tabularnewline
\hline 
{\footnotesize{}10} & {\footnotesize{}$f_{1}\wedge f_{3}$} & {\footnotesize{}$1010$} & {\footnotesize{}2} & {\footnotesize{}4} & {\footnotesize{}$-1$} & {\footnotesize{}$+1$}\tabularnewline
\hline 
{\footnotesize{}11} & {\footnotesize{}$f_{0}\wedge f_{1}\wedge f_{3}$} & {\footnotesize{}$1011$} & {\footnotesize{}3} & {\footnotesize{}1} & {\footnotesize{}$-1$} & {\footnotesize{}$-1$}\tabularnewline
\hline 
{\footnotesize{}12} & {\footnotesize{}$f_{2}\wedge f_{3}$} & {\footnotesize{}$1100$} & {\footnotesize{}2} & {\footnotesize{}5} & {\footnotesize{}$-1$} & {\footnotesize{}$+1$}\tabularnewline
\hline 
{\footnotesize{}13} & {\footnotesize{}$f_{0}\wedge f_{2}\wedge f_{3}$} & {\footnotesize{}$1101$} & {\footnotesize{}3} & {\footnotesize{}2} & {\footnotesize{}$-1$} & {\footnotesize{}$-1$}\tabularnewline
\hline 
{\footnotesize{}14} & {\footnotesize{}$f_{1}\wedge f_{2}\wedge f_{3}$} & {\footnotesize{}$1110$} & {\footnotesize{}3} & {\footnotesize{}3} & {\footnotesize{}$-1$} & {\footnotesize{}$-1$}\tabularnewline
\hline 
{\footnotesize{}15} & {\footnotesize{}$f_{0}\wedge f_{1}\wedge f_{2}\wedge f_{3}$} & {\footnotesize{}$1111$} & {\footnotesize{}4} & {\footnotesize{}0} & {\footnotesize{}$+1$} & {\footnotesize{}$+1$}\tabularnewline
\hline 
\end{tabular}
\end{table}

\subsection{The Geometric Product of Euclidean Basis Blades\label{subsec:euclidean-geometric-product}}

The geometric product of any two vectors $u,v$ is $uv=u\cdot v+u\wedge v$.
For a single vector $u\wedge u=0\Leftrightarrow u^{2}=u\cdot u$.
When the two vectors are orthogonal then $u\cdot v=0\Leftrightarrow uv=u\wedge v=-v\wedge u=-vu$.
A Euclidean GACF $\mathcal{F}\left(\boldsymbol{F}_{1}^{n},\mathbf{A}_{\mathcal{F}}\right)$
has an IPM $\mathbf{A}_{\mathcal{F}}$ equal to the identity matrix
with basis vectors satisfying $f_{i}\cdot f_{i}=1$ and $f_{i}\cdot f_{j}=0\,\forall i\neq j$.
This leads to the geometric product of Euclidean basis vectors satisfying
$f_{i}^{2}=1$ and $f_{i}f_{j}=-f_{j}f_{i}\,\forall i\neq j$. For
such GACF it is straight forward to compute the geometric product
of any two basis blades $G_{\mathcal{F}}(F_{r},F_{s})$ as a signed
basis blade in the form $G_{\mathcal{F}}(F_{r},F_{s})=F_{r}F_{s}=\pm F_{q}$.
We only need to find the value of $q$ and the sign $Sign_{EGP}(r,s)$
associated with the resulting basis blade $F_{q}$ given the two integers
$r,s$. As an example, take the geometric product of two basis blades
$F_{13}F_{19}=(f_{0}\wedge f_{2}\wedge f_{3})(f_{0}\wedge f_{1}\wedge f_{4})=(f_{0}f_{2}f_{3})(f_{0}f_{1}f_{4})$.
We can use the associativity of the geometric product to apply a series
of swaps between basis vectors to reach the canonical form of the
final basis blade as follows: 

\begin{eqnarray*}
F_{q} & = & +\left(f_{0}\wedge f_{2}\wedge f_{3}\right)\left(f_{0}\wedge f_{1}\wedge f_{3}\right)\\
 & = & +f_{0}f_{2}\left(f_{3}f_{0}\right)f_{1}f_{3}\\
 & = & -f_{0}f_{2}\left(f_{0}f_{3}\right)f_{1}f_{3}\\
 & = & -f_{0}\left(f_{2}f_{0}\right)f_{3}f_{1}f_{3}\\
 & = & +f_{0}\left(f_{0}f_{2}\right)f_{3}f_{1}f_{3}\\
 & = & +\left(f_{0}f_{0}\right)f_{2}f_{3}f_{1}f_{3}\\
 & = & +f_{2}f_{3}f_{1}f_{3}\\
 & = & -f_{2}f_{1}f_{3}f_{3}\\
 & = & +f_{1}f_{2}\left(f_{3}f_{3}\right)\\
 & = & +f_{1}\wedge f_{2}=F_{6}
\end{eqnarray*}

Using the corresponding IDs we note that:

\begin{eqnarray*}
id\left(f_{1}\wedge f_{2}\right)_{2}^{4} & = & 0110\\
 & = & 1011\textnormal{ XOR }1101\\
 & = & id\left(f_{0}\wedge f_{1}\wedge f_{3}\right)_{2}^{4}\textnormal{ XOR }id\left(f_{0}\wedge f_{2}\wedge f_{3}\right)_{2}^{4}
\end{eqnarray*}

This is not a coincidence because if the same basis vector $f_{i}$
is present or absent in both input basis blades it will always be
absent in the final basis blade due to the property $f_{i}^{2}=1$,
and if a basis vector is only present in one of the input basis blades
it's always present in the final basis blade. Hence we can find the
ID of the final basis blade $F_{q}$ by a bit-wise XOR operation between
the IDs of the input basis blades $F_{r},F_{s}$: 
\begin{eqnarray}
F_{r}F_{s} & = & Sign_{EGP}\left(r,s\right)F_{q}\\
\left(q\right)_{2}=id\left(F_{q}\right)_{2}^{n} & = & id\left(F_{r}\right)_{2}^{n}\textnormal{ XOR }id\left(F_{s}\right)_{2}^{n}\nonumber \\
 & = & \left(r\right)_{2}^{n}\textnormal{ XOR }\left(s\right)_{2}^{n}\\
 &  & \forall r,s\in\left\{ 0,1,\ldots,2^{n}-1\right\} \nonumber 
\end{eqnarray}

We can compute the sign of the final geometric product term using
Algorithm \ref{alg:egp-sign} or a similar variant.

\begin{algorithm}
\caption{$Sign_{EGP}\left(r,s\right)$: Computes the sign of the geometric
product $F_{r}F_{s}$ of two Euclidean basis blades $F_{r},F_{s}\in\mathcal{G}^{n,0,0}$ }
\label{alg:egp-sign}
\begin{enumerate}
\item Initialize the sign variable $S\leftarrow+1$ and the ID variables
$id_{r}\leftarrow id\left(F_{r}\right)$, $id_{s}\leftarrow id\left(F_{s}\right)$
\item For increasing $i$ from $0$ to $n-1$ do steps 3-6:
\item \quad{}If bit $i$ in $\left(id_{s}\right)_{2}^{n}$ is a $1$ do:
\item \quad{}\quad{}For decreasing $j$ from $n-1$ to $i+1$ do step
5:
\item \quad{}\quad{}\quad{}If bit $j$ in $\left(id_{r}\right)_{2}^{n}$
is a $1$ Then set $S\leftarrow-S$
\item \quad{}\quad{}If bit $i$ in $\left(id_{r}\right)_{2}^{n}$ is a
$1$ Then set it to $0$ Else set it to $1$
\item Return final result in $S$
\end{enumerate}
\end{algorithm}

Using such algorithm, we can construct a Euclidean Geometric Product
Sign lookup table having $2^{n}-1$ rows and $2^{n}-1$ columns where
each cell at row $i$ and column $j$ contains the number $Sign_{EGP}(F_{i},F_{j})$
. Although this table is specific to Euclidean metric of dimension
$n$, we can use it to compute the Euclidean geometric product of
basis blades of any dimension $m\leq n$ because of the universal
property if this method of representation. In addition, we can compute
the geometric product of basis blades having other metrics based on
the signs in this Euclidean table, as I will show shortly. An important
property for $Sign_{EGP}(i,i)$ is:

\begin{eqnarray}
F_{i}^{2} & = & F_{i}F_{i}\nonumber \\
 & = & F_{i}\left(\widetilde{F_{i}}\right)^{\sim}\nonumber \\
 & = & \left(-1\right)^{g(g-1)/2}F_{i}\widetilde{F_{i}}\nonumber \\
 & = & \left(-1\right)^{g(g-1)/2}\nonumber \\
\Rightarrow Sign_{EGP}(i,i) & = & \left(-1\right)^{g(g-1)/2},\,g=grade\left(F_{i}\right)\\
 &  & \forall i\in\left\{ 0,1,\ldots,2^{n}-1\right\} \nonumber 
\end{eqnarray}

\subsection{The Geometric Product of Orthogonal Basis Blades}

An orthogonal GACF $\mathcal{F}\left(\boldsymbol{F}_{1}^{n},\mathbf{A}_{\mathcal{F}}\right)$
has a diagonal IPM $\mathbf{A}_{\mathcal{F}}$ with basis vectors
satisfying $f_{i}\cdot f_{i}=d_{i}$ and $f_{i}\cdot f_{j}=0\,\forall i\neq j$
leading to the geometric product of orthogonal basis vectors satisfying
$f_{i}^{2}=d_{i}$ and $f_{i}f_{j}=-f_{j}f_{i}\,\forall i\neq j$.
The only difference between a Euclidean GACF and an orthogonal GACF
is that the square of a basis vector can be any real number $d_{i}$,
including negative numbers and zero. The same algorithm applied for
a Euclidean GACF can thus be used to deduce a geometric product for
such GACF with a single change to step 5 to become: ``If bit $i$
in $\left(id_{r}\right)_{2}^{n}$ is a $1$ Then set it to $0$ and
set $S\leftarrow d_{i}S$ Else set it to $1$''. We could then create
a similar lookup table for each orthogonal GACF in our problem. There
is a better alternative in this case, however, by using the geometric
product for a Euclidean GACF $\mathcal{E}\left(\boldsymbol{E}_{1}^{n},\mathbf{A}_{\mathcal{E}}\right)$
with the same dimension having basis blades $\boldsymbol{E}^{n}=\left\langle E_{0},E_{1},\cdots,E_{2^{n}-1}\right\rangle $.
If $E_{r}E_{s}=Sign_{EGP}(r,s)E_{k}$ then $F_{r}F_{s}=Sign_{EGP}(r,s)\lambda_{k}F_{k}$
where $\lambda_{k}=\prod\left(\left\langle d_{0},d_{1,}\cdots,d_{n-1}\right\rangle ,k\right)=F_{k}\widetilde{F_{k}}=\left\Vert F_{k}\right\Vert $,
called the signature of $F_{k}$, is the multiplication of all $d_{i}$
having a corresponding $1$-bit in the bit pattern $\left(k\right)_{2}^{n}=id\left(F_{k}\right)_{2}^{n}=id\left(E_{k}\right)_{2}^{n}=\left(r\right)_{2}^{n}\textnormal{ XOR }\left(s\right)_{2}^{n}$.
This leads to a save in memory by only storing $2^{n}$ scalar values
$\lambda_{k}=F_{k}\widetilde{F_{k}},\,k\in\left\{ 0,1,\cdots,2^{n}-1\right\} $
for each orthogonal GACF, then the Euclidean Geometric Product Sign
lookup table is used to compute $F_{r}F_{s}$ as follows: 
\begin{eqnarray}
F_{r}F_{s} & = & Sign_{EGP}\left(r,s\right)\lambda_{k}F_{k}\\
\left(k\right)_{2}^{n} & = & \left(r\right)_{2}^{n}\textnormal{ XOR }\left(s\right)_{2}^{n}\\
 &  & \forall r,s\in\left\{ 0,1,\ldots,2^{n}-1\right\} \nonumber 
\end{eqnarray}

When the orthogonal GACF is degenerate we have some null basis vectors
with $d_{i}=0$ and subsequently we find the basis blade signatures
$\lambda_{k}$ computed from these null basis vectors will also equal
zero. For degenerate orthogonal GACFs we have to be careful when computing
with null basis blades in some GA operations; for example when we
need to divide by the norm of a blade we must take care not to use
null blades.

\subsection{Constructing a Derived GACF\label{subsec:derived-gacf}}

Having a general GACF $\mathcal{E}\left(\boldsymbol{E}_{1}^{n},\mathbf{A}_{\mathcal{E}}\right)$
with basis vectors $\boldsymbol{E}_{1}^{n}=\left\langle e_{0},e_{1},\cdots,e_{n-1}\right\rangle $
we can use an invertible Change-of-Basis Matrix $\mathbf{C}=\left[c_{ij}\right]$
to define a new derived set of basis vectors $\boldsymbol{F}_{1}^{n}=\left\langle f_{0},f_{1},\cdots,f_{n-1}\right\rangle $
for the same linear space as $f_{i}=\sum_{j=0}^{n-1}c_{ij}e_{j}\,\forall i\in\{0,1,\cdots,n-1\}$.
If a vector $x$ is represented on the basis $\boldsymbol{E}_{1}^{n}$
by the column vector $\left[x\right]_{\boldsymbol{E}_{1}^{n}}=\left[\begin{array}{cccc}
x_{0} & x_{1} & \cdots & x_{n-1}\end{array}\right]^{T}$ and on the basis $\boldsymbol{F}_{1}^{n}$ by the column vector $\left[x\right]_{\boldsymbol{F}_{1}^{n}}=\left[\begin{array}{cccc}
y_{0} & y_{1} & \cdots & y_{n-1}\end{array}\right]^{T}$ we find that:

\begin{eqnarray}
\left[x\right]_{\boldsymbol{E}_{1}^{n}} & = & \mathbf{C}^{T}\left[x\right]_{\boldsymbol{F}_{1}^{n}}\\
\Leftrightarrow\left[x\right]_{\boldsymbol{F}_{1}^{n}} & = & \mathbf{C}^{-T}\left[x\right]_{\boldsymbol{E}_{1}^{n}}
\end{eqnarray}

In the special case that $\mathbf{C}$ is orthogonal $\mathbf{C}^{-1}=\mathbf{C}^{T}$
we get $\left[x\right]_{\boldsymbol{F}_{1}^{n}}=\mathbf{C}\left[x\right]_{\boldsymbol{E}_{1}^{n}}$.
The elements of the derived IPM $A_{\mathcal{F}}=\left[f_{i}\cdot f_{j}\right]$
can be easily calculated from the IPM $\mathbf{A}_{\mathcal{E}}=\left[e_{i}\cdot e_{j}\right]$
as follows for any invertible $\mathbf{C}$:

\begin{eqnarray}
f_{i}\cdot f_{j} & = & \left(\sum_{r=0}^{n-1}c_{ir}e_{r}\right)\cdot\left(\sum_{s=0}^{n-1}c_{js}e_{s}\right)\nonumber \\
 & = & \sum_{s=0}^{n-1}\sum_{r=0}^{n-1}c_{ir}c_{js}(e_{r}\cdot e_{s})\nonumber \\
 & = & \sum_{s=0}^{n-1}\left(\sum_{r=0}^{n-1}c_{ir}(e_{r}\cdot e_{s})\right)c_{sj}^{T}\nonumber \\
\Rightarrow\mathbf{A}_{\mathcal{F}} & = & \mathbf{C}\mathbf{A}_{\mathcal{E}}\mathbf{C}^{T}
\end{eqnarray}

Using $\boldsymbol{F}_{1}^{n}$ and $\mathbf{A}_{\mathcal{F}}$ we
can then construct a derived GACF $\mathcal{F}\left(\boldsymbol{F}_{1}^{n},\mathbf{A}_{\mathcal{F}}\right)$
relative to the given base GACF $\mathcal{E}\left(\boldsymbol{E}_{1}^{n},\mathbf{A}_{\mathcal{E}}\right)$
by means of the invertible CBM $\mathbf{C}$. To compute the geometric
product of multivectors on the derived GACF we have 3 separate cases:
\begin{itemize}
\item In the case when $\mathbf{A}_{\mathcal{F}}$ is diagonal then $\mathcal{F}$
is an orthogonal GACF and the geometric product of two multivectors
represented on $\mathcal{F}$ can be computed using the method in
the previous subsection. 
\item When $\mathbf{A}_{\mathcal{F}}$ is not diagonal but the base GACF
$\mathcal{E}$ is orthogonal and $\mathbf{C}$ is an orthogonal CBM
$\mathbf{C}^{-1}=\mathbf{C}^{T}$, the geometric product of the derived
basis blades $\boldsymbol{F}^{n}$ can be computed by extending $\mathbf{P}=\mathbf{C}^{T}=\mathbf{C}^{-1}$
and $\mathbf{P}^{T}=\mathbf{C}^{-T}=\mathbf{C}$ as two adjoint orthogonal
outermorphisms $\overline{\mathbf{P}}$ and $\overline{\mathbf{P}}^{T}$respectively.
These two outermorphisms preserve all bilinear products including
the geometric product. We can safely use $\overline{\mathbf{P}}$
and $\overline{\mathbf{P}}^{T}$ to transform bilinear products of
multivectors back and forth between the base GACF $\mathcal{E}$ and
the derived GACF $\mathcal{F}$. Any bilinear product $\star$ of
two multivectors $X,Y$ can be computed on the derived GACF $\mathcal{F}$
as:
\end{itemize}
\begin{equation}
XY=\mathbf{\overline{P}}\left[\overline{\mathbf{P}}^{T}\left[X\right]\star\overline{\mathbf{P}}^{T}\left[Y\right]\right],\,\overline{\mathbf{P}}=\overline{\mathbf{C}}^{T},\overline{\mathbf{P}}^{T}=\overline{\mathbf{P}}^{-1}\label{eq:derived-bilinear-product}
\end{equation}

\begin{itemize}
\item When $\mathbf{A}_{F}$ is not diagonal and either $\mathcal{E}$ is
not orthogonal or $\mathbf{C}$ is not an orthogonal CBM, another
method for computing the geometric product is needed, which is explained
in the following subsection.
\end{itemize}

\subsection{Constructing a Non-Orthogonal GACF\label{subsec:non-orthogonal-gacf}}

We can directly define a non-orthogonal GACF $\mathcal{F}\left(\boldsymbol{F}_{1}^{n},\mathbf{A}_{\mathcal{F}}\right)$
using a given non-diagonal symmetric real IPM $\mathbf{A}_{\mathcal{F}}$.
For a non-orthogonal GACF $\mathcal{F}$ the geometric product of
any two basis blades is not guaranteed to be a term (i.e. a weighted
basis blade) but is generally a multivector (i.e. the sum of terms
of different basis blades). If we try to make a geometric product
lookup table for such GACF, each cell in the lookup table would then
be a full multivector that may contain up to $2^{n}$ terms. This
is a lot to store in memory for a single GACF; $2^{3n}$ terms many
of which are typically zeros. A better alternative is to use a diagonalization
technique on the IPM $\mathbf{A}_{\mathcal{F}}$ to express the non-orthogonal
GACF as a derived GACF from a base orthogonal GACF $\mathcal{E}\left(\boldsymbol{E}_{1}^{n},\mathbf{A}_{\mathcal{E}}\right)$
with basis vectors $\boldsymbol{E}_{1}^{n}=\left\langle e_{0},e_{1},\cdots,e_{n-1}\right\rangle $.
This is done by finding the IPM $\mathbf{A}_{\mathcal{E}}$ of the
base orthogonal GACF and the orthogonal CBM $\mathbf{C}=\left[c_{ij}\right],\mathbf{C}^{-1}=\mathbf{C}^{T}$
that expresses the basis vectors in $\boldsymbol{F}_{1}^{n}$ as linear
combinations of the orthogonal basis vectors in $\boldsymbol{E}_{1}^{n}$
using $f_{i}=\sum_{j=0}^{n-1}c_{ij}e_{j}\,\forall i\in\left\{ 0,1,\cdots,n-1\right\} $
as explained in the previous subsection. This time we already have
$\mathbf{A}_{\mathcal{F}}$ and we need to compute $\mathbf{A}_{\mathcal{E}}$
and $\mathbf{C}$.

Noting that the IPM $\mathbf{A}_{\mathcal{F}}$ is a symmetric real
matrix, it is easy to find the real eigen values $d_{i}$ and $n$
corresponding orthonormal eigen column vectors $V_{i}$ of $\mathbf{A}_{\mathcal{F}}$
that satisfy $\mathbf{A}_{\mathcal{F}}V_{i}=d_{i}V_{i},\,V_{i}^{T}V_{j}=0\,\forall i,j\in\left\{ 0,1,\cdots,n-1\right\} $.
We can then create an orthogonal matrix $\mathbf{P}=\left[\begin{array}{cccc}
V_{1} & V_{2} & \cdots & V_{n}\end{array}\right],\mathbf{P}^{-1}=\mathbf{P}^{T}$ as a concatenation of the orthonormal column vectors $V_{i}$. The
matrix $\mathbf{A}_{\mathcal{E}}=\mathbf{P}^{T}\mathbf{A}_{\mathcal{F}}\mathbf{P}$
is actually a diagonal matrix containing the eigen values $d_{i}$
on its diagonal. Hence $\mathbf{A}_{\mathcal{E}}$ can be considered
the IPM of a base orthogonal GACF from which we derive the non-orthogonal
GACF $\mathcal{F}\left(\boldsymbol{F}_{1}^{n},\mathbf{A}_{\mathcal{F}}\right)$.
Now we can use equation (\ref{eq:derived-bilinear-product}) to compute
any bilinear product on two multivectors as before. This means that
for each non-orthogonal GACF $\mathcal{F}$ it is necessary to construct
and store the orthogonal outermorphisms $\overline{\mathbf{P}}^{T}$
and $\mathbf{\overline{P}}$ created through an eigen analysis of
$\mathbf{A}_{\mathcal{F}}$.

\subsection{Constructing a Reciprocal GACF}

Having a general non-degenerate GACF $\mathcal{E}\left(\boldsymbol{E}_{1}^{n},\mathbf{A}_{\mathcal{E}}\right)$with
basis vectors $\boldsymbol{E}_{1}^{n}=\left\langle e_{0},e_{1},\cdots,e_{n-1}\right\rangle $
and non-null basis blades, it is possible to create a special type
of derived GACF called the Reciprocal GACF $\mathcal{F}\left(\boldsymbol{F}_{1}^{n},\mathbf{A}_{\mathcal{F}}\right)$
having non-null basis vectors $\boldsymbol{F}_{1}^{n}=\left\langle f_{0},f_{1},\cdots,f_{n-1}\right\rangle $
using the relations \citep{Dorst.2009}:

\begin{eqnarray}
f_{i} & = & (-1)^{i-1}\left(e_{1}\wedge e_{2}\wedge\cdots\wedge e_{i-1}\wedge e_{i+1}\wedge\cdots\wedge e_{n}\right)\rfloor I^{-1}\label{eq:reciprocal_vector}\\
I & = & e_{1}\wedge e_{2}\wedge\cdots\wedge e_{n},\nonumber \\
I^{-1} & = & \dfrac{(-1)^{n(n-1)/2}}{I\widetilde{I}}I\nonumber \\
\Rightarrow f_{i}\cdot e_{j} & = & \delta_{j}^{i},\\
f_{i}\cdot f_{j} & = & e_{i}\cdot e_{j}\\
 &  & \,\,\forall i,j\in\{0,1,\cdots,n-1\}\nonumber \\
\Leftrightarrow\mathbf{A}_{\mathcal{F}} & = & \mathbf{A}_{\mathcal{E}}^{-1}\label{eq:reciprocal-ipm}
\end{eqnarray}

If the base GACF $\mathcal{E}$ is orthogonal the derived reciprocal
GACF $\mathcal{F}$ is also orthogonal and the above relations reduce
to the simpler form:

\begin{eqnarray}
f_{i} & = & \dfrac{1}{e_{i}\cdot e_{i}}e_{i}\,\,\forall i\in\{0,1,\cdots,n-1\}\\
\Leftrightarrow\mathbf{A}_{\mathcal{F}}=\mathbf{A}_{\mathcal{E}}^{-1} & = & \mathbf{A}_{\mathcal{E}}^{-1}
\end{eqnarray}

For a non-orthogonal base GACF $\mathcal{E}$ the reciprocal GACF
$\mathcal{F}$ is also non-orthogonal. We can then use equation (\ref{eq:reciprocal-ipm})
to compute $\mathbf{A}_{\mathcal{F}}$ and then find the orthogonal
CBM $\mathbf{P}$ and continue as described in the previous subsection.

\subsection{Computing Bilinear Products on a GACF\label{subsec:computing-bilinear-products}}

Starting with orthogonal GACFs, any bilinear product $\star$ of two
multivectors $X,Y\in\mathcal{G}^{p,q,r}$ performed on their representations
$\left[X\right]_{\mathcal{F}},\left[Y\right]_{\mathcal{F}}$ in an
orthogonal GACF $\mathcal{F}\left(\boldsymbol{F}_{1}^{n},\mathbf{A}_{\mathcal{F}}\right)$
with basis blades $\boldsymbol{F}^{n}=\left\langle F_{0},F_{1},\cdots,F_{2^{n}-1}\right\rangle $
can be implemented as:

\begin{equation}
X\star Y=\sum_{r=0}^{2^{n}-1}\sum_{s=0}^{2^{n}-1}x_{r}y_{s}\left(F_{r}\star F_{s}\right)
\end{equation}

The goal is to find the value of $F_{r}\star F_{s}$ for all $r,s\in\{0,1,\cdots,2^{n}-1\}$.
Due to the properties of the geometric product on orthogonal frames
and the definitions of the bilinear products, the bilinear product
of any two basis blades $F_{r}\star F_{s}$ is either a zero or a
single term $\lambda_{k}^{\star}F_{k}$, but never more than a single
term. Actually when the value of $F_{r}\star F_{s}=\lambda_{k}^{\star}F_{k}$
the term is equal to the geometric product of the two basis blades
$\lambda_{k}^{\star}F_{k}=G_{\mathcal{F}}\left(F_{r},F_{s}\right)=F_{r}F_{s}$.
Assuming $a=grade\left(F_{r}\right)$ and $b=grade\left(F_{s}\right)$,
the following relations list some useful GA bilinear products and
their relations with the geometric product on the basis blades: 

\textbf{Scalar Product:}

\begin{eqnarray}
F_{r}\ast F_{s} & = & \left\langle F_{r}F_{s}\right\rangle _{0}\\
 & = & \begin{cases}
0 & \left(r\right)_{2}\textnormal{ XOR }\left(s\right)_{2}\neq\left(0\right)_{2}\\
F_{r}F_{s} & otherwise
\end{cases}
\end{eqnarray}

\textbf{Left Contraction Product:}

\begin{eqnarray}
F_{r}\rfloor F_{s} & = & \left\langle F_{r}F_{s}\right\rangle _{b-a}\\
 & = & \begin{cases}
0 & \left(r\right)_{2}\textnormal{ AND NOT }\left(s\right)_{2}\neq\left(0\right)_{2}\\
F_{r}F_{s} & otherwise
\end{cases}
\end{eqnarray}

\textbf{Right Contraction Product:}

\begin{eqnarray}
F_{r}\lfloor F_{s} & = & \left\langle F_{r}F_{s}\right\rangle _{a-b}\\
 & = & \begin{cases}
0 & \left(s\right)_{2}\textnormal{ AND NOT }\left(r\right)_{2}\neq\left(0\right)_{2}\\
F_{r}F_{s} & otherwise
\end{cases}
\end{eqnarray}

\textbf{Fat-Dot Product:}

\begin{eqnarray}
F_{r}\bullet F_{s} & = & \left\langle F_{r}F_{s}\right\rangle _{0}+\left\langle F_{r}F_{s}\right\rangle _{b-a}+\left\langle F_{r}F_{s}\right\rangle _{a-b}\\
 & = & \begin{cases}
0 & a=b,\,\left(r\right)_{2}\textnormal{ XOR }\left(s\right)_{2}\neq\left(0\right)_{2}\\
0 & a<b,\,\left(r\right)_{2}\textnormal{ AND NOT }\left(s\right)_{2}\neq\left(0\right)_{2}\\
0 & a>b,\,\left(s\right)_{2}\textnormal{ AND NOT }\left(r\right)_{2}\neq\left(0\right)_{2}\\
F_{r}F_{s} & otherwise
\end{cases}
\end{eqnarray}

\textbf{Hestenes Inner Product:}

\begin{eqnarray}
F_{r}\bullet_{H}F_{s} & = & \begin{cases}
F_{r}\bullet F_{s} & ab>0\\
0 & ab=0
\end{cases}\\
 & = & \begin{cases}
0 & ab>0,\,a=b,\,\left(r\right)_{2}\textnormal{ XOR }\left(s\right)_{2}\neq\left(0\right)_{2}\\
0 & ab>0,\,a<b,\,\left(r\right)_{2}\textnormal{ AND NOT }\left(s\right)_{2}\neq\left(0\right)_{2}\\
0 & ab>0,\,a>b,\,\left(s\right)_{2}\textnormal{ AND NOT }\left(r\right)_{2}\neq\left(0\right)_{2}\\
0 & ab=0\\
F_{r}F_{s} & otherwise
\end{cases}
\end{eqnarray}

\textbf{Commutator Product:}

\begin{eqnarray}
F_{r}\underline{\times}F_{s} & = & \frac{1}{2}\left(F_{r}F_{s}-F_{s}F_{r}\right)\\
 & = & \begin{cases}
0 & Sign_{EGP}(r,s)=Sign_{EGP}(s,r)\\
F_{r}F_{s} & otherwise
\end{cases}
\end{eqnarray}

\textbf{Anti-Commutator Product:}

\begin{eqnarray}
F_{r}\overline{\times}F_{s} & = & \frac{1}{2}\left(F_{r}F_{s}+F_{s}F_{r}\right)\\
 & = & \begin{cases}
0 & Sign_{EGP}(r,s)\neq Sign_{EGP}(s,r)\\
F_{r}F_{s} & otherwise
\end{cases}
\end{eqnarray}

One exception to this pattern is the \textbf{Outer Product} that is
metric-independent, and can't be computed from the metric-dependent
geometric product. Based on the discussion in section \ref{subsec:euclidean-geometric-product}
we can assume a Euclidean GACF with the same dimension $\mathcal{N}\left(\boldsymbol{N}_{1}^{n},\mathbf{I}_{n}\right)$
with no loss of generality and then compute the outer product $F_{r}\wedge F_{s}$
from the geometric product on the Euclidean GACF $\mathcal{N}$: 

\begin{eqnarray}
F_{r}\wedge F_{s} & = & N_{r}\wedge N_{s}\\
 & = & \left\langle N_{r}N_{s}\right\rangle _{a+b}\\
 & = & \begin{cases}
0 & \left(r\right)_{2}\textnormal{ AND }\left(s\right)_{2}\neq\left(0\right)_{2}\\
N_{r}N_{s} & otherwise
\end{cases}
\end{eqnarray}

Another exception is the \textbf{Regressive Product} that we can compute
given two multivectors $X,Y$ using other bilinear products. This
particular product has several definitions in the literature. The
following is just one of them:

\begin{eqnarray}
X\vee Y & = & \left(X^{\ast}\wedge Y^{\ast}\right)^{\odot}\\
 & = & \left(XI^{-1}\wedge YI^{-1}\right)I
\end{eqnarray}

For a non-orthogonal GACF $\mathcal{F}$ a bilinear product of two
basis blades is not guaranteed to produce a single term, except for
the outer product. All the above computational relations become invalid
in this case. We can use equation (\ref{eq:derived-bilinear-product})
to compute any metric-dependent bilinear product from the two orthogonal
outermorphisms $\overline{\mathbf{P}}^{T}$ and $\mathbf{\overline{P}}$
associated with $\mathcal{F}$.

\subsection{Computing Linear Maps on GACFs}

Any Grassmann Space $\bigwedge^{n}$ is a linear space with $2^{n}$
basis blades. We can define and use general linear maps between two
Grassmann spaces $\mathbf{T}:\bigwedge^{n}\rightarrow\bigwedge^{m}$
and use any given bases $\mathcal{E}\left(\boldsymbol{E}_{1}^{n},\mathbf{A}_{\mathcal{E}}\right)$
and $\mathcal{F}\left(\boldsymbol{F}_{1}^{m},\mathbf{A}_{\mathcal{F}}\right)$
on the two spaces to create a $\left(2^{m}-1\right)\times\left(2^{n}-1\right)$
representation matrix $\mathbf{M_{T}}=\left[\begin{array}{cccc}
M_{0} & M_{1} & \cdots & M_{2^{n}-1}\end{array}\right]$ for $\mathbf{T}$ on the two bases where the column vectors $M_{k}=\left[\mathbf{T}\left[E_{k}\right]\right]_{\mathcal{F}}$
represent the transformed basis blades in $\mathcal{E}$ using the
basis blades in $\mathcal{F}$. We can then transform any multivector
$A=\sum_{i=0}^{2^{n}-1}a_{i}E_{i}\in\mathcal{G}^{p_{1},q_{1},r_{1}}$
by representing it using a column vector $\left[A\right]_{\mathcal{E}}=\left[\begin{array}{cccc}
a_{0} & a_{1} & \cdots & a_{2^{n}-1}\end{array}\right]^{T}$ and computing its transformation using simple matrix multiplication: 

\begin{equation}
\left[\mathbf{T}\left[A\right]\right]_{\mathcal{F}}=\mathbf{M_{T}}\left[A\right]_{\mathcal{E}}\label{eq:general-map-mv}
\end{equation}

A special kind of linear maps on multivectors are the outermorphisms
discussed earlier. We can fully define an outermorphism $\overline{\mathbf{f}}:\bigwedge^{n}\rightarrow\bigwedge^{m}$
as an extension of a linear map $\mathbf{f}:\mathbb{R}^{n}\rightarrow\mathbb{R}^{m}$
using two GACFs $\mathcal{E}\left(\boldsymbol{E}_{1}^{n},\mathbf{A}_{\mathcal{E}}\right)$
and $\mathcal{F}\left(\boldsymbol{F}_{1}^{m},\mathbf{A}_{\mathcal{F}}\right)$
by applying the outer product preservation property to basis vectors
$\overline{\mathbf{f}}[e_{i}\wedge e_{j}]=\overline{\mathbf{f}}\left[e_{i}\right]\wedge\overline{\mathbf{f}}\left[e_{j}\right]=\mathbf{f}\left[e_{i}\right]\wedge\mathbf{f}\left[e_{j}\right]\:\forall e_{i},e_{j}\in\boldsymbol{E}_{1}^{n}$.
I will denote the vectors $\mathbf{f}\left[e_{i}\right]\in W$ as
$h_{i}$ and their ordered set as $H_{1}^{n}=\left\langle h_{0},h_{1},\ldots,h_{n-1}\right\rangle $.
Note that the vectors in $H_{1}^{n}$ are not guaranteed to be LID
or even span the whole of $W$, so the outer product of any subset
of $H_{1}^{n}$ may be a zero blade in $\bigwedge^{m}$. This way,
we can represent an outermorphism using a large sparse $\left(2^{m}-1\right)\times\left(2^{n}-1\right)$
representation matrix $\mathbf{M}_{\overline{\mathbf{f}}}=\left[\begin{array}{cccc}
M_{0} & M_{1} & \cdots & M_{2^{n}-1}\end{array}\right]$ where $M_{k}=\left[\prod_{\wedge}\left(H_{1}^{n},k\right)\right]_{\mathcal{F}}$
and use (\ref{eq:general-map-mv}) for transforming multivectors. 

Computationally, we can exploit the sparsity of $\mathbf{M}_{\overline{\mathbf{f}}}$
to reach more efficient representations for outermorphisms. First
we note that any multivector $h_{i}^{g}=\prod_{\wedge}\left(H_{1}^{n},id\left(g,k\right)\right)$
is actually a blade of grade $g$ in $\bigwedge^{m}$ and generally
only needs $\left(\begin{array}{c}
m\\
g
\end{array}\right)$ non-zero coefficients to be represented as a linear combination of
basis blades of grade $g$ on the basis $\boldsymbol{F}_{g}^{m}$.
In addition, we can express any multivector $A\in\bigwedge^{n}$ using
its k-vectors decomposition: $A=\sum_{g=0}^{n}\left\langle A\right\rangle _{g}$
and transform each k-vector $\left\langle A\right\rangle _{g}=\sum_{k=0}^{r-1}a_{id\left(g,k\right)}E_{id\left(g,k\right)},\,r=\left(\begin{array}{c}
n\\
g
\end{array}\right)$ separately using:

\begin{eqnarray}
\mathbf{\overline{\mathbf{f}}}[\left\langle A\right\rangle _{g}] & = & \sum_{k=0}^{r-1}a_{id\left(g,k\right)}\overline{\mathbf{f}}\left[E_{id\left(g,k\right)}\right]\nonumber \\
 & = & \sum_{k=0}^{r-1}a_{id\left(g,k\right)}h_{k}^{g}
\end{eqnarray}

Using this method we need to create a set of $n+1$ transformation
matrices $\mathbf{M}_{\mathbf{f}}^{g}=\left[\begin{array}{cccc}
M_{0}^{g} & M_{1}^{g} & \cdots & M_{r-1}^{g}\end{array}\right],\,r=\left(\begin{array}{c}
n\\
g
\end{array}\right)$, one matrix per grade $g$, with column vectors $M_{k}^{g}=\left[h_{k}^{g}\right]_{\boldsymbol{F}_{g}^{m}}$
representing the $g$-vectors $h_{k}^{g}$ on the $g$-vectors basis
$\boldsymbol{F}_{g}^{m}$. We can then represent the transformations
of $\left\langle A\right\rangle _{g}$ under $\overline{\mathbf{f}}$
using the following relation, and finally recombine them into $\left[\mathbf{\overline{\mathbf{f}}}[A]\right]_{\mathcal{F}}$:

\begin{eqnarray}
\left[\mathbf{\overline{\mathbf{f}}}[\left\langle A\right\rangle _{g}]\right]_{\boldsymbol{F}_{g}^{m}} & = & \mathbf{M}_{\mathbf{f}}^{g}\left[\left\langle A\right\rangle _{g}\right]_{\boldsymbol{F}_{g}^{m}}\\
 &  & \forall g\in\left\{ 0,1,\ldots,n\right\} \nonumber 
\end{eqnarray}

Because the base linear map on vectors $\mathbf{f}$ has rank $rank_{\mathbf{f}}\leq min\left(m,n\right)$,
we will only need to transform $g$-vectors of grades $0\leq g\leq rank_{\mathbf{f}}\leq min\left(m,n\right)$
because all $h_{k}^{g}=0\:\forall k>rank_{\mathbf{f}}$. This is due
to the guaranteed linear dependence of any set of $k>rank_{\mathbf{f}}$
vectors in $\mathbb{R}^{m}$ that are images of vectors in $\mathbb{R}^{n}$
under $\mathbf{f}$. If we can't compute $rank_{\mathbf{f}}$ in advance
we can simply use $min\left(m,n\right)$ as an upper limit. For each
outermorphism we need to compute and store the matrices $\mathbf{M}_{\mathbf{f}}^{g}$.
This approach is summarized in Algorithm \ref{alg:outermorphism-matrix-construction}.

\begin{algorithm}
\caption{Computes the k-vector transformation matrices $\mathbf{M}_{\mathbf{f}}^{g}$
of an outermorphism $\overline{\mathbf{f}}:\bigwedge^{n}\rightarrow\bigwedge^{m}$
given a $m\times n$ transformation matrix $\mathbf{M}_{\mathbf{f}}$
representing its base linear map $\mathbf{f}$ on two GACFs $\mathcal{E}\left(\boldsymbol{E}_{1}^{n},\mathbf{A}_{\mathcal{E}}\right)$
and $\mathcal{F}\left(\boldsymbol{F}_{1}^{m},\mathbf{A}_{\mathcal{F}}\right)$}
\label{alg:outermorphism-matrix-construction}
\begin{enumerate}
\item Set $\mathbf{M}_{\mathbf{f}}^{0}\leftarrow\mathbf{I_{1}}$, the $1\times1$
identity matrix.
\item Set $\mathbf{M}_{\mathbf{f}}^{1}\leftarrow\mathbf{M}_{\mathbf{f}}$.
\item Either set $K\leftarrow rank\left(\mathbf{M}_{\mathbf{f}}\right)$
or set $K\leftarrow min\left(n,m\right)$
\item Construct $n$ column vector representations $v_{i}^{1}$ on $\mathcal{E}$
using the column vectors of $\mathbf{M}_{\mathbf{f}}$. Each $v_{i}^{1}$
is a sparse $\left(2^{n}-1\right)\times1$ column vector containing
non-zero entries only at rows $1,2,4,8,\ldots,2^{n-1}$.
\item For increasing $g$ from $2$ to $K$ do steps 6-11:
\item \quad{}Set $r_{g}\leftarrow\left(\begin{array}{c}
n\\
g
\end{array}\right)$
\item \quad{}For increasing $k$ from $0$ to $r_{g}-1$ do steps 8-10:
\item \quad{}\quad{}Select any two integers $r,s$ such that $\left(id\left(g,k\right)\right)_{2}=\left(id\left(g-1,r\right)\right)_{2}\textnormal{ OR }\left(id\left(g-1,s\right)\right)_{2}$.
\item \quad{}\quad{}Compute the outer product $v_{k}^{g}=v_{r}^{g-1}\wedge v_{s}^{g-1}$
as discussed in Section \ref{subsec:computing-bilinear-products}. 
\item \quad{}\quad{}Construct the column vector $m_{k}^{g}$ from $\left\langle v_{k}^{g}\right\rangle _{g}$
by selecting coefficients of basis g-blades in $v_{k}^{g}$ into rows
of $m_{k}^{g}$ in their canonical order.
\item \quad{}Set $\mathbf{M}_{\mathbf{f}}^{g}\leftarrow\left[\begin{array}{cccc}
m_{0}^{g} & m_{1}^{g} & \cdots & m_{r_{g}-1}^{g}\end{array}\right]$.
\item Return final result as the matrices $\mathbf{M}_{\mathbf{f}}^{g}$.
\end{enumerate}
\end{algorithm}

If the outermorphism is invertible, all its matrices are also invertible.
This approach has some benefits for representing a related outermorphism
like the inverse $\overline{\mathbf{f}}^{-1}$ or the adjoint $\overline{\mathbf{f}}^{T}$on
the same GACF. Taking the adjoint as an example, we can either compute
$\mathbf{M}_{\mathbf{f}}^{T}$ and apply Algorithm \ref{alg:outermorphism-matrix-construction}
to get the outermorphism matrices, or we could directly apply the
transpose operations to all matrices $\mathbf{M}_{\mathbf{f}}^{g}$
already computed for $\overline{\mathbf{f}}$ to get the outermorphism
matrices $\left(\mathbf{M}_{\mathbf{f}}^{g}\right)^{T}$ for $\overline{\mathbf{f}}^{T}$.
Similar options exist for: 
\begin{itemize}
\item The inverse of an outermorphism $\overline{\mathbf{f}}^{-1}$ and
its adjoint $\overline{\mathbf{f}}^{-T}$ for which we compute the
inverse and transposed inverse of the matrices respectively.
\item An outermorphism that extends a linear map $\mathbf{f}:\mathbb{R}^{n}\rightarrow\mathbb{R}^{m}$
that is a linear combination of other linear maps: $\mathbf{f}\left[x\right]=\sum_{i}a_{i}\mathbf{f}_{i}\left[x\right],$
where $\mathbf{f}_{i}:\mathbb{R}^{n}\rightarrow\mathbb{R}^{m}$ are
linear maps on the same linear spaces. In this case we apply the same
linear combination to the matrices of $\overline{\mathbf{f}}_{i}$.
\item An outermorphism that is the composition of other outermorphisms:
$\mathbf{f}\left[x\right]=\mathbf{f}_{2}\left[\mathbf{f}_{1}\left[x\right]\right]$,
where we use matrix multiplication between the matrices of $\overline{\mathbf{f}}_{i}$.
\end{itemize}
Within the same GACF, any Outer Product-preserving linear operation
on vectors can be converted to an outermorphism matrix representation.
For example expressions like the projection of a vector on a blade
$\mathbf{L}\left[x\right]=\left(x\rfloor B\right)B^{-1}$or the versor
product on a vector $\mathbf{L}\left[x\right]=\left(-1\right)^{grade\left(A\right)}AxA^{-1}$
can be extended as outermorphisms. Given some GACF $\mathcal{E}\left(\boldsymbol{E}_{1}^{n},\mathbf{A}_{\mathcal{E}}\right)$,
we can construct the column vectors $m_{i}$ of a linear transformation
matrix $\mathbf{M}_{\mathbf{L}}$ for any such expression by applying
the expression $\mathbf{L}\left[x\right]$ to the basis vectors of
$\boldsymbol{E}_{1}^{n}$:

\begin{eqnarray}
m_{i} & = & \left[\mathbf{L}\left[e_{i}\right]\right]_{\boldsymbol{E}_{1}^{n}}\\
 &  & \forall i\in\left\{ 0,1,\ldots,n-1\right\} \nonumber 
\end{eqnarray}

We can then construct the matrix representation using Algorithm \ref{alg:outermorphism-matrix-construction}.
For an Automorphism (an orthogonal outermorphism) we can either use
the above outermorphism matrix representation that would then have
orthogonal matrices, or we can use the Versor multivector representation
$A=\prod_{i=1}^{k}a_{i}$ where $a_{i}$ are $k$ non-null vectors,
and the Versor Product $A\ovee X$ described earlier to compute the
automorphism, the latter being more efficient in many cases. If we
have the versor as a multivector $A$ we can find the column vectors
$m_{i}$ of the corresponding linear map representation matrix $\mathbf{M}_{\mathbf{f}}$
using:

\begin{eqnarray}
m_{i} & = & \left(-1\right)^{grade\left(A\right)}\left[Ae_{i}A^{-1}\right]_{\boldsymbol{E}_{1}^{n}}\\
 &  & \forall i\in\left\{ 0,1,\ldots,n-1\right\} \nonumber 
\end{eqnarray}

We can also find the versor multivector $A$ given an orthogonal matrix
that represents an orthogonal linear map. This can be done using Householder
Operators to find the Householder vectors $a_{i}$ \citep{Davis_2006,0521880688,9781107004122}
then compute their geometric product to get the desired versor $A=\prod_{i=1}^{k}a_{i}$.

\section{Summary and Conclusions}

Software developers and engineers are natural Computational Thinkers.
Introducing an elegant and sophisticated mathematical language like
Geometric Algebra to software developers requires initially to focus
on the abstract concepts and their relations more than the mathematics.
To really understand the structure of Geometric Algebra the software
developer should be familiar with some important conceptual abstractions
of metric linear spaces not commonly taught in linear algebra courses.
Only then that the software developer can use GA-related constructs
like the outer product and the contraction to understand the elegant
GA structure and the role of each of its components. Software developers
better learn by doing; they need to watch abstract mathematical ideas
come to life on computer displays. Creating a GA-based software library
is the best way for a software developer to learn the mathematical
details of GA. This article provided a Computational Thinking-based
introduction to Geometric Algebra targeting software developers. The
main three parts of this article introduced concepts of metric linear
systems, and then used them to construct the main structural elements
of GA in the second part. The third part aimed at providing enough
mathematics to implement a GA-based software library either for learning,
prototyping, or production purposes. In addition, the interested reader
can find enough resources in the references for more information on
the concepts and techniques presented in this article.

I believe the future of widely accepting GA as a universal mathematical
language for Geometric Computing depends on how the scientific computing
and software engineering communities appreciate GA as a powerful language
for developing Geometric Computing software systems. Making GA implementations
into valuable and enjoyable software systems for the public domain
is possible only through the efforts of good software developers who
understand and use GA in their own creative Computational Thinking
way. Targeting these communities should be a top priority for the
GA community to gain more popularity for their GA-based models. I
recommend for the GA community to communicate more with software developers
on both academic and practical levels. This would also make the GA
community more aware of the practical problems facing GA-based software
implementations that would require more research into GA-based algorithms
and techniques.

\bibliographystyle{IEEEtran}
\bibliography{GA-Refs}

\end{document}